# The outburst triggered by the Deep Impact collision with Comet Tempel 1

Sergei I. Ipatov [1,a,*] and Michael F. A'Hearn [2]

[1] *Department of Physics, Catholic University of America, Washington DC, 20064, USA*
[2] *Department of Astronomy, University of Maryland, College Park MD, 20740, U.S.A.*

[a] The work was initiated at University of Maryland

**ABSTRACT**

Time variations in velocities and relative amount of observed particles (mainly icy particles with diameter $d$<3 μm) ejected from Comet 9P/Tempel 1 are studied based on analysis of the images made by *Deep Impact* (DI) cameras during the first 13 minutes after the collision of the DI impactor with the comet. Analysis of maxima or minima of plots of the time variations in distances of contours of constant brightness from the place of ejection allowed us to estimate the characteristic velocities of particles at several moments in time $t_e$ of ejection after impact for $t_e$≤115 s. Other approaches for estimates of the velocities were also used. All these estimates are in accordance with the same exponential decrease in velocity. The estimates of time variations in the relative amount of ejected particles were based also on results of the analysis of time variations in the size of the bright region of ejected material. At $t_e$~10 s, the morphology of the ejecta (e.g. the location and brightness of the brightest pixel) changed and the rate $r_{te}$ of ejection of observed material increased. Between 1 and 3 seconds after the impact and between 8 and 60 seconds after the impact, more small bright particles were ejected than expected from crater excavation alone. An outburst triggered by the impact could cause such a difference. The sharp (by a factor of 1.6) decrease in the rate of ejection at 55<$t_e$<72 s could be caused by a decrease in the outburst that began at 10 s. Analysis of observations of the DI cloud and of the outbursts from some comets testifies in favour of the proposition that there can be large cavities, with material under gas pressure, below a considerable fraction of a comet's surface. Internal gas pressure and material in the cavities can produce natural and triggered outbursts and can cause splitting of comets.

[1] * E-mail: siipatov@hotmail.com

## 1 INTRODUCTION

On 2005 July 4, a 370 kg impactor collided with Comet 9P/Tempel 1 at a velocity of 10.3 km $s^{-1}$ (A'Hearn et al. 2005). It was an oblique impact, and the angle above the horizon was about 20-35°. Evolution of the cloud of ejected material was observed by *Deep Impact* (DI) cameras, by space telescopes (e.g. *Rosetta, Hubble, Chandra, Spitzer*), and by over 80 observatories on the Earth. Ejections similar to the DI ejection can take place when a small celestial body collides with a comet at a high velocity. Therefore, studies of the ejection of material after the DI impact are important for understanding the collisional processes in the Solar system.

### 1.1 Velocities of ejected material

Velocities of ejected dust particles studied in previous publications utilised ground-based observations and observations made by spacecraft other than *Deep Impact*. The values of the projection of the velocity $v_{le}$ of the leading edge of the dust cloud of ejected material onto the plane perpendicular to the line of sight at several moments in time and of the mass of ejected material obtained from different observations by different authors are presented in Table 1. The velocity of the particles that dominate the cross-section is considered. Estimates of $v_{le}$ made by different authors at $t$~1-2 h after impact can differ by a factor of several and were mainly about 100-200 m $s^{-1}$ (see Table 1). For greater times, the differences between the values of $v_{le}$ were smaller ($v_{le}$ equaled 150-200 and 200-260 m $s^{-1}$ for observations made 4 and 20-24 h, respectively). A few authors estimated $v_{le}$ for more than one moment in time and used the same approach for different times. These estimates allow one to make conclusions about the tendencies of time variations in $v_{le}$ (but not about the exact values of $v_{le}$). Barber et al. (2007) obtained the growth of $v_{le}$ from 125 to 260 m $s^{-1}$ at time $t$ from 1.8 to 20 h. According to Lara et al. (2007), there was a decrease of $v_{le}$ from 230 to 150 m $s^{-1}$ at $t$ from 15 to 40 h. Observed velocities of gas (e.g. CN) were greater than those of dust, and gas could accelerate dust particles.

The above velocities of DI particles are similar to those of particles ejected from several other comets. The velocity of the dust cloud formed at the 2007 October 24 outburst of Comet 17P/Holmes was about 200 m $s^{-1}$ (Montalto et al. 2008). Similar velocities were observed in comae of different comets. A spherically symmetric outer shell of Comet 17P expanded at 430 m $s^{-1}$ (Meng et al. 2008).

[**Table 1]**

### 1.2 Analysis of images made by DI cameras

In contrast with the papers that analysed ground-based observations and observations made by spacecraft other than DI, in the present paper, we estimate velocities of observed ejected particles based on various images made by DI cameras during the first 13 minutes. Analysing three images made by the DI camera at time $t$~8-15 s, Ipatov and A'Hearn (2006) concluded that the projection of velocity of the brightest material onto the plane perpendicular to the line of sight was ~100 m $s^{-1}$.

The images made by the DI spacecraft during the first second after impact and at a later time were presented and discussed by A'Hearn et al. (2005), Ernst, Schultz & A'Hearn (2006), Melosh (2006), Ernst & Schultz (2007), Schultz et al. (2007), and other scientists. In contrast with the papers cited above, our studies were based mainly on the analysis of time variations in contours of constant brightness in DI images. We discuss the relative amount of material and typical velocities of particles ejected at various times, mainly after that initial, fast puff of hot material. For studies of such amounts and velocities, other authors used other observations, theoretical models, and laboratory experiments. Their results are discussed below.

### 1.3 Total mass and sizes of ejected particles

The total mass of ejected dust particles with diameter *d* less than 2, 2.8, 20, and 200 μm was estimated by different authors (see Table 1) to be about $7.3\times10^4$ – $4.4\times10^5$, $1.5\times10^5$ – $1.6\times10^5$, $5.6\times10^5$ - $8.5\times10^5$, and $10^6$ - $1.4\times10^7$ kg, respectively. Measurements based on observations in the first hour or two are likely dominated by icy particles, and the amount of ice ejected was about $4.5\times10^6$ - $9\times10^6$ kg (Keller et al. 2005; Küppers et al. 2005). A'Hearn & Combi (2007) noted that the amount of material ejected with velocity greater than the escape velocity was $10^7$ kg, including 5 to $8\times10^6$ kg of ice. Theoretical estimates of escaping ejecta mass do not exceed $1.4\times10^5$ kg for most types of soil (Holsapple & Housen 2007). At SPH simulations of a DI-like impact, the total mass of material ejected with the escape velocity was $3\times10^6$ kg for nonporous case and $5\times10^5$ kg for porous case (Benz & Jutzi 2007). Note that even the mass for nonporous case is smaller than the amount of escaping ejecta obtained based on observations. Maximum velocities of ejected particles are considered to be approximately proportional to $d^{-1}$ (Koschny & Grün 2001; Jorda et al. 2007).

In the papers by Lisse et al. (2006), Schleicher, Barnes & Baugh (2006), and Meech et al. (2005), the maximum in emitted particle surface area was in the 2-10, 1-5, 1-2 μm particle diameter range, respectively. Spectral modeling showed (Sunshine et al. 2007) that the water ice in the ejecta from Comet Tempel 1 was dominated by 1±1 μm diameter, pure particles, which were smaller than dust particles (2-10 μm). Jorda et al. (2007) concluded that particles with $d<2.8$ μm represent more than 80 per cent of the cross-section of the observed dust cloud. The results cited above show that velocities discussed in our paper correspond mainly to particles with $d<3$ μm, which probably constitute only a few per cent of the total ejected material.

### 1.4 Relation between brightening rate and ejected mass

Meech et al. (2005) concluded that from impact to 1 min after, the comet brightened sharply. Then for the next 6 min, the brightening rate was more gradual. However, at 7 min after impact, the brightening rate increased again, although not as steeply as at first. This rate remained constant for the next 10 to 15 min, at which point the comet's flux began to level off. In the smallest apertures (radius ≈ 1 arc sec), the flux then began to decrease again ~45 min after impact. Observations made in the Naval Observatory Flagstaff Station showed (A.K.B. Monet, private communication, 2007) that there were two episodes of rapid brightening – (1) during the first three minutes after impact and (2) from 8$^{th}$ to 18$^{th}$ minute after impact. Barber et al. (2007), Keller et al. (2007), and Sugita et al. (2005, 2006) obtained a steady increase in the visible flux from the comet until it reached a maximum around 35 min, 40 min, and an hour post-impact, respectively. The brightness monotonically decreased thereafter. The increase in brightness takes longer than the estimated crater formation time, which is considered to be about 3-6 min (Schultz, Ernst & Anderson 2005).

Since the icy particles may be more highly reflective than the refractory particles, they could dominate the velocities that we are measuring. Spectra by Sunshine et al. (2007) imply that relatively pure ice grains were present, while very preliminary measurements of the albedo of the grains by King, A'Hearn & Kolokolova (2007) showed high albedo in particles of the early ejection. Küppers et al. (2005) and Keller et al. (2007) supposed that the relation between flux and ejected mass may be non-linear, either because increasing optical depth of impact ejecta limits the flux from newly produced material or because the size distribution of the ejecta changes with time. In their opinion, the probable cause of the long-lasting brightness increase is sublimation and fragmentation of icy particles within the first hour after impact.

The above discussion shows that flux may not depend linearly on the ejected mass, and that in the present paper we probably mainly study velocities of icy particles, which are larger than velocities of typical crater ejecta.

### 1.5 Problems considered in different sections

More detailed description of previous studies can be found in the initial version of the paper on http://arxiv.org/abs/0810.1294. In Section 2, we discuss the images and contours of fixed brightness that are analysed in our paper. We study the time variations in maximum brightness and in location of the brightest pixel in an image without many saturated pixels. We discuss some factors (e.g. saturated pixels and cosmic ray signatures) that may not allow one to determine accurately the brightness and position of the real brightest spot in an image. Section 2 illustrates the problems that can arise during the work with images (e.g. with images that contain many saturated pixels) and can be interesting to those scientists who work with spacecraft (e.g. DI) images. In Sections 3 and 4, we discuss the ejection of material observed during the first 3 seconds and at 4-800 s, respectively. Based on analysis of DI images, in these sections we estimate typical velocities of ejected material at several moments in time. For such estimates of velocities, we suppose that the same ejected material corresponded to two different maxima (or minima) of the time variations in distances of two different contours of constant brightness from the place of ejection. Based on these estimates and on the estimates of velocities presented in Section 3, in Section 5 we construct the models for calculation of the time variations in rates and velocities of ejection. In Sections 6, 7, and 9, these time variations are studied and compared with those obtained for theoretical models and at experiments. For a quick look on the results of our studies of rates and velocities of ejection, only these three sections can be read. Rays of ejected material are studied in Section 8. If it is not mentioned specially, below we consider the projections of velocities onto the plane perpendicular to a line of sight, but not real velocities.

## 2. IMAGES CONSIDERED, CONTOURS OF FIXED BRIGHTNESS, AND BRIGHTEST PIXELS

### 2.1 Images considered

For this study, we use reversibly calibrated (RADREV) images from both cameras located on board the *Deep Impact* flyby spacecraft. The data are available at the Small Bodies Node of the Planetary Data System[2] together with a discussion of the calibration procedures. The images used were taken up to 13 minutes after impact. Several series of images considered are described in Table 2. In each series, the total integration time for each image and the number of pixels in an image were the same. At the time of impact, both cameras were taking images as rapidly as possible in a continuous sequence. Beginning between 5 and 10 s after impact, the spacing between images gradually increased. For the maximum-speed images, the range is essentially constant, so all images can be treated as having the same scale, with the Medium Resolution Instrument (MRI) having a scale of $P_{MRI}$=87 meters per pixel and High Resolution Instrument (HRI) having a scale 5 times smaller, or 17 m per pixel. The scale is proportional to the distance

[2] http://pdssbn.astro.umd.edu/holdings/dif-c-mri-3_4-9p-encounter-v2.0/dataset.html and http://pdssbn.astro.umd.edu/holdings/dif-c-hriv-3_4-9p-encounter-v2.0/dataset.html

*R* between the cameras and the comet. For the later images, the pixel scale can decrease significantly with time (Fig. 1), and this is taken into account in the analyses. Phase angle (Sun-Target-Spacecraft) varied from 62.9$^o$ to 71.6$^o$ during 800 s. Such variation and photometric errors (Klaasen et al. 2008; Li et al. 2007) do not influence the conclusions of our paper and were not considered, but they may be included in our future models. For considered images, Klaasen et al. (2008) concluded that errors of absolute calibration were less than 5 per cent and errors of calculation of relative brightness were even smaller. The discussion of how to avoid some specific problems with calculation of peak brightness is presented in Sections 2.2-2.3.

**[Table 2]**

**[Figure 1]**

We analyse here only images taken through a clear filter. For all images we use the mid-time of the exposure, an important point for the earliest images where the time from impact is noticeably different at the start and end of the exposure. (Note that some authors have used the start of the exposure time interval.) Times are all measured from the time of impact, i.e. the image in which the first sign of an impact occurs. This relative time is much better known than the absolute time (which is uncertain by about 2 seconds) but is still limited by not knowing when the impact occurred within the image that first shows the impact, i.e. an uncertainty of about 50 msec full range. As in other DI papers, original images were rotated by 90$^o$ in anti-clockwise direction.

In DI images, calibrated physical surface brightness (hereafter CPSB, always in W m$^{-2}$ sterad$^{-1}$ micron$^{-1}$) is presented. Below sometimes we designate it simply as brightness. In the series *Ma*, *Ha*, and *Hc* (see Table 2), we considered the differences in brightness between images made after impact and a corresponding image made just before impact (at time $t$=-0.057 s for MRI images and at $t$=-0.629 s for HRI images) in order to eliminate the difference between the brightness of the nucleus and the coma. In other series, we considered current images.

### 2.2 Saturated and 'hot' pixels and cosmic ray signatures

In this subsection, we discuss some factors that could spoil some DI images in such a way that they may not truly represent a real picture, and the brightest pixel in an image may not correspond to the brightest point of the DI cloud. The discussion shows that for images made at greater times $t$ after impact (i.e. at smaller distances between DI cameras and the place of ejection) it is more difficult to accurately calculate the peak brightness and the coordinates of the brightest pixel.

The first factor discussed is the problem of saturated pixels. Discussions presented in our paper are based on analysis of RADREV images in supposition that such images give true information about the peak brightness. However, in some DN (uncalibrated) images, there were large regions of saturated pixels, which may not allow one to calculate accurately the peak brightness. In the FITS headers of calibrated images, the number of 'possibly saturated pixels' (with DN≥11,000) and the number of 'likely saturated pixels' (with DN≥15,000) are presented. The linear size of the region with DN>11,000 was greater than that of the region inside the closest contour (CPSB=1.75 for the series *Mb* and CPSB=3 for other series) considered in Section 5 usually by a factor of 1.4-1.5, 1.5-1.7, and 1.9 for the series *Mb*, *Hb*, and *Hc*, respectively. In the case of saturated pixels, one cannot be sure that the brightness and coordinates of the brightest pixel in a DI image represent correctly the brightest spot of the cloud. The problem of saturated pixels is discussed in more detail in Section 2.3.

The second factor studied is the problem of the pixels that became 'hot' at small distances between DI cameras and the place of ejection and were not considered during the calibration process. At $t \geq 665$ s, coordinates of the brightest pixel were the same for a group of HRI images, but differed for different groups of images. The values of DN for the pixel of some group that is the brightest pixel for another group were usually greater by >1000 than those for nearby pixels (for all such pixels DN>10,000), i.e. some saturated pixels were 'hot'. At $737 \leq t \leq 802$ s for the series *Mb* or at $665 \leq t \leq 772$ s for the series *He*, coordinates of the brightest pixel were exactly the same for several images. It can mean that some pixels became 'temporarily hot' when the distance *R* between the spacecraft and the nucleus became small and the brightness in DN increased for corresponding DN (uncalibrated) images. Therefore, the increase in *Br* at these time intervals could be caused not only by the increase in the peak brightness of the cloud, but also by the behavior of 'temporarily hot' pixels at small *R,* and the latter increase might not take place (or could be smaller) in the real case. The reviewer noted that such pixels did not become 'hot' (in the sense of pixels with increased dark current and resulting higher count rate than other pixels at the same flux), but such effect was caused by a "memory" for previous overexposure (not all charge from the overexposure removed), an effect frequently observed with CCD detectors. It can be also possible that such 'temporarily hot' pixels were the result of the work of the calibration code because such pixels are inside large regions with brightness equal to 255 in 8-bit images received from the spacecraft, and the brightness of the pixels was calculated based on the brightness of pixels outside this region.

Signatures of cosmic rays on DI images can also prevent to find the true location of the brightest pixel in an image. The brightness of the pixel that was the brightest pixel at exposure ID EXPID=9000990 (and differed by more than 3000 DN from that for close pixels) did not differ much from that for close pixels if we analyse other DN HRI images. Therefore, we suppose that the brightest pixel at EXPID=9000990 belonged to a small signature of a cosmic ray, and the jump in *Br* up to 1.4 at $t$=529 s in Fig. 2a was caused by the signature. The pixel corresponding to this maximum was not detected belonging to a signature of a cosmic ray when we used the codes for removal of cosmic ray signatures considered by Ipatov, A'Hearn & Klaasen (2007). The expected number of signatures of cosmic rays in an image consisted of 1024×1024 pixels is about 8-16 at INTIME=1 s (Ipatov et al. 2007) and about 5-10 at INTIME=0.6 s (the latter value of INTIME is for images from the series *Hb*). A pixel of a cosmic ray signature can be the brightest pixel if it is located in the region corresponding to the bright part of the cloud of ejected material. The size of such region in pixels (and so the probability that a cosmic ray signature corresponds to the peak brightness) is greater for smaller *R* and for HRI images than for MRI images.

In calculations of the peak brightness $B_p$ for images consisting of 1024×1024 pixels, we did not consider the pixels which coordinates $x$ or $y$ were equal to 512 or 513 (for numeration beginning from 1) because these pixels in RADREV images could be by a factor of 1.1-1.2 brighter than other close pixels. The brightness of some pixels with such coordinates could be even greater than the maximum brightness of all pixels which coordinates differed from 512 and 513. At $x$ and $y$ equal to 512 or 513, these bright pixels cannot be detected as too bright pixels with the use of quality and bad pixel maps because they look like the neighboring pixels in these maps. If we did not exclude pixels with such coordinates, then the brightest pixel often was located on the border of two quadrants (each quadrant consists of 512×512 pixels). In series *Hc,* we did not exclude pixels constituting a brighter line at $x \approx 256$.

**2.3 Brightness of pixels in images with saturated pixels**

Some images contain saturated pixels, and in this case one can not reliably conclude anything about the brightness and location of the brightest pixel. The adjacent, non-saturated pixels set a lower limit to the brightness of the saturated pixels. Beyond that one can not usefully infer the brightness of a saturated pixel. The number of saturated pixels was small for images from the series *Ma* and *Ha* (at time *t* up to 109 s). Images belonging to other series usually included relatively large regions of saturated pixels. Here and in Section 2.6, we discuss how saturated pixels can influence the data which were used in our studies.

We subtracted the pre-impact brightness from some series, but not from all (see Section 2.1). That contributes to the brightness difference between images of different series. The difference between subtracted and non-subtracted images could have smaller influence on the peak brightness in calibrated images than the time of integration. The brightness of most pixels was almost the same for non-subtracted images made by the same instrument at almost the same time, but belonged to different series. However, our analysis of DI images showed that the values of the peak brightness (and the brightness of some other pixels which brightness is close to the maximum or minimum brightness) in calibrated DI images made at almost the same time can be different for different series of images if these images include large regions of saturated pixels. For example, typical values of the peak brightness $B_p$ for the series *Hb* are usually greater by a factor of ~1.6 than those in the series *Mb* for images made at approximately the same time (i.e. at the same $Br_p$, the values of $B_p$ can differ by a factor of 1.6 for these two series of images).

In order to illustrate the differences in brightness for different series of images, in this paragraph we discuss the brightness of pixels in three images made at *t*~139-146 s. The CPSB values of the peak brightness $B_p$ are equal to 3.39 and 5.67 (i.e. differ by a factor of 1.67) for the series *Hb* and *Hd*, respectively. At almost the same time, $B_p$ equals 2.2 for the series *Mb*. This value of $B_p$ is smaller by a factor of 2.6 than that for the series *Hd*. The brightness of pixels is almost the same for two RADREV images at CPSB~1-3 and CPSB~0.3-1 if we compare pairs (*Hb* and *Hd*) and (*Hb* and *Mb),* respectively. For the *Hd* image with EXPID=9000951 (at INTTIME=0.1 s), the pixels that are close to the pixel corresponded to the peak brightness are brighter than those for the *Hb* image with EXPID=9000950 (at INTTIME=0.6 s). The pixels which brightness is close to the minimum brightness are less bright for the *Hd* image than for the *Hb* image. The minimum brightness is equal to 0.0067 and 0.0036 for the series *Hb* and *Hd*, respectively. The brightness of the median pixel is almost the same (is equal to 0.32) for these two images. The above three images were received from the DI spacecraft as 8-bit images. For such images, the values of brightness can vary from 0 to 255. For the 8-bit HRI image at EXPID=9000950, the region with brightness equal to 255 is close to the region inside the contour CPSB=3 in the corresponding RADREV image, to the region of saturated pixels with DN>15,000, and to the region of essentially not calibrated pixels (the size is ~1 km). Therefore, it is not possible to calculate accurately the actual brightness inside this region. For the *Hb* image, the number of pixels with DN>15,000 was greater by a factor of 5 and 6 than that for the *Hd* and *Mb* images, respectively. For the MRI image with EXPID=9000951 (at INTTIME=0.3 s), the region inside the contour of CPSB=1.5 was close to the regions inside the contours CPSB=3 for the HRI images, and the brightness of the median pixel in the image array equaled 0.0069, i.e. was smaller by a factor of 47 than that for the HRI images. The minimum brightness in this MRI image was negative.

Let us discuss why in the case of large regions of saturated pixels, the peak brightness and the minimum brightness could differ for different images made at almost the same time. For the series *Mb*, *Hb*, *Hd*, and *He*, the size of images is 1024×1024, and 8-bit images were received from the DI spacecraft. For some 8-bit images, there were large regions with brightness equal to 255. For the corresponding RADREV images, brightness of the same pixels belonging to different RADREV images made at almost the same time usually was almost the same, but the peak brightness could be different. The problem is that one can not reliably conclude anything about the brightness and location of the brightest pixel if there is a large region of pixels with brightness equal to 255 in an 8-bit image received from the spacecraft. The estimates for the brightest pixel were based on the brightness of pixels located outside the region corresponded to brightness equal to 255 in the 8-bit image. For images made at approximately the same time, the size of the region depends on the integration time INTTIME and on whether HRI or MRI images are considered. This is one of the reasons why the peak brightness could be different for different series of images. As a value of INTTIME and a scale of meters per pixel (for MRI images, the scale was greater by a factor of 5 than for HRI images) were used in the calibration process, they could affect the resulting RADREV image also in some other ways. There could be much less problems with estimating the true peak brightness if the corresponding integration time (and so the region of saturated pixels) is smaller than that for the considered DI image with a large region of saturated pixels. The estimates could be better if the procedure of compression allows to obtain a smaller region of pixels with brightness equal to 255 or if original (not compressed 8-bit) images are received on the Earth.

Our analysis of images showed that the difference in the peak brightness $B_p$ was greater for different series of images than for images belonging to the same series. For different series of images, the dependencies of $B_p$ on time $t$ at $t$>27 s are non-crossing lines located relatively far from one another. At intervals of IMPACTM (i.e. of times $t$ when images were made) presented in Table 2, the intervals of the values of $B_p$ are (2.19, 2.32), (3.4, 4.7), (5.5, 5.9), and (4.9, 5.4) for the series *Mb, Hb, Hd,* and *He* (all these series consisted of non-subtracted images), respectively. For the series *Ma* and *Ha,* the values of $B_p$ are almost identical at the same moments in time. For example, the ratio of the values of $B_p$ is equal to 1.04 for these two series at $t$≈5.2 s.

The relative brightness $Br_p$ of the brightest pixel in an image made at time $t$ is presented in Fig. 2 for different $t$. It is considered that $Br_p$=1 at $t$=4 s. The close values of $B_p$ for images belonging to the same series and the overlapping of considered time intervals for different series of images (e.g. series 1 and 2) allowed us to calculate the relative peak brightness $Br_p$ for different series. Let us illustrate such calculations for the case when $B_p=c_{11}$ and $B_p=c_{12}$ for images made at times $t_{11}$ and $t_{12}$ for series 1, and $B_p=c_{21}$ and $B_p=c_{22}$ for images made at $t_{21}\approx t_{11}$ and $t_{22}$ for series 2. If we know the relative brightness $Br_p=Br_1$ at $t_{12}$ for series 1, then we can obtain $Br_p=Br_1(c_{11}/c_{12})$ at $t_{11}\approx t_{21}$ for both series and $Br_p=Br_1(c_{11}/c_{12})(c_{22}/c_{21})$ at $t_{22}$ for series 2. At the first comparison, series 1 included the observation at $t_{12}\approx 4$ s. For images belonging to different series, the plots $Br_p(t)$ are more close to each other than the plots $B_p(t)$. The plots $Br_p(t)$ partly eliminate the errors caused by different sizes of regions of saturated pixels for different series.

If there are any saturated pixels, the brightest pixel is always going to be a saturated one so one can not usefully interpret the brightest pixel. Therefore, in the case of saturated pixels, even for the series *Ma* and *Ha* (with images almost without saturated pixels), the brightness and position of the brightest pixels sometimes were not calculated accurately. For other series, the regions of saturated pixels were much greater than for the series *Ma* and *Ha,* and the errors of calculation of the brightness and location of the brightest pixel at $t$>109 s were much greater than at smaller *t*. It may be possible that the errors can be of the order of the differences in the values obtained for different series, and they can be small for the series *Ma* and *Ha*. We would like to mention that the plots of $Br_p$ vs. time (e.g. local minima and maxima) were similar for different series of images, including the series *Ha* and *Ma*. Zones of saturated pixels were quite different for different series, but time variations in $Br_p$ were similar. Therefore, the plots in Fig. 2 can show some major features of real variations in $Br_p$, may be even at $t$>100 s (see also discussion in Section 2.6).

The value of $Br_p$ decreased at $1<t<8$ s (Fig. 2). This decrease could be caused mainly by the decrease in temperature of ejected material. Similar decreases in brightness and temperature were observed in experiments. A small increase in $Br_p$ at $10<t<30$ s could be caused by the increase in reflectance of ejected material. The increase in reflectance could be associated with an increase in the number of small icy particles due to the destruction of larger particles in the cloud or/and due to the increase in the fraction of icy particles among new ejected material. The latter increase could be associated with the outburst that mainly began at 10 s (see Section 6.1).

It may be possible that for the real cloud there was a local minimum of $Br_p$ at $100<t<150$ s and the amplitude of variations in $Br_p$ at $100<t<500$ s did not exceed $0.11Br_p$ (the values of $Br_p$ in Fig. 2 were in the range of 0.95-1.06). Such relatively constant values of $Br_p$ might be caused by that the optical thickness of the region of the DI cloud corresponding to the brightest pixel mainly exceeded 1.

One can not be sure that the tendency towards an increase in $Br_p$ with $t$ at $t$~150-800 s, and especially at $t$~650-800 s, presented in Fig. 2 really took place with the observed rate of increase. At least partly, such increase could be caused by the increase of the region of saturated pixels with time. However, some increase could take place, because some authors consider that the number of small icy particles increased at that time due to destruction of larger particles in the cloud (and icy particles were brighter than other particles).

The value of CPSB of the brightest pixel at $t$~4-800 s was greater than a typical value of CPSB of the comet nucleus before impact by a factor of ~5-6. The brightest pixel corresponded to a smaller projected area of the ejecta at closer spacecraft-comet distance. It could cause greater fluctuations of *Br* at greater *t*. The observed square region corresponding to one MRI pixel is greater by a factor of 25 than that corresponding to one HRI pixel. Therefore, fluctuations of *Br* for HRI images could be greater than those for MRI images.
**[Figure 2]**

## 2.4 Position of the brightest pixel

Time variations in the position of the brightest pixel in a considered image relative to the position of such pixel in MRI images at $t$=0.001 and $t$=0.06 s are presented in Fig. 3a. The latter position is denoted as the place ‘I’ of impact. The point ‘E’ corresponding to the place of main ejection (the brightest pixel at $t$=0.165 s for the series *Ma*) is located 1 MRI pixel (5 HRI pixels) below the point ‘I’. These places were discussed by Ernst et al. (2006) and Ernst & Schultz (2007). For the series *Ha*, the point E is the brightest pixel at $t$=0.215 s. In Fig. 3, we present the differences in brightness between a current image from the series *Ma* (or *Ha,* or *Hc*) and an image before impact. In Figs. 4-8, the position of the brightest pixel at a current time *t* is presented by a cross (the largest cross in Figs. 4-6 shows the place of ejection). Usually coordinates of the brightest pixel for a subtracted image are the same as those for a corresponding original image, but they could be different for some images (e.g. for the image of the series *Ma* at $t$=0.282 s).

The $x$-shift of the brightest pixel from the points 'I' and 'E' was mainly about 5-15 HRI pixels at $t$~0.2–12 s and about 20-26 HRI pixels at $t$~13–36 s. $Y$-coordinate of the brightest pixel mainly decreased (and the absolute value of $y$-shift increased as $y$<0) with time at $t$<3 s, was approximately the same at $t$~4-12 s, but at a greater $t$ it became closer to 0. At $t$=3.3 s, the $y$-shift equaled to 8 MRI pixels (which correspond to 40 HRI pixels and 700 m), i.e. the mean velocity of the brightest pixel was a little more than 200 m $s^{-1}$.

**[Figure 3]**

The angle φ of the direction from the brightest pixel at $t$=0.215 s (i.e. from the place "E" of ejection) to the brightest pixel at a current time is presented in Fig. 3b for HRI images from the series *Ha* and *Hc*. This direction corresponds to the main direction of ejection of the brightest material. For $x$-axis, we have φ=0. As it is seen from figs. 5-6 in (Schultz et al. 2007), the angle $\varphi_i$ characterizing the direction of the impact was about $-60^o$, and the angle of direction to the downrange plume was about $-70^o$ at $t$~0.3-0.6 s. During the first 12 s, we have $\varphi<\varphi_i$ (Fig. 3b). At 12<$t$<13 s, there was a jump in $x$ and $y$ coordinates of the brightest pixel and the increase in φ by about $50^o$. Just after the above jump, there was a small increase in $Br_p$ (see Fig. 2). About 2-5 s could be needed for material to pass ~500 m from the place of ejection to the brightest place of the cloud observed at $t$~5-35 s. At 13<$t$<55 s, the angle φ was mainly smaller (and closer to $\varphi_i$) for greater $t$. As the number of saturated pixels was small for the series *Ha* and *Ma*, the position of the brightest pixel was found relatively accurately at $t$<100 s, and we are sure that the jumps in the direction from the place of ejection to the brightest pixel (to the brightest spot of the DI cloud) at ~10 s and ~60 s really took place.

## 2.5 Contours of fixed brightness and optically thick regions

Our studies of velocities of ejected material and the relative amount of material ejected per unit of time were based (see Sections 4-7) on the analysis of time variations in distances from the place of ejection to contours of CPSB=const. In order to make one estimate of velocity at one moment in time, we analysed contours for two values of CPSB in images made at many times (not in a single image). DI images were calibrated in such a way that the brightness of comet's surface outside the region of the cloud of ejected material did not differ with distance $R$ between the cameras and the nucleus. Therefore, we compare the sizes of regions inside contours in kilometers.

Figs. 4-6 show the contours CPSB=const for the differences in brightness between *Ma* or *Ha* images and a corresponding image (from the same series) made before impact. In contrast with the above figures, Figs. 7-8 demonstrate CPSB contours for images (from the series *Hb* and *Mb*), but not for their differences in brightness with another image made before impact. Maximum values of CPSB are about 4.5-4.8 for the series *Ha* and exceed 4.2 (exceed 4.7 at $t$≤3.3 s) for the series *Ma*. For the series *Mb*, the maximum values are about 2. Therefore, the values of CPSB for contours presented in Fig. 8 (the series *Mb*) are different from those in Figs. 5-7 (the series *Ha* and *Hb*). In Fig. 5, one can see that the position of the brightest pixel was on the image of the comet before and after the jump in the direction from the place of ejection to the brightest pixel at $t$~12-13 s. So a limb of the comet could not affect this jump.

**[Figure 4, Figure 5, Figure 6, Figure 7, Figure 8]**

The brightness in an image depends not only on thickness of a dust cloud, but also on reflectance and temperature of material. For most images from the series *Ma* and *Ha*, maximum values of CPSB are greater than 3 by a factor of >1.5. As the brightness of an optically thick region is almost the same for different parts of the region, only part of the region inside the contour CPSB=3 could be optically thick. In Figs. 4-6, the maximum distance of this contour from the place of ejection was less than 1.5 km and usually did not exceed 1.3 km. It means that the size of the region of essential opaque probably did not exceed 1 km. Our studies do not contradict to the results by Harker et al. (2007), who concluded that the optical depth is less than unity for all the times considered by the ground-based IR observations, but our results do not agree with the theoretical models by Holsapple & Housen (2007), who predicted a central opaque region of constant brightness extending to about 5 km height above the comet for the first 1.5 h.

**2.6 Relative size of the bright region and independence of conclusions of the paper from saturated pixels**

In Sections 4-5, we analyse the distances *L* from the place of ejection to the edge of the contour CPSB=const, including the distances *L* for a bright region. The definition and calculation of the bright region are described below. The studies presented in Section 5 used the results of analysis of size of this region. In calibrated DI images with large regions of saturated pixels, the values of *L* in km (and the values $B_p$ of the peak brightness; see Section 2.3) at approximately the same time can differ for different series of images. Therefore, for more accurate studies of the rate of ejection, we calculated the relative linear size $L_r$ of the bright region, which better characterizes the real size of the region than *L*. The distance *L* from the place of ejection to the contour CPSB=const is measured on the plane perpendicular to the line of sight and is smaller than the real distance *R* from this place to the particles constituting the contour. The dependence of *L* for the bright region on a series of images usually was relatively small (about a few per cent), except for *Mb* series. Therefore, there is no difference whether to base our conclusions on the values of *L* or $L_r$.

It was considered that $L_r$=1 at *t*=1 s. In order to calculate $L_r$ for other values of *t*, starting from the series *Ma* or *Ha*, we compared the values of *L* at approximately the same time for two series, considering that values of $L_r$ are the same at the same time for different series of images. For such comparison of two series of images, the calculations of $L_r$ for the 'second' series at time $t_2$ (if $L_r$ is known for the 'first' series at time $t_1$) were made similar to the calculations of $Br_p$ in Section 2.3, but the values of *L* were used instead of $B_p$. $L_r$ was calculated based mainly (exclusive for the series *Mb*) on the values of *L* for the contour CPSB=3. For all HRI images, the region of saturated pixels practically does not exceed the region inside this contour, and so saturated pixels should not prevent the relatively accurate calculation of the size of the region inside the contour (see also discussion in Section 2.3), and the time variations in the size of the region inside the contour CPSB=3 should characterize the time variations in the size of the bright region. The conclusions of our paper (see Sections 4-10) concerning the rate of ejection and the triggered outburst were based on analysis of contours with CPSB≤3 and can be obtained without the use of analysis of the brightest pixel. The accordance of the times of the characteristic variations in the location and brightness of the brightest pixel with the times of the variations in the rate of ejection (e.g. the jumps in the rate at ~10 s and ~60 s) that are based on analysis of the contours (see Sections 6.1, 7.2, and 8) testifies in favour of that the main tendencies of time variations in location and brightness of the brightest pixel were correct. This accordance is also an argument in favour of the correctness of the approach that we used for studies of the rate of ejection. For example, at the same time ~10-13 s there was the jump in the direction from the place of ejection to the brightest pixel and the increase of the peak brightness, the rate of ejection, and the upper-right ray of ejection (see Section 8). It shows that the increase of the peak brightness really could take place at that time, and that the tendencies of time variations in $Br_p$ can correspond to those in the real brightness (at least at $t$<20 s) although the concrete values of the peak brightness may not be correct due to saturated pixels.

The relative size $L_r$ of the considered bright region has local maxima at $t$~2-4 s and $t$~20 s. This size and sizes of other regions presented in Fig. 10 increased with time at $t$~150-800 s and especially at $t$>7 min. This increase is in accordance with the increase in the brightening rate obtained at the ground-based observations discussed at the beginning of Section 1.4.

### 2.7 Differences between velocities of particles and velocities of contours of fixed brightness

Actual velocities of particles are greater than the velocities of contours CPSB=const for several reasons: (1) We see only a projection of velocity onto the plane perpendicular to a line of sight; the real velocity can be greater than the projected velocity by a factor of 1.5-2. Richardson et al. (2007) supposed that the ejecta were likely nearly in the plane of the sky and that the de-projection factor was about 2; in principle, much larger factors than 2 were also possible. (2) If the same amount of material moves from distance $D_1$ from the place of ejection to a greater distance $D_2$, then the number of particles in a line of sight (and so the brightness) decreases (at $R>>D$) by a factor of $D_2/D_1$ (the farther is ejected material from the crater, the smaller is a spatial density, and for an abstract model of continuous ejection with a fixed velocity and fixed rate of production of dust, the brightness is proportional to $D^{-1}$). (3) Ejected particles became cooler with time, and so they became less bright. On the other hand, the light from the impact illuminated the dust which was near the comet before impact. The velocities considered in the present paper likely correspond mainly to small icy particles (see Sections 1.3-1.4), and typical velocities of large particles were smaller than these velocities. The approaches used to estimate the characteristic velocities of observed particles are discussed in Section 3-4. In our opinion (see Section 5.3), the variations in velocities of particles during their motion in the vicinity of several km from the place of ejection were small compared with the velocities considered in our paper (at least at ejection time $t_e \leq 100$ s).

## 3 THE EJECTION OF MATERIAL OBSERVED DURING THE FIRST THREE SECONDS AFTER IMPACT

In this section, we discuss the velocities and relative rates of ejection of material observed during the first three seconds after impact. Only MRI images from the series *Ma* are analysed. For such images, the distance from the DI cameras to the place of impact, and, thus, the image scale, can be considered to be approximately constant. For all images considered in Section 3 and in Figs. 4 and 9, we study the difference in brightness between a current image and the image made at $t$=-0.057 s.

### 3.1 Velocities of material observed during the first second

Images from the series *Ma* consisted of 64×64 pixels, but only 32×32 pixels are presented in Fig. 9. (In Figs. 4-8, all pixels described in Table 2 are presented.) In Fig. 9a, white region corresponds to CPSB≥3, and in Fig. 9b it corresponds to CPSB≥0.5, but both sub-figures present the same images. CPSB of the brightest pixel in an image has peaks at $t$~0.22-0.28 s (CPSB=9.7) and at $t$~0.52-0.64 s (CPSB=9.5). The first peak corresponds to the ejection of high temperature material, and the second one corresponds to the peak of the amount of observed material (small particles) ejected per unit of time. According to Melosh (2006), the hot plume cooled down very rapidly, from 3500 to 1000 K in only 420 ms. Note that the two images at 0.22 and 0.28 s are both saturated at the brightest pixel, so it's real brightness can be greater.

**[Figure 9]**

*3.1.1 Velocities of hot material*

In images made at $t$=0.34 and $t$=0.4 s, there are two spots of ejected material corresponding to two different ejections. The material which formed the brightest pixel of the lower spot at $t$=0.34 s was located about 8 pixels (i.e. 0.7 km) below the brightest pixel at $t$=0.165 s. The velocity of this material exceeded 0.7/(0.34-0.165)≈4 km $s^{-1}$. At $t$=0.165 s, the bright spot was small and had a centre at the point 'E' (the point is discussed in Section 2.5), so most of material which formed the lower spot in Fig. 9 at $t$=0.34 s could have been ejected after $t$=0.165 s. The lower part of the contour CPSB=3 of the lower spot was located 10, 7, and 2 pixels below the point 'E' at $t$=0.34, 0.282, and $t$=0.224 s, respectively. The difference in 5 pixels (435 m) for the latter two times corresponds to the projection $v_p$ of velocity onto the plane perpendicular to the line of sight equal to 0.435/(0.282-0.224)=**7.5 km s$^{-1}$** and to the beginning of the main ejection of the material of the contour at $t_e$=**0.2 s**. A real velocity $v$ (not a projection) of material could exceed 10 km $s^{-1}$. This velocity was mentioned earlier by several other authors, e.g. by A'Hearn et al. (2005). Holsapple & Housen (2007) noted that in the absence of an internal source of energy, conservation of energy limits the mass with $v$>10 km $s^{-1}$ to less than the mass of the impactor. They argue that the high velocities at the beginning must be due to extra acceleration.

In the image made at $t$=0.4 s, the lower spot is by a factor of 3 less bright than that at $t$≤0.34 s. The flash which caused the lower spot can be associated with the vaporization of the impactor and part of the comet. The second flash which caused the upper spot may be associated with the first eruption of material at the surface. Schultz et al. (2007) supposed that the initial downrange plume is likely derived from the upper surface layer (~projectile diameter), and A'Hearn et al. (2008) concluded that $H_2O$ ice is well within 1 m, and probably within 10 cm, of the surface.

*3.1.2 Velocities of cooler material*

The brightness of the brightest pixel in an image increased monotonically at 0.34≤$t$≤0.58 s, and the vertical size of a region with CPSB>3 increased from 4 to 13 pixels (from 350 to 1130 m) at 0.34≤$t$≤0.82 s. Therefore, there was a considerable continuous ejection of material at ejection time 0.34≤$t_e$≤0.58 s. If we consider the motion of the contour CPSB=3 for the lower edge of the upper spot at $t$ between 0.34 and 0.46 s, then we obtain $v_p$=1.5 km $s^{-1}$ for this bright material. (This is an approximate estimate, as coordinates of CPSB=3 at these times differed by only 3 pixels.) These values of $v_p$ are smaller than those obtained for the first ejection. Supposing that the second ejection began mainly at $t_e$~0.24-0.28 s (below we use mainly **0.26 s**), we obtain that the material which constitutes the contour CPSB=1 at $t$=0.46 s moved with **$v_p$≈3 km s$^{-1}$** (twice faster than the material constituting the contour CPSB=3 at $t$=0.46 s). If moving with the same velocity, at $t$=0.58 s this material would be located at a distance $D$ from the place of impact by a factor of 5/3 (for $t_e$=0.28 s) greater than at $t$=0.46 s and would have CPSB=0.6 for the abstract model with brightness decreased as $D^{-1}$. At the estimated distance, we have CPSB=0.5. Such difference (0.6 instead of 0.5) can be caused by a rough model (actual ejection was continuous) and by the decrease in brightness of a particle with time after the ejection (with a decrease in its temperature). All the above estimates show that some material ejected at $t_e$<0.5 s had velocities of about several km $s^{-1}$. We were able to make some of the above estimates of velocity of ejected material because we could estimate the time of ejection of material constituting the contour of a fixed brightness (and we considered the velocity as the ratio of the distance of the contour from the place of ejection to the time elapsed after ejection of the material).

### 3.2 Material ejected with velocity greater than 1 km s$^{-1}$

The contour CPSB=0.1 in Fig. 4 moved with $v_p$~1 km s$^{-1}$ during 1≤$t$≤3 s. (At any moment in time, the contour corresponded to parts of the cloud consisted of different particles, most of which moved with a greater velocity than the contour.) Therefore, real typical velocities of particles at the distances from the place of ejection to this contour probably exceeded 1.5 km s$^{-1}$ and could be ~2 km s$^{-1}$. Part of such high-velocity material could have been ejected during the first second. Ground-based observations made a few hours after impact did not show a considerable amount of material ejected with velocities $v_p$>1 km s$^{-1}$. The maximum velocities of the outer part of the cloud observed from the ground were about 600 m s$^{-1}$. It shows that the total mass of material with $v_p$>1 km s$^{-1}$ was small compared with the mass of all ejected material. In our model considered in Sections 5-6, the fraction of observed particles with such velocities was 1-2 per cent. Some of the high-velocity particles had evaporated and dispersed by the time Earth-based observers saw the cloud.

## 4 ESTIMATES OF VELOCITIES OF EJECTED MATERIAL BASED ON DI IMAGES MADE DURING 800 S

In this section, we estimate projections $v_p$ of the characteristic velocities of ejection of observed DI particles at several moments $t_e$ of ejection. Such pairs of $v_p$ and $t_e$ are marked below in bold and are used in Section 5 for construction of our model of ejection. Most of our estimates of the pairs were based on the analysis of distances $L(t)$ from the place of ejection to contours CPSB=const at different moments in time. The number of images analysed was much larger than the number of the moments in time at which velocities were estimated. Velocities of the contours (i.e. d$L$/d$t$) were not calculated and were not used in our studies of velocities of ejected particles. The velocities of particles obtained in Section 4 are in accordance with the velocities calculated in Section 8 based on the analysis of rays of ejected material, i.e. based on a quite different approach.

We analysed maxima or minima of plots of time variations of $L(t)$ presented in Fig. 10, where $L(t)$ is the distance of a contour of constant brightness from the place of ejection in an image made at time $t$. In this figure, different designations correspond to different values of CPSB. Below we consider the model for which all observed particles ejected at the same time had the same velocities.

For the series *Ma*, we supposed that $L=L_y=L_i$ was equal to the distance from the place of ejection to the contour CPSB=const down in $y$-direction. For other series, in Fig. 10 we present the difference $L=2L_x=2L_i$ between maximum and minimum values of $x$ for the contour. $L_i$ characterizes the distance from the place of ejection to the contour (in $x$ or $y$ directions).

In Fig. 10, for a fixed value of CPSB, a plot $L(t)$ usually has two local maxima and two local minima. We supposed that the particles constituted the regions of the DI cloud corresponding to two contours were the same if the values of $L(t)$ for the contours correspond to local minima (or maxima) in Fig. 10. These two contours were characterized by different values of CPSB and were located at different distances from the place of ejection in two images made at different moments in time $t$. Based on the values of $L_i$ ($L_1$ and $L_2$) for two contours at the times ($t_1$ and $t_2$) corresponded to such local minima (or maxima), we calculated the characteristic velocity $v_{pc}=(L_2-L_1)/(t_2-t_1)$ and the ejection time $t_e=t_1-L_1/v_{pc}$. These are the main formulae used for calculations of velocities of particles. The obtained pairs of characteristic velocity $v_{pc}=v_p$ and $t_e$ are summarized in Table 3. For this approach, we use results of studies of many images in order to obtain one pair of $v_p$ and $t_e$. Examples of more detailed estimates are presented below.

[**Figure 10]**

**[Table 3]**

In this paragraph, we present an example of calculations of the pair of $v_p$ and $t_e$. We analysed the plots corresponding to the contours CPSB=3 and CPSB=1. For the contour CPSB=1 and the series *Hc*, the second local maximum of $L=2L_x=7.34$ km was at $t=56$ s. It is not clear what time it is better to choose for the time corresponding to the maximum of *L* for the curve for the contour CPSB=3 at the series *Ha* and *Hc* in Fig. 10, as the values of *L* are almost the same ($L\approx2$ km) at $t\sim16$-47 s (see also Fig. 6c). Let us suppose that particles constituting the contour CPSB=3 at $t=31$ s (this is the middle of the interval [16, 46]) and the contour CPSB=1 at $t=56$ s are the same. Considering that (7.3-2.1)/2=2.6 km were passed in 56-31=25 s, we obtain $v_p\approx$**105 m s$^{-1}$** and $t_e\approx$**21**=31-10 s (1050/105=10). For *y*-direction, we have $L=L_y\approx1.56$ km for CPSB=3 at $t=30$ s and $L_y\approx4$ km for CPSB=1 at $t=56$ s; these data correspond to $v_p\approx$**100 m s$^{-1}$** and $t_e\approx$**15 s**. Larger velocities can also fit the above observations. Using similar approach, but other maxima and minima in Fig. 10, we obtained several other pairs of $v_p$ and $t_e$ presented in Table 3. Calculations for other pairs of $v_p$ and $t_e$ can be found in the first version of the paper located on http://arxiv.org/abs/0810.1294.

Note that each estimate of velocity $v_p$ at time $t_e$ of ejection made with the use of the approach discussed in the above paragraphs of this section was based on the analysis of contours of constant brightness belonging to many images (but not on analysis of a single image). Each estimate is a result of studies of two curves of the time variations in the distances from the place of ejection to two contours of constant brightness. For such estimates, we did not use any dependence of variation in brightness with distance *D* from the place of ejection, i.e. we did not use any theoretical model of dilution of the cloud. We only analysed the images using the approach discussed at the beginning of the section.

As we discuss in Sections 5.3-5.4, such factors, as destruction, sublimation, and acceleration of particles, do not affect much our estimates of velocities because we consider the motion of particles during no more than a few minutes. Our model takes into account the dilution of the cloud with distance from the place of ejection because different contours of constant brightness are located at different distances. In our model at a fixed rate of ejection, the brightness decreases approximately linearly with an increase in distance from the place of ejection.

The estimates presented below are based on another approach. We supposed that the material corresponded to the contour CPSB=3 at $t_1=8$ s moved down along the *y*-axis from the distance $D_1=875$ m with a fixed velocity $v_p$. (Here *D* is the distance from the place of ejection down along the *y*-axis.) We found that the brightness at distance $D_2=D_1+(t_2-t_1)v_p$ in an image made at $t_2=12.25$ s better corresponds to the brightness $Br=3\cdot D_1/D_2$ at $v_p$=**240 m s$^{-1}$**. This velocity corresponds to $t_e$**=4.4 s** (=8-875/240). The point ($v_p$=240 m s$^{-1}$ at $t_e$=4.4 s) was obtained for the model of *Br* proportional to $D^{-1}$. This point is close to the curve $v_p(t_e)$ connecting the points calculated based on the plots presented in Fig. 10 (see Fig. 11). Therefore, different approaches used for estimates of velocities $v_p$ do not contradict each other, and an approximate decrease of *Br* as $1/D$ can be used as an initial approximation (at least at $t_e\sim4$ s).

The lower part of the contour CPSB=0.03 in Fig. 8b (at $t=139$ s) was located about 22.5

km from the place of ejection. All particles of this part of the contour had velocities $v_{py}$>160 m s$^{-1}$. For $x$-direction and the same contour, $L$=47 km and $v_{px}>L/2t=L_x/t\approx$170 m s$^{-1}$. There was also material outside of the contour CPSB=0.03. Therefore, there could be many particles with $v_p$~200 m s$^{-1}$. The values of $v_p$~100-200 m s$^{-1}$ are in accordance with the ground-based observations of velocities presented in Table 1.

Initially we did not plan to spend much time for the studies based on analysis of the contours of constant brightness, but we reconsidered when we saw that doing so would facilitate study of the main features of the ejection and of the "physics of the processes of ejection", and would lead to important conclusions, such as those concerning the role of the triggered outburst in the DI ejection (see Sections 6-9). Complex models depend on many factors, so, if we began our studies with complex models, we could find what theoretical models best fit the observations, but we still might not be able to understand the role of the triggered outburst. The results of studies with simple models will be used to construct more complex models, which approach facilitates the understanding of the process of ejection and the role of different factors on the evolution and observable form of the cloud of ejected material.

## 5 MODELS USED FOR CALCULATION OF THE TIME VARIATION IN RELATIVE RATE OF EJECTION

### 5.1 Velocity of ejection

Here we describe our models used for calculation of the velocities and relative rates $r_{te}$ of ejection at different times $t_e$ of ejection. Such models were not considered by Ipatov & A'Hearn (2008a). In our studies, we analysed the sizes of the bright regions in DI images made at times $t$ and used the obtained relationships between these times $t$ and the times $t_e$ of ejection of material located at the edge of the bright region in an image made at time $t$ (see below). For most images, the regions inside the contours CPSB=3 were considered as bright regions (see Section 2.6). We also used the estimates of velocities obtained in Sections 3-4 and presented in Table 3. For calculation of these velocities, we divided distance by time. Therefore, we calculated the mean velocities of particles during the first few minutes or seconds of their motion. We believe that these velocities were close to the velocities of ejection because the variations in velocities of observed particles during the first minutes of the motion of the particles were relatively small, at least at $t_e\leq$115 s (see Section 5.3).

For theoretical models, Housen, Schmidt & Holsapple (1983) and Richardson et al. (2007) obtained that ejection velocity $v$ is proportional to $t_e^{-\alpha}$, where α is between 0.6 (the theoretical lower limit corresponding to basalt) and 0.75 (the theoretical upper limit). They considered that the cratering event is primarily governed by the impactor's kinetic energy at α=0.6 and by momentum at α=0.75. According to Holsapple (1993), α=0.71 for sand and dry soil and α=0.644 for water, wet soil, and soft rock. The designations (e.g. α) in the above papers were different from those in our paper. For the above four values of α, in Table 4 we present the exponents of the time dependencies of the relative volume $f_{et}$ of the material ejected before time $t_e$ and the relative rate $r_{te}$ of ejection, and the exponents of the velocity dependence of the relative volume $f_{ev}$ of material ejected with velocities greater than $v$. These exponents were obtained based on Table 1 of the paper by Housen et al. (1983).

**[Table 4]**

The pairs ($v_p$ and $t_e$) presented in Table 3 are close to the exponent dependence with α about 0.7-0.75. In Fig. 11, these values of $v_p$ are marked as $vy_{\rm obs}$ and $vx_{\rm obs}$, and exponent dependencies are presented by dotted lines. The values of $v_p$ obtained in Section 8 are marked as $v_{ray}$ and satisfy the same exponent dependence as the values from Table 3. The above values of α are in accordance with the theoretical estimates cited above, but were not taken from those estimates. If it is not mentioned specially, we use such values of α in our paper.

For calculations of the relative rate of ejection (see Section 5.2), we need to know the relation between the time $t_e$ of ejection of particles located at the edge of the bright region in an image made at time $t$ and the time $t$. We also need to find the values of $v_p$ for these particles. Therefore, we considered the model for which the typical projection $v_p$ of velocity of *particles* constituting the edge of the bright region (which is greater than velocity of the edge) in an image made at time $t$ equaled to $v_{\text{expt}}=c_e\times t^{-\alpha}$, where $c_e$ is some constant. The calculation of the bright region is discussed in Section 2.6. If it is not mentioned specially, in our paper we consider projections of velocity on the plane perpendicular to the line of sight.

In Fig. 11, we present the plots of $v_{\text{expt}}=v_p=c\times(t/0.26)^{-\alpha}$ for 4 pairs of α and $c$. The values of $vy_{\min}=L_y^*/(t\text{-}0.26)$ ($vy_{\min}=L_y^*/t$ at $t$<0.3 s) and $vx_{\min}=L_x^*/(t\text{-}0.26)$ show the minimum velocities (in km $s^{-1}$) needed to reach the edge of the bright region in a DI image made at time $t$ from the place of ejection for the motion in $y$-direction (for the series *Ma)* or in $x$-direction (for other series), respectively. As noted in Section 3.2, we suppose that the second ejection began mainly at $t_e$≈0.26 s. We considered that $L_y^*=0.95\times L_r$ and $L_x^*=0.5\times1.076\times L_r$, where $L_r$ is the relative linear size of the bright region (see Section 2.6). The values of $L_y^*$ and $L_x^*$ are in km, and $L_r$ is dimensionless. For the contours CPSB=3 at $t$=1 s, we have $L=L_y=L_y^*=0.95$ km for the series *Ma* and $L=2L_x=2L_x^*=1.076$ km for the series *Ha*. The ratio $L_x/L_y$ varied with time.

**[Figure 11, Figure 12]**

Taking into account that the time needed for particles to travel a distance $L_x^*$ in $x$-direction is equal to $dt=1.076L_r/(2v_{\text{expt}})$, we find the time $t_e=t\text{-}dt$ of ejection of material of the contour of the bright region in an image made at time $t$. Using the obtained relationship between $t$ and $t_e$ (Fig. 12) and considering that the projection $v_{\text{model}}$ of velocity of ejection at time $t_e$ of ejection equals $v_{\text{model}}(t_e)=v_{\text{expt}}(t)=c\times(t/0.26)^{-\alpha}=c\times(t_e/0.26k_e)^{-\alpha}$, we can obtain the dependencies of $v_{\text{model}}$ on $t_e$ for different values of $c$ and α. The ratio $k_e=t_e/t$ mainly increased with $t_e$ at $t_e$>1 s, and most of the values of the ratio were between 0.4 and 0.8 (see Fig. 12). In Fig. 11, we present the plots of $v_{\text{model}}$ vs. $t_e$ for four pairs of α and $c$. Actually, we analysed plots $v_{\text{model}}(t_e)$ for a greater number of pairs (α and $c$) trying to find the plots that best fit the pairs of $v_p$ and $t_e$ presented in Table 3. Note that velocity distributions over $t$ and $t_e$ were different.

The plot $v_{\text{model}}(t_e)$ is close to the exponential dependence. The values of α for which $v_{\text{model}}(t_e)$ be**st** fits the pairs of $v_p$ and $t_e$ obtained in Sections 3-4 and presented in Table 3 are also (as for the data from Table 3) about 0.7-0.75. It testifies in favour of that the supposition $v_{\text{expt}}=c_e\times t^{-\alpha}$ is approximately true. Estimates of the pairs ($t_e$ and $v_p$) presented in Table 3 were made for $t_e$≤115 s. In the model *VExp* considered in Sections 5-6, we suppose that the nearly exponential dependence of $v_{\text{model}}$ on $t_e$ is the same for any $t_e$, including $t_e$>115 s. Another model is discussed in Section 7.

Besides α, an exponential decrease in velocity is also characterized by a coefficient $c$. Though we use the same α as in theoretical models (e.g. Housen et al. 1983), the duration of ejection and the amount of observed material ejected with $v_p$>100 m $s^{-1}$ in our model are greater than for typical theoretical models. This amount depends on the time variations in the rate of ejection. The variations obtained in our studies (see Section 6) are different from those for theoretical estimates and can be explained by the triggered outburst.

## 5.2 Relative rate of ejection and volume of ejecta

Based on the obtained time variations in velocities of ejected particles and on the time variations in the size of the bright region, we calculated the time variations in the relative rate of ejection. The description of the calculations is presented below in this subsection, and the obtained dependencies are analysed in Sections 6-7.

The number of particles ejected per unit of time is considered to be equal to $c_{te} \cdot r_{te}$, where dimension of $r_{te}$ is $s^{-1}$, and $c_{te}$ is a constant. Here $r_{te}$ corresponds to the particles that were ejected at $t_e$ and reached the shell with radius $L_r$ at time $t$. The volume $V_{ol}$ of a spherical shell of radius $L_r$ and width $h$ at $h \ll L_r$ is proportional to $L_r^2 h$, and the number of particles per unit of volume is proportional to $r_{te} \cdot (L_r^2 \cdot v)^{-1}$, where $v$ is the velocity of the material moving from the centre of the sphere. The number of particles in a line of sight, and so the brightness *Br,* are approximately proportional to the number of particles per unit of volume multiplied by the length of the segment of the line of sight inside the DI cloud, which is proportional to $L_r$. Actually, the line of sight crosses many shells characterized by different $r_{te}$, but as a first approximation we supposed that $Br$ is proportional to $r_{te}(v \cdot L_r)^{-1}$. For the edge of the bright region, $Br \approx$const. Considering $v = v_{\text{expt}}$, we calculated the *relative* rate of ejection as $r_{te} = L_r \cdot t^{-\alpha}$. Based on this dependence of $r_{te}$ on time $t$ and on the obtained relationship between $t$ and $t_e$, we constructed the plots of dependencies of $r_{te}$ on $t_e$ (Fig. 13).

**[Figure 13, Figure 14, Figure 15]**

The rates $r_{te}$ were used to construct plots of the relative volume of ejecta (the relative number of observed ejected particles) launched before $t_e$ vs. $t_e$ (Fig. 14) and of the relative volume of ejecta with velocity $v_p > v_{\text{model}}$ vs. $v_{\text{model}}$ (Fig. 15). While constructing Figs. 14-15, we considered only the particles ejected before $t_{e803}$, where $t_{e803}$ is the time of ejection of the particles constituting the edge of the bright region in an image made at time $t$=803 s (the last considered image). Figs. 12-15 were obtained for the model *VExp,* for which $v_{\text{model}}(t_e) = v_{\text{expt}}(t) = c \times (t/0.26)^{-\alpha}$ at 1 s < $t_e < t_{e803}$. We didn't normalize the plots in Figs. 14-15 for all ejected material (as it was done in theoretical estimates) because in our model we could estimate the rate of ejection of material only before $t_{e803}$ and didn't know when the ejection finished. Figs. 13-15 characterize the ejection not of all ejected material, but only of the bright particles that reached a distance $R \geq 1$ km from the place of ejection. We did not analyse more close contours because there are large regions of saturated pixels at a smaller distance at $t$>110 s. The values of $r_{te}$ were calculated based on the size of the bright region of the DI cloud (which depends on the sum of cross-sections of ejected particles) for the model for which sizes of particles do not depend on $t_e$. It is considered that typical masses of ejected particles increase with time for the normal ejection. The brightness of a particle of diameter $d$ is proportional to $d^2$, and its mass is proportional to $d^3$. Therefore, the ratio of the real rate of ejection to $r_{te}$ is proportional to $d$. It increases with $t_e$ for the normal ejection.

## 5.3 Accelerations and variations in velocities of moving particles

In our models, we did not take into account the acceleration of particles by moving gas and destruction of particles. These factors are important for consideration of the evolution of the cloud during several hours. Below in this section, we present arguments in favour of that the influence of the factors on velocities of particles was not considerable during a few minutes of motion of particles. Our estimates of velocities were based on analysis of images of the regions of the cloud located no more than a few kilometers from the place of ejection. Considered particles moved in such regions during a time less than a few minutes. In particular, the time was smaller (sometimes considerably) than 13 minutes, for which we analysed DI images.

Holsapple & Housen (2007) supposed that ejected particles could be accelerated by the dust-gas interaction. The reviewer noted that the background coma (from regular activity) may be also important for gas content of the cloud. In our opinion, this acceleration cannot change the main results of the paper because the variations in velocity caused by the accelerations were relatively small (compared with the observed velocities) during the considered motion of observed particles.

Richardson et al. (2007) obtained that the gas accelerations of DI particles were about 0.04-0.4 mm $s^{-2}$. The upper limit of this acceleration corresponds to 1 μm particles and to the increase in velocity by 0.24 m $s^{-1}$ during 10 minutes. Richardson et al. noted that the accelerations of DI particles were smaller than those of the particles leaving a comet under normal circumstances because the dust density in the vicinity of the ejecta plume and the mass-loading on the outward flowing gas were high.

Earth-based observations of the velocity $v_{le}$ of the leading edge of the DI cloud made by Barber et al. (2007) showed that there was the increase in the velocity by 135 m $s^{-1}$ during ~18 hours (between the second and 20$^{th}$ hours after impact). For a constant acceleration, it corresponds to the increase in the velocity by 1.2 m $s^{-1}$ during 10 minutes. These estimates of variations in velocities during 10 min are greater than those based on the accelerations obtained by Richardson et al. (2007), but are much smaller than the velocities considered in our paper.

By July 8 (four nights after impact), the inner ~15,000 km were no brighter than prior to impact (Knight et al. 2007). Therefore, particles with velocities $v$<40 m $s^{-1}$ were not practically observed at that time. Part of these low-velocity particles could be sublimated. At constant acceleration, the increase in velocity by 40 m $s^{-1}$ during 4 days corresponds to the increase in $v$ by <0.5 m $s^{-1}$ per hour. A reviewer noted that if the absence of slow particles observed by Knight et al. is indeed due to acceleration, it was most likely not linear in time. In his opinion, most of the acceleration happened within a few nucleus radii because densities of gas were too low at greater distances, and the terminal velocity had been already reached after one hour.

The ratio of the surface of a sphere with radius $R$=10 km to the area $S_1$ of cross-section of a particle of radius $r$=1 μm equals to $4\cdot10^{20}$. According to Mumma et al. (2005), during the first 20 minutes after the impact, on average about $8\cdot10^{27}$ water molecules were ejected per second. The pre-impact rate was about $6\cdot10^{27}$ water molecules $s^{-1}$. The number of HCN molecules was smaller by a factor of ~500 than that of water molecules. The above data show that after impact ~$2\cdot10^7$ water molecules passed through the area $S_1$ during 1 s at $R$=10 km (~$2\cdot10^3$ water molecules at $R$=$10^3$ km). The mass $m_1$ of a particle of radius $r$=1 μm is $4.2\cdot10^{-12}$ g at density equal to 1 g $cm^{-3}$, i.e. it is greater by a factor of $1.4\cdot10^{11}$ than the mass of water molecule (which is $3\cdot10^{-23}$ g). At $R$=10 km, the total mass of water molecules passing through the area $S_1$ during 1 s was $1.4\cdot10^{-4}\cdot m_1$ (for some directions, it was greater than this value because DI ejection was different for different directions). For the above data at $R$=10 km and the velocity of gas relative to dust equal to 250 m $s^{-1}$, using the law of angular momentum conservation, we obtain the acceleration of a particle to be ~0.035 m $s^{-2}$ (it corresponds to the increase in velocity by 2 m $s^{-1}$ per minute). This estimate was made for the model for which all molecules that cross the area $S_1$ collide with a particle. At a distance $R$ of not more than a few km, large concentration of particles can prevent frequent collisions of molecules with a particle. The motion and evolution of the DI cloud depended on many factors. For example, gas molecules can be produced by sublimation of ejected icy particle. In recent grant applications, we described how it is possible to estimate the acceleration of observed particles by gas, based on analysis of DI images (including look-back images), but it will need more complicated studies than those in the present paper. As the mass of a particle and its surface are proportional to $r^3$ and $r^2$, respectively, the increase of particle velocity due to fast moving gas is mainly proportional to $r^{-1}$.

Let's discuss whether the below model of the influence of gas force on the motion of a dust particle at different distances $R$ from the place of ejection can be used for studies of the DI ejection. In this model, the amount of gas in a spherical shell does not depend on $R$, and the force $F_{gd}$ acting on a dust particle is considered to be approximately proportional to $(v_g-v_d)\times R^{-2}$, were $v_g$ and $v_d$ are velocities of gas and dust, respectively. As for moving dust particles $\mathrm{d}R=v_d\mathrm{d}t$, the distance $R$ grows faster than time $t$ (after the impact) if $v_d$ increases. Considering an integral of $F_{gd}$ (of acceleration for a fixed mass of a particle) over $t$ and then transforming it into an integral over $R$, we have $\int v_d^{-1}(v_g-v_d)R^{-2}\mathrm{d}R$. The integral over $R^{-2}$ gives $R^{-1}$. For a very simple model with almost constant values of $v_g$ and $v_d$, the increase in velocity during the time interval $(t_1, t_2)$ is proportional to $(t_1^{-1}-t_2^{-1})$, and the ratio $k_{vg}$ of the increase in velocity during the time between 1 and 4 h to the increase during the time between 4 and 16 h equals (1-1/4)/(1/4-1/16)=4. For increasing $v_d$, the ratio $(v_g-v_d)/v_d$ decreases, and the value of $k_{vg}$ is greater than that for a constant ratio. If we consider $v_{le}$ to be equal to 110, 175, and 230 m s$^{-1}$ at 1, 4, and 16 h, then the ratio $k_{vg}$ for these velocities is only ~1.2 (=(175-110)/(230-175)). For most other values of $v_{le}$ from Table 1, the ratio is even smaller. It shows that the above model cannot be used even after 1 h and that there should be no considerable increase in velocity during the first minutes of the motion of particles ejected with velocity $v_p>10$ m s$^{-1}$. This result is in accordance with the conclusion by Richardson et al. (2007) that accelerations of DI particles are smaller than those for the particles ejected under normal circumstances. In order to estimate the increase in velocity of particles due to gas (especially for small times), it is needed to consider much more complicated models than the above model. The latter model probably can be used for some cases when the number of ejected particles is relatively small.

Our estimates of velocities presented in Section 4 were based on analysis of the contours located mainly at a distance $L<10$ km from the place of ejection. In our studies of the relative rate of ejection (see Section 5.2), we analysed the projection $L$ (onto the plane perpendicular to the line of sight) of the distance $R$ from the place of ejection to the edge of the bright region, which was less than 2 km. At velocity of 100 m s$^{-1}$, a distance $R$=3 km is passed in only 30 s. Velocities considered in our paper are mainly greater than a few tens of meters per second. The results presented in the four above paragraphs testify in favour of that the variations in velocities of particles under the influence of gas during a time no more than a few minutes did not exceed a few meters per second, i.e. were smaller than velocities of particles considered in our paper.

Based on analysis of DI images, we obtained (see Sections 4 and 5.1) an exponential decrease in velocity $v_p$ from 7500 m s$^{-1}$ to 20 m s$^{-1}$ during the first 100 s. In Section 4, we divided a distance of a few kilometers by time. Therefore, we calculated the mean velocities during the motion of particles inside the region located not more than a few kilometers from the place of ejection, and the values of $v_p$ included the contribution of the increase in velocity during the motion of particles. We do not think that for considered particles of the same size, the increase in $v_p$ due to gas drag depended much on $t_e$ at $t_e\leq100$ s. (For the normal ejection, typical sizes of particles increased with $t_e$, and so their accelerations due to gas were smaller at greater $t_e$.) Therefore, the increase in velocity of particles by moving gas during the motion of particles inside the region with radius of a few kilometers from the place of ejection should be smaller than the velocity of 20 m s$^{-1}$ at $t_e\approx100$ s, and it is more probable that the increase did not exceed a few m s$^{-1}$ (see the discussion in the previous paragraphs).

The same estimates of $v_p$~20-25 m s$^{-1}$ at $t_e$~73-115 s (see Table 3) were obtained based on analysis of contours located at different distances from the place of ejection. For some of these estimates of $v_p$, the motion was analysed at intervals of the distances from 1.5 to 4 km. For other estimates, the interval of 1.5-12 km was used. Variations in the distances by 8 km and times $t$ by 200 s did not cause variations in estimates of $v_p$ (at least by more than 5 m s$^{-1}$, which is the accuracy of estimates of $v_p$ in Table 3). It can be one more argument in favour of that the increase in velocities of considered particles due to acceleration by gas did not exceed a few meters per second during a few minutes.

The increase in velocities of particles after they left the regions considered in Section 4 could be much greater than the increase inside the regions. For some particles, the increase during 14 h could exceed 100 m s$^{-1}$ (see Section 1.1), but in the present paper we study the motion of a particle during not more than a few minutes. (The time needed for a particle to reach a considered contour can be much less than 13 min, during which analysed DI images were made.)

### 5.4 Sublimation of particles and variations in their size distribution

Cochran et al. (2007) concluded that size distribution of particles did not change during the first two days after impact (as they saw no change in the color of the dust), and that grains usually did not fragment in the coma. In our opinion, the Cochran's result could be true because it is not a hypothesis, but a result of analysis of observations. We analysed the motion of particles during time intervals that did not exceed a few minutes. This time is much less than the two days considered by Cochran et al. The Cochran's results are in contrast with the Spitzer observations, where they report a "bluening" of the scattered light (Lisse et al. 2006), and with Schleicher et al. (2006), who noted that the material was redder in color than the general inner coma in the first 15 min.

The above discussion about Cochran's observations testifies in favour of that velocities and size distributions of particles probably did not change much during a few minutes. Therefore, the influence of sublimation on the velocities and size distributions might not be considerable during the few minutes. Biver et al. (2007) concluded that it took about 4 h for water particles to sublimate. Groussin et al. (2010) also supposed that sub-micron particles sublimated in a few hours, and that a massive external source of energy (sunlight) was required for sublimation of ejected icy grains. Though the sublimation rate was highest at the beginning, sublimation could last a few hours, and probably the consideration of sublimation cannot change the conclusions of our paper based on analysis of the motion of particles during a few minutes.

Studies of acceleration of particles and variations of their size distribution during a few minutes can be a subject of another paper. We do not think that the above factors can change considerably the conclusions of our paper because, in our opinion, variations of velocities due to these factors were smaller than considered velocities. These factors can be important if one will analyse look-back images made at times from 45 to 75 min after impact.

## 6 TIME VARIATIONS IN RATES AND VELOCITIES OF EJECTION OBTAINED FOR EXPONENTIAL TIME DECREASE IN THE EJECTION RATE

### 6.1 Time variation in ejection rate

As in Sections 5.1-5.2, in Section 6 we consider the model *VExp* with exponential decrease in ejection velocity. Below we analyse the plots of dependencies of relative rate $r_{te}$ of ejection on time $t_e$ of ejection (Fig. 13). The calculation of the plots is discussed in Section 5.2. The maximum value of $r_{te}$ at $t_e$>0.3 s is supposed to be equal to 1. Because of high temperature and brightness of ejected material, the real values of $r_{te}$ at $t_e$<1 s (especially, at $t_e$<0.2 s) could be smaller than those in Fig. 13. In this figure, there was a local maximum of $r_{te}$ at $t_e$~0.5 s. It is considered that typical sizes of ejected particles increased with $t_e$, and the fraction of small particles in the normal ejecta decreased with $t_e$. Therefore, the ratio of the actual rate of ejection of particles of all sizes to the values of $r_{te}$ presented in Fig. 13 mainly increased with time. Due to this increase, the exponent corresponding to the plot passing via the values of $r_{te}$ at $t_e$=1 and $t_e$=300 s obtained in our model was smaller (equaled to -0.6) than the minimum theoretical value, which was equal to -0.25. The exponent -0.25 corresponds to α=0.75 (see Table 4).

For most suitable pairs of α and *c* (that best fit the velocities presented in Table 3), we obtained a local maximum of relative rate $r_{te}$ of ejection at $t_e=t_{elm}\approx 9$ s (see Table 5). At *t*~12-13 s, there was a jump in the direction from the place of ejection to the brightest pixel (see Section 2.4). Typical projections $v_p$ of ejection velocities at $t_e$~10 s were ~100-200 m $s^{-1}$ (see Fig. 11). The local peak of the rate of ejection at $t_e$ about 9-10 s explains the velocities presented in Table 1.

**[Table 5]**

In the model *VExp* (see Section 5.2), the rate $r_{te}$ of ejection increased on two intervals and decreased on two intervals of $t_e$ (Fig. 13), while theoretical dependence of $r_{te}$ on $t_e$ is always monotonic with exponents presented in Table 4. Therefore, theoretical models do not explain the local maximum of the rate of ejection at $t_e$ about 9-10 s. At $t_e$~0.3-0.5 s, $r_{te}$ can be partly approximated by $r_{te}=c_r\times(t_e-c_t)^{0.2}$, where $c_r$ and $c_t$ are some constants. The exponent 0.2 corresponds to α=0.6 (see Table 4). For 1<$t_e$<3 s and 8<$t_e$<60 s, our model plot $r_{te}(t_e)$ is located above a monotonic exponential curve connecting the values of $t_e$ at 1 and 300 s. In our opinion, such plot could be partly caused (especially at $t_e$~10 s, when there was a local maximum of ejection rate) by a considerable additional ejection (below we call it 'outburst') initiated by the impact. The triggered outburst was suggested by Ipatov & A'Hearn (2008b, 2009, 2010). Two types of the outburst are discussed in Section 7. A reviewer noted that the fact that the observed escaping ejecta is larger than the predicted mass for most of the predictions (see Section 1.3) is one more argument in favour of the outburst. He also paid attention to that a change in the size distribution of the ejecta (due to e.g. layering of the subsurface of the comet) could be an alternative explanation for the observed $r_{te}$ vs. $t_e$. The explanation takes into account that the theoretical ejection rate is proportional to ejected mass, while the measured ejection rate is (approximately) proportional to the total cross section of the ejected material.

At $t_e$~55-72 s, the ejection rate $r_{te}$ decreased by a factor of 1.6 during 17 s (Fig. 13). Such decrease corresponds to the exponent equal to -1.7, whereas the exponent was -0.67 at $t_e$~28-55 s. In images made at *t*~55-60 s, there was a sharp change in the direction from the place of ejection to the brightest pixel (Fig. 3b). The above sharp changes in the direction and $r_{te}$ could be mainly caused by a sharp decrease of the 'fast' outburst (see discussion in Section 7). Note that the peak brightness at 100<*t*<200 s was smaller than at 20<*t*<90 s (Fig. 2). As it is seen from Fig. 10, this local minimum of the peak brightness is associated with the local minimum of a size of a bright region. For a smaller size of a bright region, the number of particles in a line of sight to the brightest pixel is smaller. The correlation between variations in the peak brightness and in the size of the bright region can testify in favour of that the peak optical thickness was less than 1 at that time (or can be due to calibration of saturated pixels).

In Fig. 13 at $t_e$~70-300 s, $r_{te}$ was approximately proportional to $t_e^{-0.6}$. The smaller decrease of $r_{te}$ at $t_e$~400-800 s (than at $t_e$~70-300 s) could be caused by the smaller decrease of the outburst, or/and by material falling back on the comet, or/and by the change in composition (to more icy) of ejected material, or/and by the destruction of larger particles. As $r_{te}$ is approximately proportional to $t_e^{0.2}$ at $t_e$~0.3 s and proportional to $t_e^{-\alpha}$ with α>0 at $t_e$~1-500 s (exclusive for $t_e$~10 s), we can conclude (using also data from Table 4 and taking into account that more solid material corresponds to a smaller value of α) that the material of the comet ejected at $t_e$~0.3 s could be more solid than that at $t_e$>1 s, and especially at $t_e$>70 s.

In the model *VExp*, about a half of material observed at $R$>1 km was ejected during the first 60-120 s (Fig. 14 and Table 5). At $t_e$>100 s, theoretical curves of the relative volume (mass) $f_{et}$ of all material ejected before $t_e$ vs. $t_e$ increased with $t_e$ sharper than our obtained curves. Such difference can be caused by that the mean sizes of ejected particles increased with time, but mainly small particles were observed and considered in our model.

The studies presented in Section 6 were made for the model *VExp* with exponential decrease in ejection velocity at times of ejection 1 s < $t_e$ < $t_{e803}$. In Section 7, we discuss that at $t_e$>70 s time variations in the velocity could be smaller for the outburst than for the normal (without the outburst) ejection. In this case, the fraction of observed material ejected after 70 s is greater than that for the model *VExp*.

**6.2 Amounts of material ejected with different velocities**

Theoretical and our model curves of the relative volume $f_{ev}$ of particles ejected with velocity greater than $v=v_{\text{model}}$ vs. $v$ are presented in Fig. 15. The marks in the figure correspond to the values of $f_{ev}$ obtained for the model *VExp* based on the analysis of observations of *small* particles, and the lines correspond to theoretical estimates for *all* particles. Besides theoretical exponents from Table 4, in Fig. 15 we present also the proportionality to $v^{-2.25}$ obtained in experiments (Gault, Shoemaker & Moore 1963; Petit & Farinella 1993; Holsapple & Housen 2007). In our model, the values of the relative volume $f_{et}$ of observed material ejected before $t_e$ (Fig. 14) and $f_{ev}$ are equal to 1 for the material ejected before the time $t_{e803}$ of ejection of particles corresponding to the bright region in the image made at $t$=803 s.

Comparison of the curves presented in Fig. 15 shows that the fraction of *observed* particles with 30<$v_p$<800 m s$^{-1}$ (among observed particles with all velocities) was greater than the theoretical estimates of the fraction of *all* particles with such velocities. The maximum difference in the fractions was at $v_p$~30-60 m s$^{-1}$. The difference could be partly caused by that there could be a lot of outburst particles with velocities $v_p$ greater than 30 m s$^{-1}$. Theoretical models show (e.g. Holsaple & Housen 2007) that most of material ejected as a result of the DI collision had velocities less than 1.7 m s$^{-1}$. Therefore, mean velocities of the observed outburst particles were greater than mean velocities $v_p$ of all ejected particles. Another reason of the difference between observations and theory is that larger particles could be mainly ejected with smaller velocities (at greater times), but only small particles were mainly observed. As we discuss in Sections 1.3 and 5.3, the acceleration of particles by gas was not considerable during a few minutes of motion of particles and did not affect much the velocities $v_p$≥20 m s$^{-1}$ that we measured based on the analysis of images. So the acceleration was not the main cause of the excess of particles with $v_p$≥30 m s$^{-1}$ in the model *VExp* based on the DI observations.

Let us estimate the total mass $M_t$ of observed ejected material. Melosh (2006) obtained that the mass $M_1$ of droplets ejected during the first second was about $4\cdot10^3$ kg. Material ejected during the first second was less icy, but hotter than other ejected material. Therefore, it is not clear whether it was brighter or not than the later ejecta. At $M_1$=$4\cdot10^3$ kg and the fraction $f_1$ of observed material ejected during the first second obtained for the *Vexp* model (see Table 5), we have $M_t=M_1/f_1\sim10^5$ kg. The value of $M_t$ is less than the total mass $M_{tot}$ of all ejected particles because the minimum velocities $v_{e803}$ considered in our models (see Table 5) are greater than the escape velocity $v_{es}\approx1.7$ m s$^{-1}$ by a factor of more than 4, and our estimate of $M_t$ corresponds only to small particles that mainly contribute to the brightness of the cloud. The obtained value of $M_t\sim10^5$ kg is in accordance with most estimates of the total mass of ejecta with $d<3$ μm presented in Table 1. This is a confirmation of the previous conclusion (see Section 1.3) that diameters of the particles that mainly contribute to the brightness of the cloud are less than 3 μm. Particles considered in the model *VExp* did not fall back on the comet because $v_{\min}=v_{e803}>v_{es}$, where $v_{e803}$ is the model velocity of ejection of particles that constitute the edge of the bright region in the image made at $t$=803 s.

Below we compare our estimates of the amount and fraction of observed (small) particles ejected with velocity $v_p>100$ m s$^{-1}$ with similar theoretical estimates made for all particles. For the model *VExp* at α≈0.71, about 4 per cent of observed material were ejected during the first second with $v_p>500$ m s$^{-1}$. As we noted in the previous paragraph, the amount of this material can be estimated as $\sim4\cdot10^3$ kg, i.e. 1 per cent of observed ejected material could be $\sim10^3$ kg. At α=0.71 and $c$=2.5, the fractions of material ejected with $v_p\geq200$ m s$^{-1}$ (at $t_e\leq6$ s) and $v_p\geq100$ m s$^{-1}$ (at $t_e\leq14$ s) equaled to 0.13 and 0.22, respectively. At these fractions and $M_t\sim10^5$ kg, the mass of observed particles ejected with velocities $100\leq v_p\leq200$ m s$^{-1}$ was $\approx10^4$ kg, and that with $v_p\geq100$ m s$^{-1}$ equaled to $\approx2\cdot10^4$ kg. Similar estimates were obtained at α=0.75 (see the fractions of particles ejected with different velocities presented in Table 5).

For calculations of a DI-like impact with the SPH code, less than 1 per cent of all ejected material got velocities $v>100$ m s$^{-1}$ (Benz & Jutzi 2007). For all models presented by Richardson et al. (2007) in their fig. 19, the total mass of material ejected with $v>100$ m s$^{-1}$ was about $10^3$ kg and was less than 1 per cent of the total ejected mass for most values of effective strength $S$ considered. Their models considered particles of all sizes, but nevertheless their theoretical estimates of the mass (in kg) of particles with $v>100$ m s$^{-1}$ were much smaller than our estimates made only for small particles. The first reason of the greater fraction of high velocity ejected particles in our model compared with theoretical models is that our estimates were based on observations of a cloud of small particles, and theoretical estimates were made for all particles. It is considered (see e.g. Petit & Farinella 1993) that smaller ejected particles have greater velocities. The triggered outburst is another reason of the difference. As we discuss in Section 7, the ejection of particles with velocities of ~100 m s$^{-1}$ due to the 'fast' outburst could last for at least tens of seconds, and the duration of the normal ejection with such velocities was much shorter. For the model discussed below, the fraction of particles ejected with velocity of ~100 m s$^{-1}$ is greater than for the model *VExp.*

## 7 THE MODEL OF A SUPERPOSITION OF OUTBURSTS AND THE NORMAL EJECTION

### 7.1 'Fast' and 'slow' outbursts

In Sections 5-6, we analysed the model *VExp,* for which velocity $v_e$ of ejection varies exponentially with time and is the same for all particles ejected at the same time $t_e$. The velocity $v_e$ used in this model was calculated in Section 4 as the mean velocity of the motion from the place of ejection to a distance of a few kilometers. In an image made at time $t$, the brightness $Br$ at distance $L=(t-t_e)\times v_e$ from the place of ejection was supposed to be proportional to $r_{te}/(L\times v_e)$, where $r_{te}$ is the relative amount of material ejected per unit of time at $t_e$. That is, at distance $L$, the surface of a sphere is proportional to $L^2$, but the length of the part of the line of sight that crosses the region with brightness greater than $Br$ is considered to be proportional to $L$.

Below we discuss another model, for which the relation $v_e = c_{ve} \times t_e^{-\alpha}$ is valid for the normal ejection with ejection rate $r_{te}$, there is a superposition of the outburst and the normal ejection, and the velocities of the outburst can differ from $v_e$. In this model, the brightness $Br$ at distance $L$ is proportional to $r_{te}/(L \times v_e) + r_{teof}/(L \times v_{eof}) + r_{teos}/(L \times v_{eos})$, where $r_{teof}$ is the rate of the 'fast' outburst ejection with velocity $v_{eof}$ at time $t_{eof}$ of ejection, and $r_{teos}$ is the rate of the 'slow' outburst ejection with velocity $v_{eos}$ at $t_{eos}$. If below we do not specify the type of outburst, then we consider $v_{eo}$ as the mean outburst velocity. Sometimes we use $v_e$ and $t_e$ if do not specify the type of ejection.

The 'fast' outburst could be caused by the ejection of particles from the cavities (reservoirs) located at some distances below the surface of the comet (see also Section 9.2). These cavities contained icy material under gas pressure. Some cavities could be deep, and the ejection of particles from the interior of the cavities could be due to pressure of the gas that was inside the cavities. Velocities of such particles could be greater than velocities of ejected walls of cavities, which were close to velocities of other ejected material of the crater. The outburst ejection of material from some of these cavities could be greater for specific directions. Therefore, the role of the outburst in the direction from the place of ejection to the brightest pixel could be greater than in the total ejection rate.

The first relatively large cavity probably was excavated at $t_e \approx 4$ s, and the increase in the rate of ejection during the next 0.5 s was ~0.2 of the normal ejection rate. At $t_e \approx 8$ s, a greater cavity was excavated. This cavity could be deep because the excavation from the cavity could last for at least a few tens of seconds. The direction from the place of ejection to the brightest pixel in images made at $t$~13-55 s (see Fig. 3b) probably depended much on the ejection from this cavity. The beginning of the main excavation of the cavities at $t_e$~4-8 s shows that the cavities were not close to the surface of the comet (their upper boarders may be located about a few meters below the surface). With the increase of the crater, more cavities could be excavated.

Velocities of different particles ejected at the 'fast' outburst (or at the 'slow' outburst, or at the normal ejection) could be different even for the same ejection time. Some small particles that were inside a cavity under gas pressure could have initial velocities not much less than those of gas. Other particles could be captured by gas from cavity's walls and could have smaller velocities. We suppose that the mean values of velocity $v_{eof}$ at the 'fast' outburst could be ~100 m s$^{-1}$. The reasons of this mean value of $v_{eof}$ are the following: (1) A great number of particles with $v_p$~100 m s$^{-1}$ were observed by ground-based telescopes one hour after the impact. (2) Our estimates of velocities $v_p$ are almost the same (are equal to 100 m s$^{-1}$) for different $t_e$ at $10 \le t_e \le 20$ s (see Table 3). In principle, besides small particles, much larger objects (e.g. parts of walls of the cavities with size of up to a meter) could be also ejected during the 'fast' outburst. Typical velocities of these objects were probably much smaller than those of small particles. If such meter-sized objects existed, most of them (as >90% of all ejected mass; see Richardson et al. 2007) probably did not reach distances from the comet greater than 1 km and moved inside a saturated region in DI images. So it would be difficult to find such objects in images.

The 'slow' outburst ejection could be similar to the ejection from a 'fresh' surface of a comet. The duration of ejection of particles from the fresh surface of the crater could be much longer than the time of formation of the crater and the duration of the normal ejection. Actually only the 'fast' outburst is a real outburst, but below it is convenient to denote both types of additional ejection triggered by the impact as an outburst. The gas pressure inside cavities in the comet that pushed out particles could differ for different cavities and different particles, initial velocities of particles could be different, and there could be no strict boarder between velocities for the ‘fast’ and ‘slow’ outbursts.

In our opinion, the DI triggered outbursts could be produced by the ejection of material from the whole surface of the DI crater (in contrast with the normal ejection only from the edges of a crater), and there could be regions in the crater with greater and faster ejection than from other regions. The rate of ‘slow’ outburst ejection could mainly increase while the size of the crater was growing. There can be conglomerations of solid ice inside a comet. Therefore, the fraction of icy particles in the ejected material could vary with time. Such conglomerations could also affect the time variations in the rate of ejection of observed particles as the brightness of the DI cloud was mainly due to small icy particles. After the end of growth of the DI crater, the outburst could decrease because some material fell back into the crater, and part of icy material located close to the surface of the crater had been already ejected. The area of the DI crater was small compared with the surface of the comet. Therefore, though the ‘slow’ outburst could last for many days, its contribution to the total ejection from the comet was negligible a few days after the impact.

In Section 4, we estimated the mean values of velocities $v_p$ of ejected particles during their motion from the place of ejection to a distance of a few km. At $73 \le t_e \le 115$ s, the velocities were equal to 20-25 m $s^{-1}$ and practically did not depend on $t_e$. We did not estimate $v_p$ at times greater than 115 s. It may be possible that at $t_e \ge 70$ s a considerable fraction of material was ejected due to the ‘slow’ outburst. It may be possible that $v_p$ did not vary much with time $t_e$ of ejection and was about ~20-25 m $s^{-1}$ at $t_e$~70-800 s. In Section 5.3, we noted that the increase in velocity of particles due to the dust-gas interaction probably did not exceed a few meters per second during their motion with velocity of a few tens of m $s^{-1}$ along a distance of a few km. The increase was greater if particles left the surface of the comet with a velocity of no more than a few meters per second, because greater time was needed to pass the same distance with smaller velocities. At $v_p$=20 m $s^{-1}$ for the distance of the edge of the bright region from the place of ejection equal to $L$=1500 m, we obtain $dt=t-t_e=L/v_p$=75 s.

The outburst probably did not dominate the ejection of observed material during the first 100 s, because in this case it may be difficult to explain the observed exponential decrease in the velocities of ejected particles obtained in Section 4. At some time intervals (e.g. after the end of formation of the crater), the mean velocity of outburst particles could decrease with time due to the increase in the fraction of ‘slow’ outburst particles. If at some time interval the mean outburst velocities decreased with time approximately exponentially, then a major part of observed particles could have been ejected at the outburst, and the time variations in velocities of the normal ejection and those of the outburst could be characterized by similar exponents, but by different coefficients $c$, in order to satisfy the estimates of velocities presented in Table 3.

### 7.2 Time variations in rate of ejection and duration of ejection

In this subsection, we use the plots obtained for the model *VExp* for the estimates of time variations in the rate of ejection for the model for which velocities of ejection could be different for the normal ejection and for the outburst. In Fig. 13 at $\alpha$=0.75 and $c$=3, the relative rate $r_{te}$ of ejection decreased as $t_e^{-1}$ during 1-4 s, increased by a factor of 1.2 during 4-4.5 s, and decreased as $t_e^{-0.5}$ at $4.5 \le t_e \le 7.3$ s. For $7.8 \le t_e \le 10.8$ s, $r_{te}$ was greater by a factor of 1.1-1.15 than at $t_e$=7.3 s. It decreased as $t_e^{-0.78}$ at $12.4 \le t_e \le 54$ s. For $4 \le t_e \le 110$ s, the average decrease of $r_{te}$ was proportional to $t_e^{-0.62}$. The decrease in the normal ejection could be greater than $t_e^{-0.62}$ because at $t_e$=110 s some material probably was ejected due to the outburst.

Starting from 4 s and using the proportionality of $r_{te}$ to $t_e^{-0.62}$, we obtain that the estimated value of $r_{te}$ at 10 s is smaller by a factor of 2 than the value presented in Fig. 13. This factor equals to 1.3 if we start from 7.3 s and compare the values of $r_{te}$ at 10.8 s. In Table 3, $v_e \approx 100$ m s$^{-1}$ at $t_e \sim 10$-20 s, i.e. $v_{eof} \approx v_e$. The above estimates suggest that at $t_e$=10 s about half of the ejection could be due to the 'fast' outburst (see the above comparison of $r_{te}$ at 4 and 10 s), but only ~1/3 of the outburst was due to the additional outburst that started at ~8 s (see the comparison at 7.3 and 10.8 s). The latter 'fast' outburst was less uniform than other outbursts and the normal ejection, and it had a considerable influence on the position of the brightest pixels in images and on several other parameters of ejection (see e.g. Sections 2.4 and 6.1).

For theoretical models, the rate of ejection varies as $t_e^{-\beta}$ with $\beta \le 0.25$ (see Table 4). In Fig. 13, the rate decreases more sharply than even at $\beta$=0.25. It is mainly caused by a decrease of the fraction of small particles among all ejected particles (at least for the normal ejection) with time $t_e$ of ejection. If we compare $t_e^{-0.62}$ (the dependence based on the values of $r_{te}$ at 4 and 110 s in Fig. 13) and $t_e^{-0.25}$ (the theoretical dependence of $r_{te}$ on $t_e$ presented in the first line of Table 4), then the fraction of small observed particles in the rate of ejection of all particles decreases (as $t_e^{-0.37}$) by a factor of 2.3, 5, and 12.9 with the increase of $t_e$ by a factor of 10, 100, and 1000, respectively.

There was a sharp decrease in the ejection rate at $t_e \sim 55$-75 s in Fig. 13. For this time interval, the decrease was greater by about a factor of 1.3 than it should be if it was the same as for the previous time interval. If the decrease was fully caused by the end of the normal ejection, then >2/3 of the brightness of the edge of the bright region corresponding to $t_e \sim 60$ s was due to the outburst. If the decrease was caused only by the decrease of the outburst, then ≥1/4 (=0.3/1.3) of the brightness at $t_e \approx 55$ s was due to the outburst. It is more probable that the above decrease of ejection was mainly caused by the decrease of the 'fast' outburst that began at $t_e \approx 10$ s, and no new large cavities were excavated after 50-60 s. The direction from the place of ejection to the brightest pixel quickly changed with time in images made at $t \sim 12$-13 s, and at $t \sim 55$-60 s it returned to the value that was at $t \le 12$ s. Such time variation in the direction also testifies in favour of the above conclusion that the decrease in the rate was mainly due to the outburst. In Section 9.3, we discuss the arguments which testify in favour of that the duration of the normal ejection exceeded 1 min. During some time interval, the rate of ejection of small particles at the 'fast' outburst could be compared with that of the normal ejection. The total mass of ejected particles with $d$<2.8 μm could exceed $10^5$ kg (see Table 1). Therefore, the amount of such particles ejected due to the 'fast' outburst could exceed $10^4$ kg by a factor of several.

At $t_e \sim 55$-75 s, the contribution of the 'fast' outburst to the rate of ejection of observed particles was greater than its contribution to the brightness of the DI cloud because the 'fast' outburst velocities were greater than those of the normal ejection at that time. (The brightness was used for construction of Figs. 10-15.) In Fig. 11, we have $v_p \approx 50$ m s$^{-1}$ at $t_e \approx 60$ s. If at $t_e \approx 60$ s the characteristic outburst velocities of particles were 100 m s$^{-1}$ (i.e. about twice greater than velocities of the normal ejection) and the contribution of the outburst to the brightness of the edge of the bright region was about the same as the contribution of the normal ejection, then (taking into account that brightness is inversely proportional to velocity) we obtain that the rates of ejection of small observed particles for the 'fast' outburst were greater by a factor of two than those for the normal ejection. The above estimates are for the 'fast' outburst. For smaller characteristic outburst velocities (i.e. for a greater fraction of 'slow' outburst particles), the contribution of the outburst to the rate of ejection would be smaller than that for the ‘fast’ outburst. Though the above estimates are very rough, they show that at some time the contribution of the ‘fast’ outburst to the rate of ejection of small observed particles could be comparable to that of the normal ejection. At $t_e$>25 s, the velocities of particles ejected due to the ‘fast’ outburst were probably greater than those of the normal ejection. The ‘fast’ outburst with velocities ~100 m $s^{-1}$ probably could last for at least several tens of seconds, and it could significantly increase the fraction of particles ejected with velocities ~100 m $s^{-1}$, compared with the normal ejection and even with the estimates for the model *VExp* with an exponential time decrease in characteristic velocity of ejected particles.

In Fig. 10, one can see the increase in $L_3$ (which is usually close to the distance from the place of ejection to the edge of the bright region on DI images) at $t$~150-800 s. We do not think that it was caused by an increase in the observed ejection rate, as the crater probably did not grow during all this time. The increase in $L_3$ could be caused by the small decrease in characteristic ejection velocities $v$ with time (as brightness $Br$ is proportional to $v^{-1}$) and/or by the increase in the production of small particles in the DI cloud (as greater time is needed for particles to pass the same distance at smaller velocity).

The particles that mainly contributed to the brightness of the DI cloud were small, and more massive particles had smaller velocities (see references in Section 1.3). For the model *VExp* at $t_e$<13 min, characteristic velocities of ejected particles were greater than 7 m $s^{-1}$ (see Fig. 11) and exceeded the escape velocity of 1.7 m $s^{-1}$ by at least a factor of several, and so these small particles did not fall back on the comet. It is seen from fig. 6 from (Holsaple & Housen 2007) that particles with $v_p$<1 m $s^{-1}$ did not reach $R$=1 km. Theoretical estimates presented in fig. 2 of the same paper show that less than 3 per cent of ejected material had velocities greater than 1 m $s^{-1}$. Richardson et al. (2007) also concluded that most ejected particles never got more than a few hundred meters off the surface of the comet. Therefore, most of ejected material, including most of large particles, made a small contribution to the brightness of the part of the cloud located at a distance $R$>1 km from the place of ejection. Our studies of the rate of ejection were based on analysis of regions of the cloud at $R$>1 km (for $t$>1 s). The above discussion shows that the role of the particles with velocities smaller than the escape velocity in the value of $L_3$ was small, though these slow particles made up the major fraction of the ejected mass. Figs. 13-15 describe the ejection only of those particles that reached the edge of the considered bright region. If ejection finished, for example, at $t_{\rm end}$~6-7 min, then all bright regions observed up to $t$~13 min must be caused by particles ejected before $t_{\rm end}$, and it is difficult to imagine that there was a considerable ejection of small bright particles with a wide range of velocities (including small velocities) at the same time.

The existence of rays of excessive ejection close to the nucleus in images made up to 13 min (see Section 8) also testifies in favour of the ejection of particles at $t_e$~10 min. This excessive ejection probably was due to the ‘fast’ outburst. The contours CPSB=const on the DI images are easily explained by the continuous ejection of material during at least 10 minutes after collision. For example, the time $t_{e803}$ of ejection of particles constituting the edge of the bright region in an image made at $t$=803 s exceeded 10 min for all models considered in Table 5. It is difficult to conclude about the times of the end of the normal ejection and the outburst. In principle, the ejection time could be smaller than 10 min because variation in brightness of the cloud might not depend only on the rate of ejection, and the relation between flux and ejected mass was non-linear (see Section 1.4). In any case, we suppose that the normal ejection was shorter than the outburst, and its duration did not exceed a few minutes.

According to Cochran et al. (2007), there was no considerable fragmentation of icy grains that increased the brightness of the cloud (for the same total mass of the cloud). As the total brightness of the DI cloud increased during the first 35-60 min (see Section 1.4), the Cochran's conclusion may show that duration of the triggered outburst could exceed 35 min. The long ejection is in accordance with the conclusion by Harker et al. (2007) that the best-fit velocity law necessitates a mass production rate that was sustained for duration of 45-60 min after impact.

## 8 RAYS OF EJECTED MATERIAL

The bumps on the left and right edges of some contours CPSB=const (see e.g. Figs. 8a-c) were produced by the rays of ejected material (i.e. more material was ejected in some directions; see e.g. Schultz et al. 2007). It is considered that the rays are caused by internal sources of energy of the comet released after impact. The effect of an oblique impact could also play a role in the asymmetry of the cloud of ejected material (Richardson et al. 2007), but it could not give such sharp rays as the observed rays. In our opinion, the rays of ejected material could be caused mainly by the excess of ejection in some directions during the triggered outburst. Together with hydrodynamics of the explosion, the ejection of particles from the former cavities with gas under pressure and from icy conglomerates in the crater could affect the formation of the rays. The rays were probably caused by a greater amount of small (~1 μm) particles ejected in a few directions. The ejection of massive particles, which contribute less to the brightness of the DI cloud, could be more uniform for different directions than the ejection of small icy particles.

For considered bumps of contours CPSB=const, the values of the angle $\psi$ between the upper direction and the direction to a bump measured in a clockwise direction are presented in Table 6. The bumps are seen on most of the images made during the first 13 minutes.

**[Table 6]**

In this and the next four paragraphs, we study time variations of the *upper-right bump*. For this bump of the contour CPSB=1, at most times the values of $\psi$ were close to 65-70$^o$ (e.g. in Figs. 6b,d, 7b,d, 8d-f), but $\psi\approx60^o$ in Fig. 8a (at $t$=80 s) and $\psi\approx80^o$ in Fig. 7c (at $t$=140 s). Note that for CPSB=1, there was a local minimum of $L(t)$ at $t$~80-140 s in Fig. 10. The upper-right bump was usually accompanied by the upper-upper-right bump with $\psi\approx$40-45$^o$ in most images ($\psi\approx50^o$ in Fig. 8e at $t\approx$350 s). These two bumps usually had similar sizes and can be considered as two parts of a M-type bump. At $t$=2.7 s, the upper-right bump is seen for CPSB=1, but is not seen for CPSB≤0.3. Using the obtained relationship between $t_e$ and $t$ for CPSB=3 and taking into account that particles constituting the contour CPSB=1 were ejected before those for CPSB=3, we obtain that already at $t_e\approx$1 s there could be excessive ejection in this direction. At $t$~5-14 s ($t_e$~3-8 s), the bump is not practically seen for the considered contours of constant CPSB. After $t$=15 s, the upper-right bump for the contour CPSB=1 began to increase with time. Probably, the outburst that began at $t_e$~10 s caused the changes of the direction from the place of ejection to the brightest pixel in images made at $t$~12-13 s (see Fig. 3), the local increase in the peak brightness (Fig. 2b) and the rate of ejection (Fig. 13), and the upper-right bump (a few seconds were needed for particles to reach the contour CPSB=1).

Studies of the bumps allowed us to estimate velocities of ejected particles at several times of ejection. For the contours CPSB=1, the upper-right bump is seen much better at 25≤$t$≤43 s (Fig. 6d) than at $t$≤21 s (Fig. 6b). The sharpest bumps are seen for CPSB=1 at $t$=43 s in Fig. 6b, for CPSB=1.5 at $t$=39 s in Fig. 7a, and for CPSB=0.5 at $t$=66 s in Fig. 7b. The distance between the two last mentioned bumps is passed at velocity of ~**120** (=3300/(66-39)) **m s$^{-1}$**. Material of these contours moving with such velocity was ejected at $t_e$~**20 s**. In Fig. 7c ($t$=142 s), the bump of the contour CPSB=0.5 is clearly seen. It is located at distance $D$≈4200 m from the place of ejection. Note that $D$ is the projection of the distance onto the plane perpendicular to the line of sight, and the real distance is larger than $D$. Considering that $t_e/t$≈0.3 (as for CPSB=0.5 and $t$=66 s), we obtain $t_e$≈142·0.3≈**43 s** and $v_p$≈4200/(142-43)≈**42 m s$^{-1}$**. Therefore, at $t_e$~40 s the excess of ejection in the upper-right direction could still be considerable. The above pairs ($t_e$ and $v_e$) are in a good agreement (see Fig. 11) with the data obtained in Sections 3-4 with the use of quite different approaches and presented in Table 3. This agreement testifies in favour of the correctness of our estimates of velocities of ejection made by different methods.

It is seen from Fig. 13 that at 1<$t_e$<3 s and 8<$t_e$<60 s our model plot of the rate of ejection was located upper than the line of monotonic exponential decrease. Therefore, the greater was the rate of ejection due to the outburst, the greater were the rays of ejected material.

The upper-right bump is also seen for some outer CPSB contours at $t$~300-770 s. For example, the bump of the contour CPSB=1 is seen in Fig. 7e at $t$=529 s, though the distance of the contour CPSB=1 from the place of impact is only 2 km. The contour CPSB=1 in an MRI image at $t$=772 s also has the same bump. The bumps in images made up to $t$~13 min testify in favour of that there was the excessive ejection of particles to a few directions at $t_e$~10 min.

In the series *Mb* at $t$=139 s for the upper-right bumps of the contours CPSB=0.1 ($D$≈15 km) and CPSB=0.03 ($D$≈25 km), $v_p$ was greater than 110 and 180 (≈25000/139) m s$^{-1}$, respectively. At $t$=191.5 s and CPSB=0.03, the bump is located at $D$≈30.5 km, and these values of $t$ and $D$ correspond to $v_p$≥160 m s$^{-1}$. These estimates show that velocities of some particles constituting the rays were ~100-200 m s$^{-1}$.

A small *right bump* (which became down-right with time) is seen at some contours in Figs. 5-8. For this bump, ψ increased from 90$^o$ in Figs. 5b-c (at $t$~4-8 s) to 110-120$^o$ in Figs. 6d, 7d, 8f (at $t$~25-400 s). At $t$~4-12 s (i.e. before the jump in the direction to the brightest pixel), the right bump was mainly greater than the upper-right bump, but later it was not well seen.

The 'M'-type (i.e. double) *left bump* is clearly seen for three outer contours in Fig. 8a (at $t$=78 s), for two contours (CPSB=0.1 and CPSB=0.03) in Fig. 8b ($t$=139 s), and only for the contour CPSB=0.03 in Fig. 8c ($t$=191.5 s). Therefore, considerable excessive ejection in this direction was not long (<100 s). The end of this ejection can be associated with the relatively sharp decrease in ejection rate at $t_e$~55-70 s (Fig. 11) and with the sharp variation in the direction from the place of ejection to the brightest pixel in images made at $t$~55-60 s (Fig. 3b). The left bump is also seen in Figs. 6d, 7a-b (at $t$~25-66 s). For the upper and lower parts of the M-bump, we obtained ψ≈260$^o$ and ψ≈245$^o$, respectively. For outer contours in Figs. 7b and 8b-c, these parts are a little smaller than the upper-right and upper-upper-right bumps. The direction from the place of ejection to the lower part of the left M-bump (ψ≈245$^o$) is opposite to the upper-right bump (ψ≈70$^o$). Both directions are approximately perpendicular to the direction of impact. Note that one of the rays of ejected material obtained in the experiment by Schultz et al. (2007) and presented in their fig. 31 was also perpendicular to the direction of impact. As there was no outburst in their experiment, there could be an excess of the normal DI ejection (may be together with the outburst excess) in the directions perpendicular to the direction of impact. At $t_e$>100 s, instead of the left M-bump there was the *down-left M-bump*, which is less clearly seen than the left bump at smaller $t_e$. For two parts of the down-left M-bump, the values of ψ are about 230-235$^o$ and 210-220$^o$. The down-left bump is still seen in images at $t$~400-760 s. The above discussion shows that directions of excessive ejection could vary with time.

The *upper bump* of the outer contour is clearly seen at $t$~139-411 s in Fig. 8b-f (especially, in Fig. 8c). At $t$~25-42 s, the contour CPSB=1 has the same bump (Fig. 6d). The angle $\psi$ varied from about 0 in Fig. 8b (at $t$=139 s) to -25° in Fig. 8f (at $t$=411 s). It was -15° in Fig. 6d (at $t$~8-21 s). Note that the values of $\psi$ for the upper-right bump at $t$~140 s are also different from those at much smaller or larger $t$, and there was the minimum size $L_3$ of the region inside the contour CPSB=3 in images made at time close to 140 s (see Fig. 3). The direction from the place of ejection to the upper bump was not far from the direction opposite to the impact direction (i.e. the bump corresponds to the excessive ejection backwards to the impact direction), but was not exactly perpendicular to the line connecting down-left and upper-right bumps. The upper bump is not well seen in all contours in Fig. 8a (at $t$=78 s). Therefore, the upper bump of CPSB=0.03 in Fig. 8c ($t$=191.5 s) consisted mainly of particles ejected at $t_e$>80 s, and the excessive ejection backwards to the impact direction was mainly after 80 s, though it could be also found in images made at $t$~8-42 s. Schultz et al. (2007) concluded that uprange ejecta plume directed back out the initial trajectory (during the first 10 s) and at very late stage (700 s). In our studies, the upper bump was more pronounced if it consisted of particles ejected at $t_e$>100 s. For experiments described by Hermalyn et al. (2008), at the middle of ejection time interval, the velocities of material ejected in the uprange direction were smaller than in the downrange direction. The DI images are in agreement with these experiments.

In Figs. 7b and 8c for the upper-right bump at a distance from the place of ejection ~3-10 km, the values of CPSB for the bump were greater by a factor of 1.3 than those for a close no-bump region. For the left bump, this factor was ~1.1-1.3. The direction from the place of ejection to the brightest pixel was down-right-right at $12<t<55$ s and down-right at other values of $t$. It did not coincide with the directions of rays mentioned above (see e.g. Figs. 5-6).

# 9 DISCUSSION

## 9.1 Differences between the DI ejection and theoretical models, experiments, and natural outbursts

Conditions of ejection of material from Comet Tempel 1 were different from those for experiments and theoretical models. The difficulties in having different gravity, velocities, sizes in laboratory experiments compared to Deep Impact are partly overcome by use of scaling laws involving non-dimensional quantities (see e.g. Housen & Schmidt 1983, Holsapple 1993). The great difference in projectile kinetic energy introduces challenges when scaling the laboratory results to DI conditions, e.g. some materials will vaporize that otherwise would remain in solid or liquid form (Ernst & Schultz 2007). Holsapple & Housen (2007) concluded that for the normal cratering mechanism only a negligible amount of mass ejected had velocities of the order of 100's of m $s^{-1}$ and velocities of 100's of m $s^{-1}$ that were observed are due to the particles which were accelerated by vaporization of ice in the plume and fast moving gas. The fraction of water vaporized at the impact is considered to be ~0.2 per cent of the total amount of water ejected (DiSanti et al. 2007).

According to Biver et al. (2007), the amount of water released at the DI impact was about 0.2 days of normal activity, but that during the natural outburst on 22-23 June, 2005 was about 1.4 days of normal activity (i.e. was larger than at the DI burst). At the natural outburst, water was in the form of gas, so the outburst was not as bright as the burst after impact. In Section 7, we discuss that a considerable fraction of the brightness of the DI cloud could be due to the triggered outburst (probably except for the first few seconds after impact) and the outburst could increase the duration of ejection of material and the velocities of ejected particles and caused the jumps in time variation in the rate of ejection.

A few other differences of the DI ejection from experiments are the following: gravity on the comet (0.04 cm $s^{-2}$) is much smaller than that on the Earth (9.8 m $s^{-2}$), and masses of projectiles in experiments were small. Diameters of particles that made the main contribution to the brightness of the DI cloud are considered to be less than 3 μm, and sizes of sand particles in experiments were much larger (~100 μm) than those of the observed DI particles. The observed DI cone of ejected material was formed mainly by small particles, which had higher velocities than larger particles.

For an oblique impact, on the down-range side of the ejecta plume, ejection velocities are higher and particle ejection angles are lowered compared with a vertical impact (Richardson et al. 2007). For all models considered by Richardson et al. (2007) and Holsapple & Housen (2007), most of the mass was ejected with $v$<3 m $s^{-1}$. If we extrapolate a plot of $v_{\rm model}$ at α=0.71 in Fig. 11 to greater $t_e$, we obtain $v$=3 m $s^{-1}$ at $t_e$~3000 s = 50 min. Actually, the ejection with $v$~3 m $s^{-1}$ due to normal cratering took place at times $t_e$ which did not exceed a few minutes, together with the ejection with greater velocities due to the outburst, which might continue for a long time. Besides the outburst, the differences between theoretical estimates and observed velocities are partly caused by that in the model considered by Richardson et al. (2007), all particles ejected at the same time had the same velocities and were ejected at the same distance from the place of impact. In our opinion, at the same time DI particles could be ejected with different velocities and at different distances from the center of the crater.

We studied the motion of small particles with velocities greater than the escape velocity at $t$<13 min. These particles constituted a small part of all ejected material. Our studies were based on analysis of the contours which correspond to the material located at a distance $R$ greater than 1 km from the place of ejection. While analyzing DI images, Richardson et al. (2007) and Holsapple & Housen (2007) considered the motion mainly of particles that were ejected with small velocities $v_e$ and fall back on the comet (i.e. they studied the motion of quite different particles than we). Holsapple & Housen (2007) analysed ejecta trajectories at $v_e$≤1.8 m $s^{-1}$ and the locus of particles at 7 specific times from 15 min to 2 h (located at a distance of a few kilometers from the place of ejection). At that time, particles considered in our studies were much farther from the comet. Richardson et al. (2007) studied the plume base; it was of order 150-350 m in diameter at time 9 to 13 min after the impact. They concluded that >90 per cent of the ejected mass never gets more than a few hundred meters off the surface of the comet, and has been redeposited within 45 min after the impact. We did not analyse the ejection of slow-moving particles and did not make any conclusions based on the particles that were located at $R$<1 km in images made at 1 s<$t$<13 min.

In the model *VExp*, all particles were supposed to be ejected with the same velocity at each considered time (other models are discussed in Section 7). We plan to make computer simulations of the brightness of the cloud produced by particles ejected at different times with different velocities and at different ejection rates in order to choose such time variations in the rates and velocities and such velocity distributions that best fit the observations. The integration of the motion of particles will be made similar to (Ipatov & Mather 2006). We will study the combination of models of the normal crater ejection (when velocity decreases exponentially with time) and of the triggered outbursts with relatively small variations in velocity.

### 9.2 Outbursts from different comets

Outbursts from different comets were observed. They testify in favour of the existence of cavities with gas under pressure and the relatively close location of the cavities to the surface of a comet. The triggered DI outburst was one of many other outbursts of comets.

The total mass of material ejected at the 2007 October 24 outburst of Comet 17P/Holmes (~1-4 per cent of the nucleus mass of the comet, i.e. $(1\text{-}3)\times10^{11}$ kg) was much greater than that at the DI collision. Schleicher (2009) concluded that production of OH decreased by a factor of 200-300 during 124 days after the outburst of Comet 17P/Holmes in 2007, but it was still greater than before the outburst. It shows that the ejection of material from a 'fresh' surface of a comet can make a noticeable contribution to the total ejection from the comet for many days. Schleicher (2009) suggested that the explosion occurred at greater depth in Holmes than in other comets. Possibly, the explosion at such depth can explain large (up to 125 m $s^{-1}$) on-sky velocities of 16 large (with effective radii between ~10 and ~100 m) fragments of Comet 17P/Holmes observed by Stevenson, Klena & Jewitt (2010).

Outbursts from comets caused by internal processes could last for weeks or months, much longer than for the DI outburst caused by the impact. Paganini et al. (2010) observed the outburst activity of Comet 73P/Schwassmann-Wachmann 3 in 2006 May. They showed a decrease in gaseous productivity of this comet by a factor of 2 in about a week. Prialnik, Benkhoff & Podolak (2004) concluded that the outburst of Comet 1P/Halley could take place during a few months when the comet moved at a distance greater than 5 AU from the Sun.

Prialnik et al. (2004) supposed that crystallization of amorphous ice in the interior of the porous nucleus, at depths of a few tens of meters, caused the release of gas. The role of crystallization of amorphous ice in bursts of comet activity was discussed in several other papers. A few references and examples of such bursts are presented by Prialnik (2002). Boehnhardt (2002) concluded that if the gas pressure cannot be released through surface activity, the tensile strength of the nucleus material can be exceeded and fragmentation of the comet occurs. Internal gas pressure is considered to be one of the main reasons of splitting of comets (Boehnhardt 2004, Fernandez 2009). Ishiguro et al. (2010) concluded that the 2007 outburst of Comet 17P/Holmes was caused by an endogenic energy source. Reach et al. (2010) supposed that the explosion of this comet was due to crystallization and release of volatiles from interior amorphous ice within a subsurface cavity: once the pressure in the cavity exceeded the surface strength, the material above the cavity was propelled from the comet. Mechanism of activity of Comet Tempel 1 was considered by Belton et al. (2007). Belton et al. (2010) concluded that natural outbursts on Comet 9P/Tempel 1 were caused by that at some depth the stress of gas overwhelmed the strength and overburden pressure of cometary material. In their opinion, the events might be triggered by changing thermal stresses or other processes in surface material in response to a cooling of the surface.

Comet nuclei are assumed to be of porous structure. For example, Richardson et al. (2007) considered that the bulk density of Comet Temple 1 is ~0.4 g $cm^{-3}$. Sources of gas that can fill cavities and pores in comets include the crystallization of amorphous ice (see the above references) and the sublimation at 'internal' surfaces (Möhlmann 2002).

The above examples and the observation of the DI triggered outburst testify in favour of that cavities containing particles and gas under pressure can be located below a considerable fraction of a comet's surface. The material under pressure can produce natural and triggered outbursts and can cause splitting of comets. At a triggered outburst caused by a collision, the duration of the outburst can be short because most of the material under pressure can leave the excavated cavity quickly. Duration of some natural outbursts can be much longer.

Cometary activity of asteroid 7968 Elst-Pizarro, also known as Comet 133P/Elst-Pizarro, could be caused by the same internal processes as the triggered or natural outbursts from Comet Tempel 1, but its solid crust could be much thicker than that of Comet Temple 1. In 1996, 2002, and 2007, the object Elst-Pizarro had a comet tail for several months. This object moves in an asteroid orbit with $a$=3.161 au, $e$=0.1644, and $i$=1.386°. The orbit of this object is stable (Ipatov & Hahn 1997, 1999). Based on studies of the orbital evolution of Jupiter-crossing objects (Ipatov & Mather 2003, 2004), Ipatov and Mather (2007) supposed that the object Elst-Pizarro earlier could be a Jupiter-family comet, and it could circulate its orbit also due to non-gravitational forces.

Hsieh et al. (2010) concluded that activity of Comet 133P/Elst-Pizarro was consistent with seasonal activity modulation and took place during hemisphere's summer, when the comet received enough heating to drive sublimation. We suppose that there could be natural outbursts during the 'summer' and they could be one of the sources of observed activity of the comet. It could be possible that vaporized material formed under the crust moved outside through narrow holes for a long time. There can be a lot of ice under the crust of the object Elst-Pizarro, and this ice produced a comet tail after the crust had been damaged in some way (e.g. due to high internal pressure).

Cometary-like activity was also observed for P/2010 A2 (LINEAR), which has a typical asteroid orbit ($a$=2.29 au, $e$=0.12, and $i$=5.26°). The total amount of the dust released during eight months was estimated by Moreno et al. (2010) to represent 0.3 per cent of the nucleus mass. They supposed that some subsurface ice layer exists in this object. Several other authors (e.g. Jewitt et al. 2010 and Snodgrass et al. 2010) believe that the trail of P/2010 A2 is the result of the collision between two asteroids, not of cometary activity, because this object is close to the inner edge of the asteroid belt. In our opinion, if this object contains ice (e.g. it was captured from a comet's orbit), then the internal gas pressure could also play a role in the ejection of particles from this object, but this role should not be considerable because the velocities of the 'fast' ejection should be greater than the velocities (<1 m s$^{-1}$) obtained by Jewitt et al. (2010).

### 9.3 Formation of the *Deep Impact* crater

Let us compare results of DI observations with the models of crater formation considered by several authors and summarized by Richardson et al. (2007). They concluded that the DI crater formed in no more than 250-550 s (4-9 min) for the case of effective strength $S$=0 (gravity-dominated cratering). Crater formation time $t_{cf}$ was supposed to be proportional to $S^{-1/2}$ and to be not less than 1-3 sec at $S$=10 kPa (for strength-dominated cratering). The amount of ejected material was about $2\times10^7$ kg at $S$=0 and about $2\times10^5$ kg at $S$=10 kPa. The cumulative mass of solid particles ejected at a velocity greater than 10 m s$^{-1}$ (or any other greater value) was almost the same for different models studied by Richardson et al. (2007), though the total mass of ejected material varied considerably for different values of effective strength $S$. Therefore, even for the normal cratering, the observational estimates of the total mass of fast moving small dust particles do not allow one to make reliable conclusions on the values of $S$. Our studies were based on analysis of material ejected with velocities >10 m s$^{-1}$, and so they do not allow one to estimate the total mass of all ejected material. The quantity of the high-velocity ejecta is greater for a smaller impact angle $I$ (Yamamoto et al. 2005). For the DI impact, $I$≈20-35°, and the quantity must be greater than for the models of impact with $I$=90° considered by Richardson et al. (2005) and Holsapple & Housen (2007).

Observations of $H_2O$ and OH showed (Küppers et al. 2005; Schleicher et al. 2006; Biver et al. 2007; Keller et al. 2007; A'Hearn 2007; A'Hearn & Combi 2007) that the amount of ejected water exceeded $5\times10^6$ kg. (Estimates made by Lisse et al. (2006) were smaller.) Such estimates of water allow one to conclude that the DI cratering event was different from the strength-dominated cratering because the total ejected mass must be small for the latter cratering. Considering that the volume of a crater equals $V=\pi\times r_c^3/3$ (where $r_c$ is the radius of the crater) and density $\rho$ is equal to 400 kg m$^{-3}$ (Richardson et al. 2007), we obtain that $\rho\times V=5\times10^6$ kg (the above estimate of the minimum amount of water) at $r_c$=23 m and $\rho\times V=7\times10^7$ kg (the maximum estimate of the total mass of ejected material at $d$<2 m presented in Table 1) at $r_c$=55 m. Even the latter value of $r_c$ is less than the estimate of a crater radius (75-100 m) made by Busko et al. (2007) on the basis of analysis of DI images. Schultz et al. (2007) obtained a little wider range for the radius: 65-110 m. They concluded that the difference between the volume of a crater and the ejected mass is due to displaced mass for the crater. The above formula for $V$ was obtained for the ratio $k_c=h_c/d_c$ of the crater depth to its diameter equal to 1/3. In experiments the ratio was between 1/4 and 1/3 (Schmidt & Housen 1987; Melosh 1989). Some scientists considered $k_c$=1/5. Laboratory data show that the values of $k_c$ are 0.12-0.27 for bowl-shaped craters on flat water ice targets and 0.16-0.26 on rocky targets (see references in Leliwa-Kopystynski, Burchell & Lowen 2008). As $r_c$ is proportional to $k_c^{-1/3}$, then the use of $k_c$=1/5 instead of 1/3 increases $r_c$ by a factor less than 1.2 (for the same $V$).

Particles ejected at the outburst probably were mainly relatively small and fast, their contribution to the brightness of the DI cloud could be much greater than to the total ejected mass, and most of the crater volume could be caused by the normal ejection. Therefore, relatively large estimates of the radius of the crater made by Busko et al. (2007) and Schultz et al. (2007) testify in favour of gravity-dominated cratering, and so in favour of the longer duration of the normal ejection.

Estimates of the amount of material ejected at the DI impact are greater for a greater diameter $d_l$ of the largest fragment of the ejected material considered in the estimates (see Table 1). At $d_l$=2 m, even the upper estimate of the total ejected mass ($7\times10^7$ kg) corresponds to the radius $r_c$ of the crater ($r_c$=55 m at $k_c$=1/3 and $r_c$=70 m at $k_c$=1/5) smaller than mean estimates of $r_c$ made by several scientists (see the above references). Therefore, in principle, it could be possible that bodies with $d$>2 m could be present in the ejected material, though such bodies were not observed. The bodies could include parts of the cavities' walls ejected at the 'fast' outburst. Diameter of the largest body definitely could not exceed 20-25 m because of the limited depth of the crater. In experiments with projectile velocity ~1-10 km s$^{-1}$, Koschny & Grün (2001) found an upper limit for the mass of the largest ejected fragment of about 1 per cent of the total mass. For the DI crater with radius $r_c$=100 m, such limit corresponds to $d$~25 m. For porous ice-silicate mixture at mass-distribution exponent equal to -1.8 (Koschny & Grün, 2001), the increase in diameter $d_l$ of the largest fragment in the distribution by a factor of 10 corresponds to the increase in the total ejecta mass by a factor of $10^{1.8}\approx63$ and the increase in the radius of the crater by a factor of $63^{1/3}\approx4$.

Holsapple & Housen (2007) obtained the time of formation of the DI crater to be about 5 min for sand-gravity scaling, 11 min for water, and much smaller for other types of soil considered (e.g. soft rocks and cohesive soil). For the normal ejection, duration of ejection greater than 10 min obtained in our studies is not in accordance with the other types of soil considered by the above authors. As the long duration of the ejection could be due to the outburst, it is not possible to make conclusions on a type of soil. We suppose that duration of the normal ejection did not exceed a few minutes, but duration of the outburst could exceed 30-60 min.

## 10 CONCLUSIONS

Our studies of the projections $v_p$ of velocities of ejected particles onto the plane perpendicular to the line of sight and of the relative amounts of particles ejected from Comet 9P/Temple 1 were based on analysis of the images made by the *Deep Impact* cameras during the first 13 minutes after impact. We studied velocities of the particles that give the main contribution to the brightness of the cloud of ejected material, i.e. mainly of particles with diameter $d$<3 μm.

In our estimates of velocities of ejected particles, we analysed the motion of particles along a distance of a few km. Destruction, sublimation, and acceleration of particles did not affect much our estimates of velocities because we considered the motion of particles during no more than a few minutes. During the considered motion of particles with initial velocities $v_p \geq 20$ m s$^{-1}$, the increase in their velocities due to the acceleration by gas did not exceed a few meters per second.

The time variations in the rates and velocities of material ejected after the DI impact differed from those found in experiments and in theoretical models. Holsapple & Housen (2007) concluded that these differences were caused by vaporization of ice in the plume and fast moving gas. Their conclusion could be true for the ground-based observations made a few hours after the impact. In our studies of the motion of particles during a few minutes, the greater role in the difference could be played by the outburst triggered by the impact (by the increase of ejection of small bright particles), and it may be possible to consider the ejection as a superposition of the normal ejection and the triggered outburst. The contribution of the outburst to the brightness of the cloud could be considerable, but its contribution to the total ejected mass could be relatively small because the fraction of *small* observed particles among particles of *all* sizes was probably greater for the outburst than for the normal ejecta. Our model of ejection considered only those particles that reached a distance $R \geq 1$ km from the place of ejection. Large regions of saturated pixels in DI images made at time $t$ after impact greater than 110 s prevented us from drawing firm conclusions about the rates of ejection of all particles.

Results of our studies showed that there was a local maximum of the rate of ejection at $t_e$~10 s with typical projections $v_p$ of velocities onto the plane perpendicular to the line of sight of about 100-200 m s$^{-1}$. At the same time, the considerable excessive ejection in a few directions (rays of ejecta) began, there was a local increase in brightness of the brightest pixel, and the direction from the place of ejection to the brightest pixel quickly changed by about 50$^{o}$. In images made during the first 12 s and after the first 60 s, this direction was mainly close to the direction of the impact.

At $1<t_e<3$ and $8<t_e<60$ s, the plot of time variation in the estimated rate $r_{te}$ of ejection of observed material was greater than the exponential line connecting the values of $r_{te}$ at 1 and 300 s. The above features could be caused by the impact being the trigger of an outburst and by the ejection of more icy material. At $t_e$~55-60 s, the ejection rate sharply decreased and the direction from the place of ejection to the brightest pixel quickly returned to the direction that was before 10 s. This could have been caused by a sharp decrease in the outburst that began at $t_e$~10 s.

The outburst ejection could have come from the entire surface of the crater, while the normal ejection was mainly from its edges. The 'fast' outburst could be caused by the ejection of material from the cavities that contained the material under gas pressure. The 'slow' outburst ejection could be similar to the ejection from a 'fresh' surface of a comet and could take place long after the formation of the crater.

Analysis of observations of the DI cloud and of outbursts from different comets testifies in favour of the proposition that there can be large cavities, with material under gas pressure, below a considerable fraction of a comet's surface. Internal gas pressure and material in the cavities can produce natural and triggered outbursts and can cause splitting of comets. The upper edge of the cavity excavated at $t_e$~10 s could be located a few meters under the surface of Comet Tempel 1.

Our studies did not allow us to estimate accurately when the end of ejection occurred, but they do not contradict a continuous ejection of material during at least the first 10 minutes after the collision. The duration of the outburst (up to 30-60 min) could be longer than that of the normal ejection, which could last only a few minutes. Our research testifies in favour of a model close to gravity-dominated cratering.

Projections of velocities of most of the observed material ejected at $t_e$~0.2 s were about 7 km $s^{-1}$. Analysis of DI observations that used different approaches showed that at $1<t_e<115$ s the time variations in the projections $v_p$ of characteristic velocity of observed particles onto the plane perpendicular to the line of sight can be considered to be proportional to $t_e^{-\alpha}$ with $\alpha$~0.7-0.75. For the model *VExp* with $v_p$ proportional to $t_e^{-\alpha}$ at any $t_e>1$ s, the fractions of observed (not all) material ejected (at $t_e \leq 6$ and $t_e \leq 15$ s) with $v_p \geq 200$ and $v_p \geq 100$ m $s^{-1}$ were estimated to be about 0.1-0.15 and 0.2-0.25, respectively, if we consider only material observed during the first 13 minutes. The 'fast' outburst with velocities ~100 m $s^{-1}$ probably could last for at least several tens of seconds, and it could significantly increase the fraction of particles ejected with velocities ~100 m $s^{-1}$, compared with the estimates for the model *VExp* and for the normal ejection. The above estimates are in accordance with the estimates (100-200 m $s^{-1}$) of the projection of velocity of the leading edge of the DI dust cloud made by other scientists and based on various ground-based observations and observations made by space telescopes.

The excess ejection of material in a few directions (rays of ejected material) was considerable during the first 100 s, and it was still observed in images made at $t$~500-770 s. This finding shows that the outburst could continue at $t_e$~10 min. The sharpest rays were caused by material ejected at $t_e$~20 s. In particular, there were excessive ejections, especially in images made at $t$~25-50 s after impact, in directions perpendicular to the direction of impact. Directions of excessive ejection could vary with time.

The size of the region of the DI cloud of essential opacity probably did not exceed 1 km.

## ACKNOWLEDGEMENTS

This work was supported by NASA DDAP grant NNX08AG25G to the Catholic University of America and by NASA's Discovery Program Mission, Deep Impact, to the University of Maryland. The authors are extremely grateful to all the science team members, numerous engineers, scientists, and supporting people for making the mission possible and successful. We are thankful to a reviewer for helpful discussion.

**Table 1.** Projection of velocity (in m $s^{-1}$) of the leading edge of the DI dust cloud of ejected material onto the plane perpendicular to the line of sight at several moments in time and the mass of ejected material (in kg) obtained from different observations.

| Source | Telescope | Resolution, km $pixel^{-1}$ | Considered time $t$ after impact | Projected velocity at $t$, m $s^{-1}$ | Total mass of ejected material with diameter $d$, kg |
|---|---|---|---|---|---|
| Harker et al. (2005, 2007) | 8.1-m Frederick C. Gillette (Gemini-N) telescope on Mauna Kea, Hawaii; MICHELLE imaging spectrograph, 7.8-13 μm | 388 | 1 h | 220 – grains with $d$~0.4 μm | $(7.3\text{-}8.4)\times10^4$ – dust with $d$<2 μm; $1.5\times10^6$ – dust with $d$<200 μm; |
| Keller et al (2005, 2007) | *Rosetta*, OSIRIS camera (0.245-1 μm) | 1500 | 4 h (1.42 – 3.73 days to estimate the mass) | 200 | $1.6\times10^5$ – dust with $d$<2.8 μm; $(1\text{-}14)\times10^6$ – dust with $d$<200 μm; $(4.5\text{-}9)\times10^6$ – water |
| Küppers et al. (2005) | *Rosetta*, OSIRIS camera | 1500 | 40 min | >110; 300 - fine dust | $4.4\times10^5$ – dust with $d$<2 μm; $10^7$ – water and dust; $4.5\times10^6$ – water |
| Meech et al. (2005) | Many ground-based telescopes | | 20 h | 200 | $10^6$ - dust with typical $d$~1 μm |
| Ipatov and A'Hearn (2006) | *Deep Impact*, MRI | 0.087 | 8-15 s | 200; 100 – the brightest material | |
| Jehin et al. (2006) | Keck I telescope (10 m) on Mauna Kea (Hawaii), High Resolution Echelle Spectrometer (HIRES) | | 4 h | 150 – dust; 400 - gas CN | |
| Lisse et al. (2006), supplemental online material | *Spitzer* Space Telescope, Infrared Spectrograph, 5-35 μm | 550 | 45 min | 200-300 | $2.2\times10^5$ – dust at $d$<2 μm $7.8\times10^5$ – dust at $d$<20 μm $9.9\times10^5$ – dust at $d$<2 mm; $5.8\times10^5$ – water |
| Sugita et al. (2005, 2006) | 8.2 m Subaru telescope and its mid-infrared detector, COMICS, 8.8-24.5 μm | | 1 h | 125 | $(5.6\text{-}8.5)\times10^5$ - dust with *d*<20 μm $(2.8\text{-}7.0)\times10^7$ - dust and bodies with *d*<2 m |
| Schleicher et al. (2006) | Hall telescope (1.1 m) and Lowell's telescope (0.8 m) | 765 | 23 h | 220 | $\leq1.3\times10^7$ - water |
| Barber et al. (2007) | United Kingdom Infrared Telescope (3.8 m) on Mauna Kea, Hawaii; spectrometer CGS4 | | 110 min 20 h | 125 260 | |
| Bauer et al. (2007) | Palomar 200-inch telescope, near-IR PHARO camera | | 75 min | 200 | $2\times10^5$ – dust at *d*~1.4 μm |
| Biver et al. (2007) | Nançay, IRAM and CSO radio telescopes, Odin satellite | | | 350 - water | $5\times10^6$ – water |
| Cochran et al. (2007) | Keck I telescope (10 m) on Mauna Kea (Hawaii), HIRES spectrograph | | 75 min | 510 - gas CN | |
| Feldman et al. (2007) | *Hubble* Space Telescope, Advanced Camera for Surveys, V filter, 0.6 μm | 16 | 75-100 min | 70-80 – the most probable; 115 – mean expansion speed; 145 – rms velocity | |
| Jorda et al. (2007) | *Rosetta*, OSIRIS camera, Orange filter (0.645 μm) | 1500 | 10 d | Peak at 190 with FWHM= 150 | $1.5\times10^5$ – dust with *d*<2.8 μm; $(1\text{-}14)\times10^6$ – dust with *d*<200 μm |
| Knight et al. (2007) | Kitt Peak Nat. Observatory (2.1 m telescope, SQIID IR camera, 1.1-3.3 μm) and Observ. Astron. Nac.-San Pedro Martir (1.5 m telescope) | 447 | 24 h | 200-230 | |
| Lara et al. (2007) | 2 m telescope at Calar Alto Observatory, instrument CAFOS, 2.8-10 μm | | 15 h<br>40 h | 230<br>150 | $\geq 1.2\times10^{6}$ – dust |
| Tozzi et al. (2007) | European Southern Observatory, La Silla and Paranal sites, near-IR | | 20 h | 115 – average with a FWHM= 75 | |
| Walker et al. (2007) | 36-inch telescope at MIRA's Bernard Oliver Observing Station | | 50 min<br>25 h | 230<br>185 | |
| Ipatov and A'Hearn, the present paper | *Deep Impact*, MRI; *Deep Impact*, HRI | 0.087; 0.017 | 13 min | 100-200 | |

**Table 2.** Series of images considered. Instrument (telescope) used, total integration time (INTTIME, in seconds), size of considered images (in pixels), exposure ID (EXPID), and time after impact (IMPACTM, in seconds). For all images CLEAR filter was used. In the series *Ma*, the image number within commanded exposure (IMGNUM) varied from 64 to 156. In the series *Ma*, *Ha*, and *Hc*, we analysed the differences in brightness between a current image and that before impact (the MRI image with EXPID=9000910 and IMGNUM=63 made at $t$=-0.057 s, or the HRI image with EXPID=9000910 and IMGNUM=5 made at $t$=-0.629). These series are marked by "(dif)". For other series, the brightness in current images was analysed.

| Series | Instru-ment | INTTIME, seconds | Size, pixels | EXPID<br>min, max | IMPACTM, seconds<br>min, max |
|---|---|---|---|---|---|
| *Ma* (dif) | MRI | 0.0514 | 64×64 | 9000910, 9000910 | 0.001, 5.720 |
| *Mb* | MRI | 0.3 | 1024×1024 | 9000942, 9001067 | 77.651, 802.871 |
| *Ha* (dif) | HRI | 0.1 | 512×512 | 9000910, 9000945 | 0.215, 109.141 |
| *Hb* | HRI | 0.6 | 1024×1024 | 9000931, 9001002 | 39.274, 664.993 |
| *Hc* (dif) | HRI | 0.6 | 512×512 | 9000927, 9000942 | 27.664, 86.368 |
| *Hd* | HRI | 0.1 | 1024×1024 | 9000934, 9000961 | 50.715, 251.525 |
| *He* | HRI | 0.5 | 1024×1024 | 9001017, 9001036 | 719.805, 771.95 |

**Table 3.** Estimates of $x$ and $y$ projections ($\mathbf{v}_{px}$ and $\mathbf{v}_{py}$) of velocity $\mathbf{v}_p$ of ejection at several times $t_e$ of ejection. Most of the estimates were based on analysis of maxima and minima of time variations in distances $L$ between the place of ejection and considered contours of constant brightness (see Section 4); the maxima or minima were reached at times $t_1$ and $t_2$ at distances $L_1$ and $L_2$ for two levels of brightness ($CPSB_1$ and $CPSB_2$) on images belonging to $Series_1$ and $Series_2$, respectively. A few other estimates were based on analysis of images made at times $t_1$ and $t_2$ (see Section 3 and the end of Section 4). The ratio $t_e/t_1$ is presented if at $t_1$ the contour close to the edge of the bright region (usually the contour CPSB=3) was considered.

| | | | | | | | | | | | | |
|---|---|---|---|---|---|---|---|---|---|---|---|---|
| $t_e$ | 0.2 | 0.26 | 1.44 | 4.4 | 10 | 15 | 17 | 21 | 73 | 90 | 100 | 115 |
| $t_1$ | 0.224 | 0.34 | 2.74 | 8 | 55.6 | 30 | 55.6 | 31 | 120 | 120 | 140 | 140 |
| $t_2$ | 0.282 | 0.46 | 4.96 | 12.2 | 100.4 | 56 | 100.4 | 56 | 170 | 170 | 410 | 410 |
| $t_e/t_1$ | 0.89 | 0.76 | 0.53 | 0.55 | | 0.5 | | 0.68 | 0.61 | 0.75 | 0.71 | 0.82 |
| $v_{px}$ | | | | | 95 | | | 105 | | 25 | 20 | |
| $v_{py}$ | 7500 | 1500 | 930 | 240 | | 100 | 105 | | 16 | | | 26 |
| $Series_1$ | | | *Ma* | | *Hc* | *Ha* | *Hc* | *Hc* | *Hb* | *Hb* | *Hb* | *Hb* |
| $Series_2$ | | | *Ma* | | *Hb* | *Hc* | *Hb* | *Hc* | *Hb* | *Hb* | *Mb* | *Mb* |
| $CPSB_1$ | | | 3 | | 1 | 3 | 1 | 3 | 3 | 3 | 3 | 3 |
| $CPSB_2$ | | | 0.3 | | 0.3 | 1 | 0.3 | 1 | 1 | 1 | 0.3 | 0.3 |
| $L_1$, km | | | 1.2 | | 15.9 | 1.56 | 4.0 | 2.1 | 0.7 | 1.5 | 1.5 | 0.72 |
| $L_2$, km | | | 3.3 | | 7.3 | 4 | 8.6 | 7.34 | 1.5 | 4 | 12 | 7.80 |
| | | | max[1] | | max | max | max | max | min | min | min | min |

Notes: max[1] – a small decrease before the first local maximum

**Table 4.** Exponents of the time dependencies of the ejection velocity $v$, the relative rate $r_{te}$ of ejection, and the relative volume $f_{et}$ of material ejected before time $t_e$, and exponents of the velocity dependence of the relative volume $f_{ev}$ of material ejected with velocities greater than $v$.

| $v$ | $r_{te}$ | $f_{et}$ | $f_{ev}$ | the cratering event is primarily governed by | material |
|---|---|---|---|---|---|
| $t_e^{-0.75}$ | $t_e^{-0.25}$ | $t_e^{0.75}$ | $v^{-1}$ | the impactor's momentum | |
| $t_e^{-0.71}$ | $t_e^{-0.13}$ | $t_e^{0.87}$ | $v^{-1.23}$ | | sand |
| $t_e^{-0.644}$ | $t_e^{0.07}$ | $t_e^{1.07}$ | $v^{-1.66}$ | | soft rock |
| $t_e^{-0.6}$ | $t_e^{0.2}$ | $t_e^{1.2}$ | $v^{-2}$ | the impactor's kinetic energy | basalt |

**Table 5.** Characteristics of our model *VExp* of ejection for several pairs of α and $c$ (see Sections 5-6). Designations: $t_{e803}$ is the value of time $t_e$ of ejection of particles constituting the edge of the bright region in an image made at $t$=803 s, $v_{e803}$ is the ejection velocity at $t_{e803}$, $t_{elm}$ is the ejection time corresponded to the local peak of ejection rate, $f_1$ is the fraction of material ejected during the first second, $f_{ev200}$ is the fraction of material ejected with velocity $v_p \geq 200$ m s$^{-1}$, $f_{ev100}$ is the fraction of material ejected with velocity $v_p \geq 100$ m s$^{-1}$ (for calculation of all fractions, only material ejected at $t_e < t_{e803}$ was taken into account); $t_{ev200}$ and $t_{ev100}$ are the values of the ejection time corresponded to $f_{ev200}$ and $f_{ev100}$, respectively; $t_{et05}$ is the time during which a half of material observed during 800 s was ejected, $v_{et05}$ is the velocity of ejection at $t_{et05}$.

| | | | | | |
|---|---|---|---|---|---|
| α | 0.6 | 0.644 | **0.71** | **0.75** | 0.75 |
| $c$ | 2 | 2 | **2.5** | **3** | 2 |
| $t_{e803}$, s | 728 | 696 | **658** | **636** | 553 |
| $v_{e803}$, m s$^{-1}$ | 16 | 11 | **8** | **7** | 5 |
| $t_{elm}$, s | 10 | 9 | **9** | **9** | 6 |
| $f_1$ | 0.03 | 0.03 | **0.04** | **0.05** | 0.07 |
| $f_{ev200}$ | 0.10 | 0.10 | **0.13** | **0.15** | 0.13 |
| $f_{ev100}$ | 0.25 | 0.22 | **0.22** | **0.25** | 0.16 |
| $t_{ev200}$, s | 8 | 6 | **6** | **6** | 3 |
| $t_{ev100}$, s | 30 | 18 | **14** | **14** | 5 |
| $t_{et05}$, s | 170 | 120 | **95** | **70** | 60 |
| $v_{et05}$, m s$^{-1}$ | 38 | 35 | **34** | **37** | 25 |

**Table 6.** Directions of rays of ejected material. ψ is the angle between the upper direction and the direction to a bump measured in a clockwise direction.

| bump | upper | upper-right | right | left down-left |
|---|---|---|---|---|
| ψ, deg | 0 → -25 | 60 → 80 | 90 → 120 | (245-260) → (210-235) |

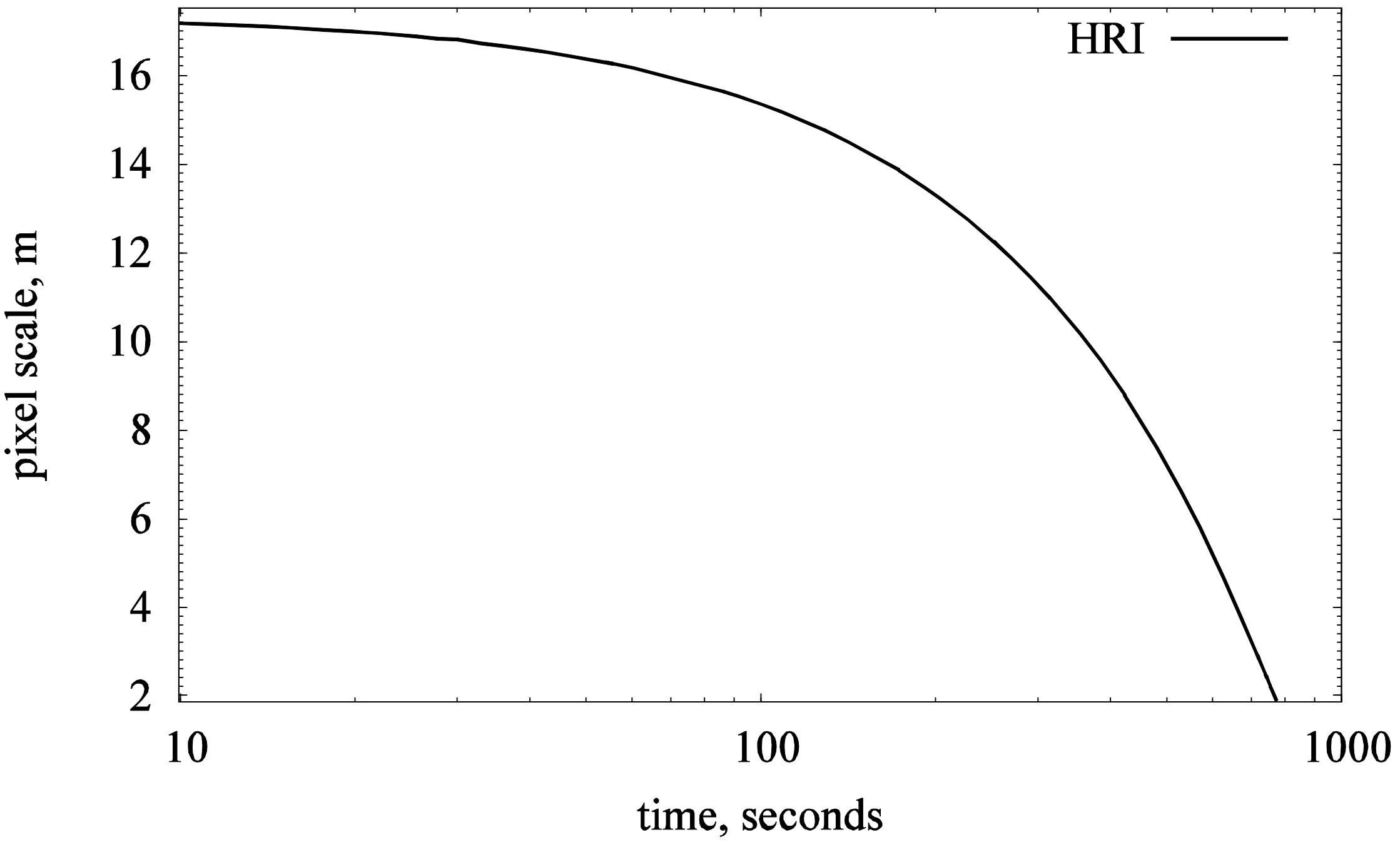

**Figure 1.** Variation in pixel scale (in meters) with time (in seconds) for HRI (High Resolution Instrument) images.

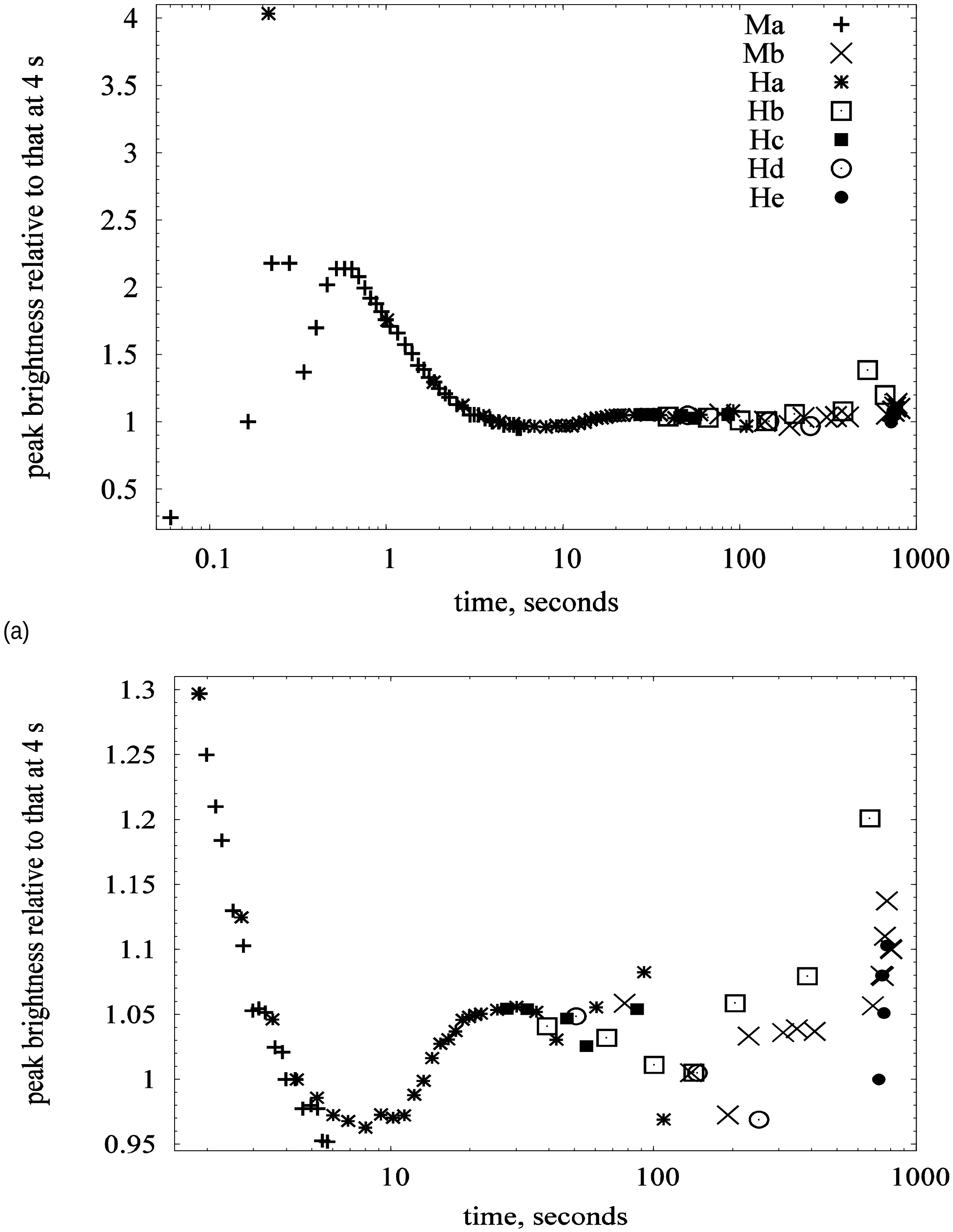

**Figure 2.** Variation in the relative brightness $Br_p$ of the brightest pixel with time $t$ (in seconds). It is supposed that $Br_p$=1 at $t$=4 s. The series of images *Ma, Mb, Ha, Hb, Hc, Hd*, and *He* are described in Table 2. Plot (b) is a part of plot (a) and allows one to see variations at $t$>10 s in more detail.

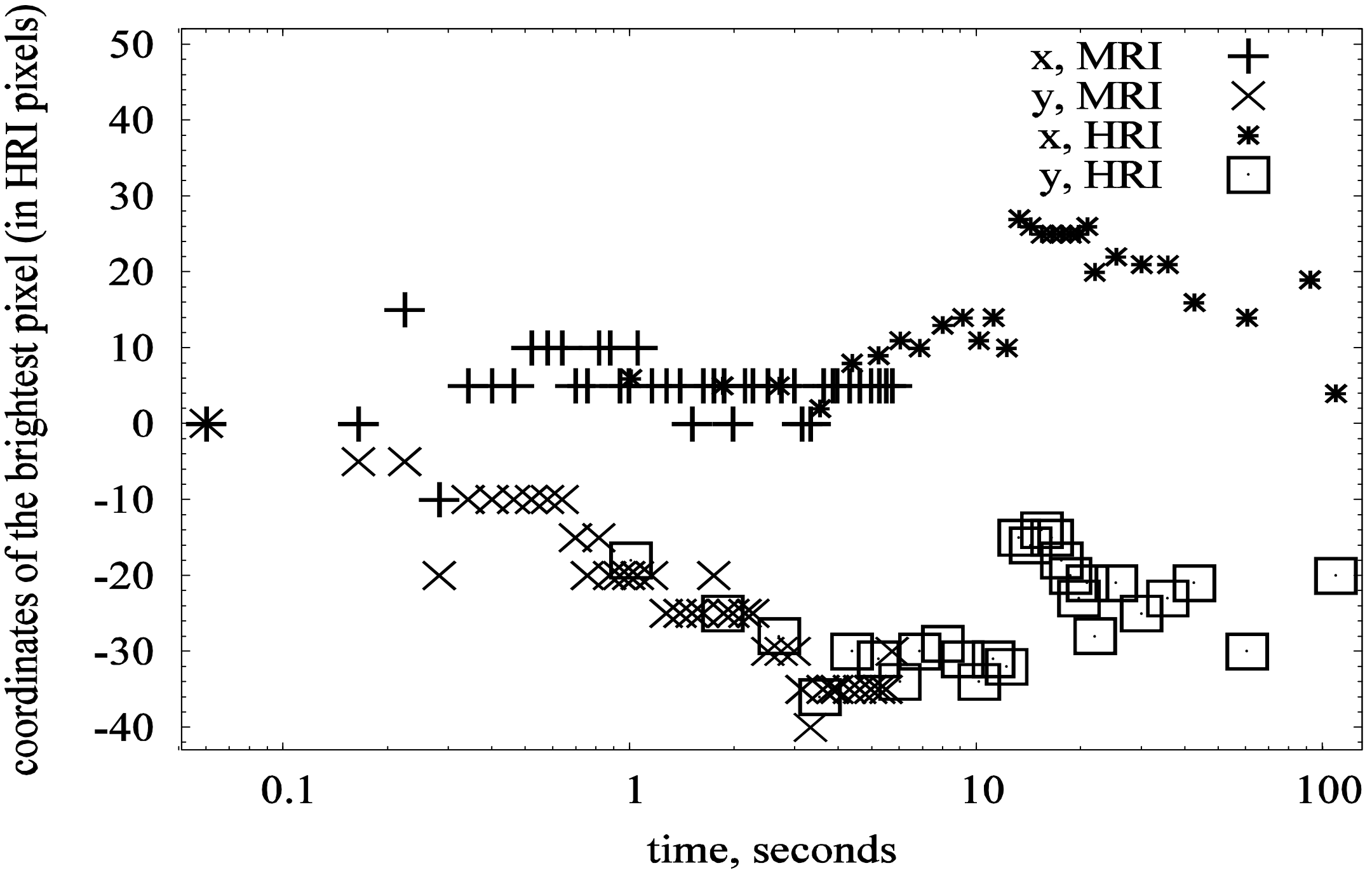

(a)

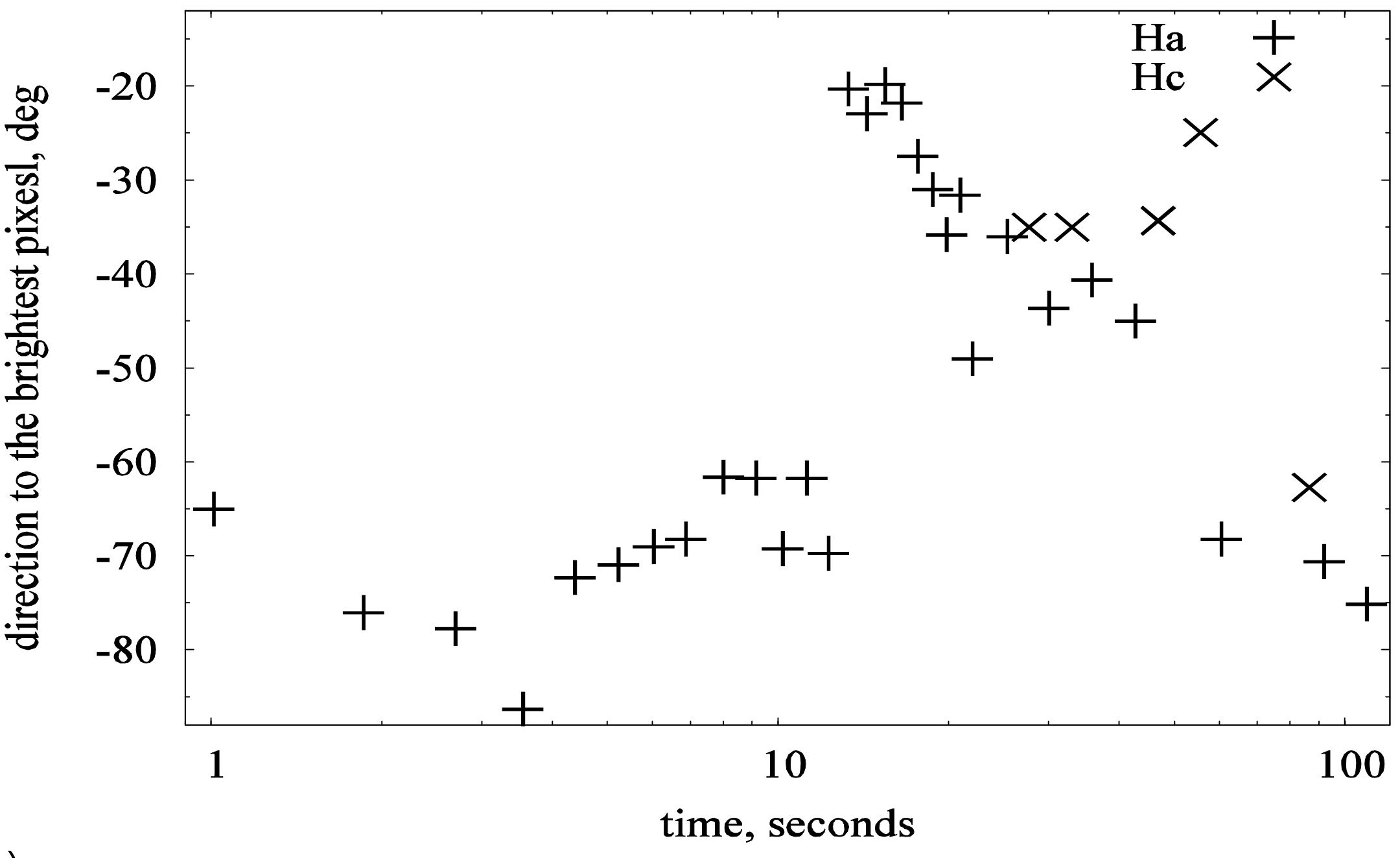

(b)

**Figure 3.** (a) Coordinates *x* and *y* of the brightest pixel in MRI (the series *Ma*) and HRI (the series *Ha*) images relative to the position of the brightest pixel in the MRI (Medium Resolution Instrument) image at $t$=0.001 s (the place of impact) at different times after impact. The difference in brightness between a current image and an image before impact was analysed. Coordinates are given in HRI pixels (i.e., the number of MRI pixels was multiplied by a factor of 5). HRI $y$-plot relative to the position of the brightest pixel in the HRI image at $t$=0.215 s (the place of ejection of material at $t$≥0.2 s) was shifted down by 5 pixels to present the position of the brightest pixel relative to the place of impact. (b) The angle (in degrees) of the direction from the brightest pixel at $t$=0.215 s to the brightest pixel at a current time for HRI images from the series *Ha* and *Hc*. The angle corresponding to the direction of the impact was about -60$^{o}$.

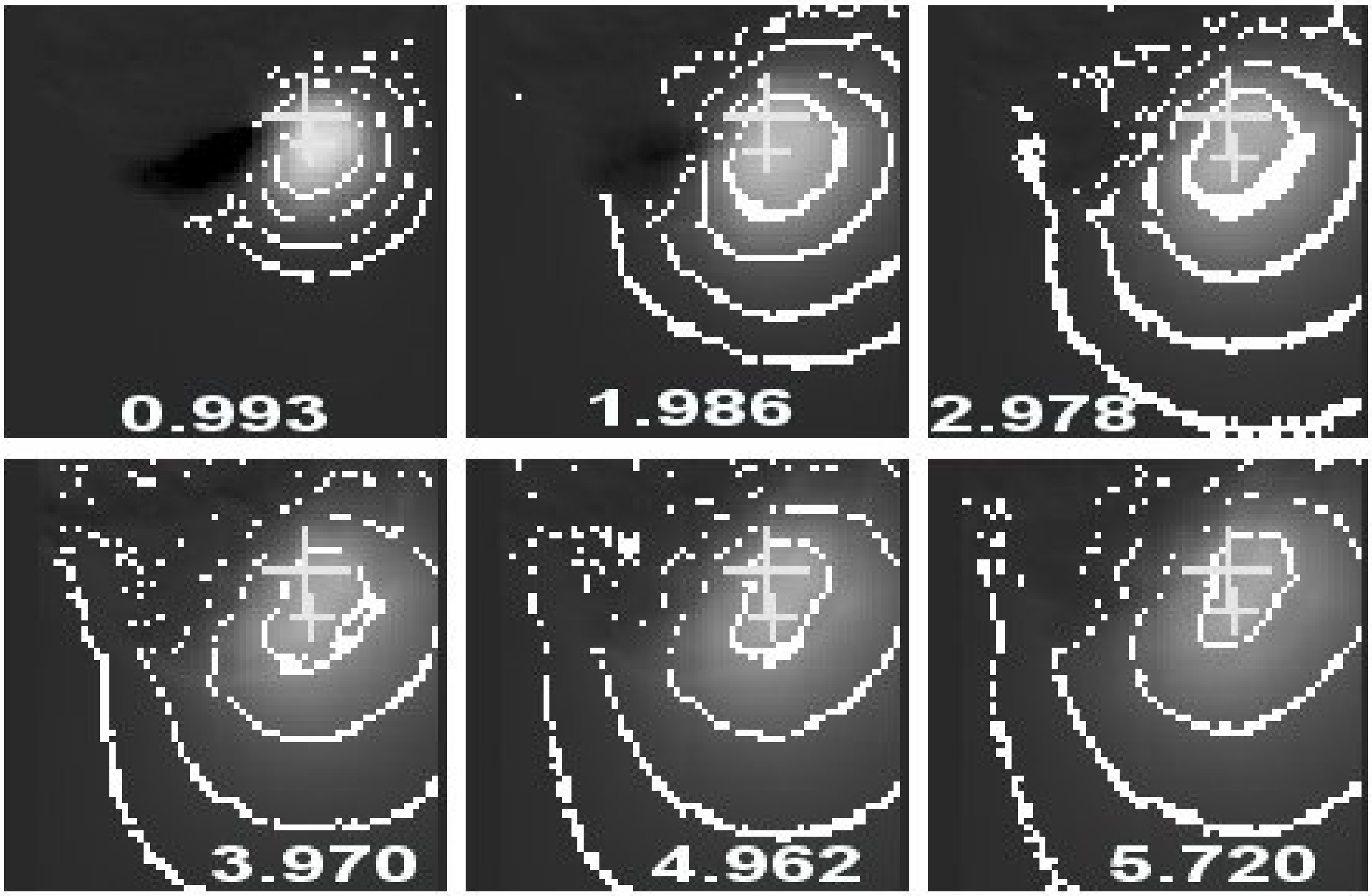

**Figure 4.** Contours for the difference in brightness between MRI images (the series *Ma*) made 0.993, 1.986, 2.978 (upper row), 3.970, 4.962, and 5.720 s (lower row) after impact and the image at $t$=-0.057 s. The contours correspond to constant calibrated physical surface brightness (CPSB, in W $m^{-2}$ $sterad^{-1}$ $micron^{-1}$) equal to 3, 1, 0.3, and 0.1, respectively. A large cross shows the position of the brightest pixel at $t$=0.06 s, and a smaller cross, at current time.

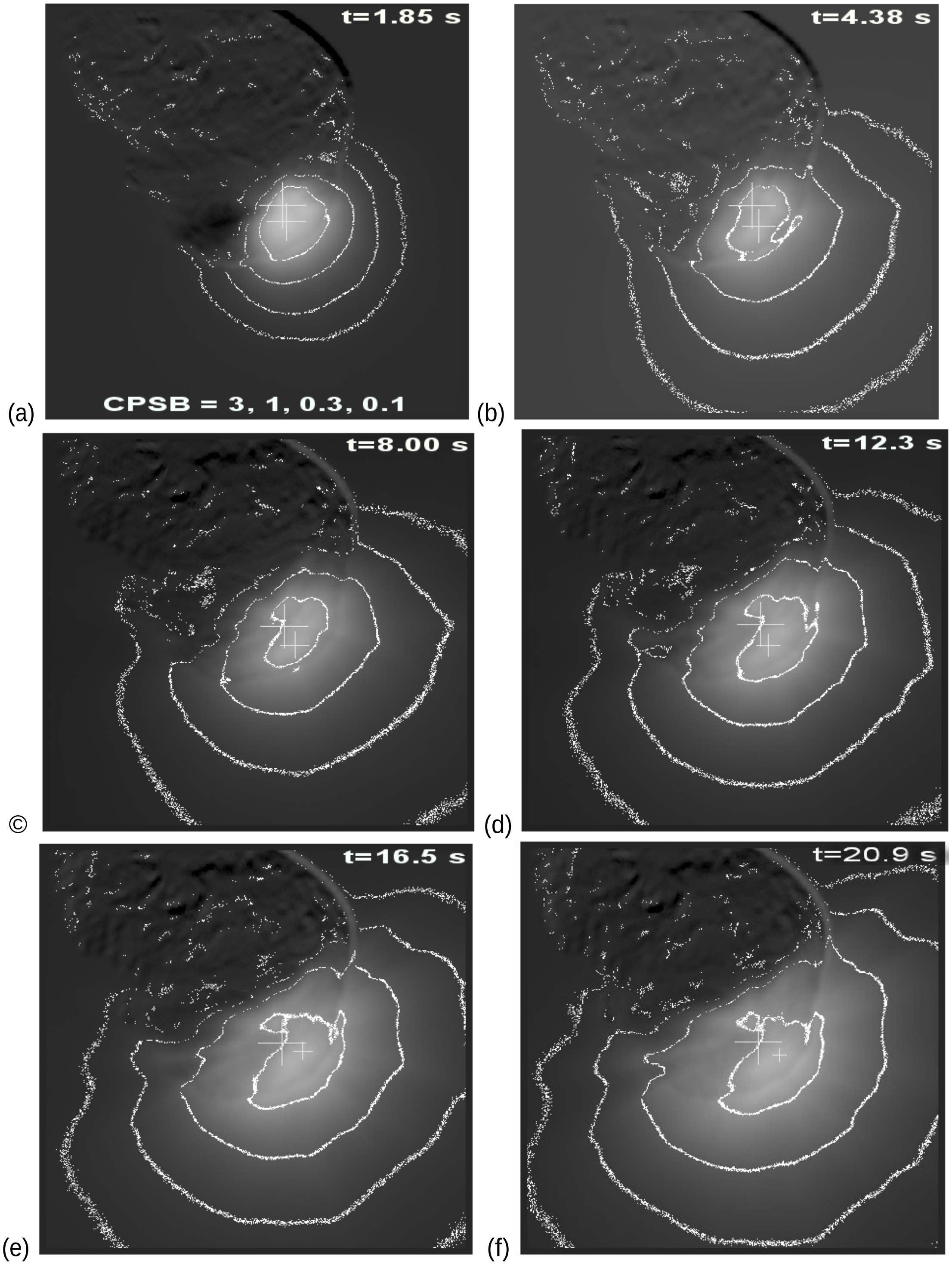

**Figure 5.** Contours corresponding to CPSB equal to 3, 1, 0.3, and 0.1 for the difference in brightness between HRI images from the series *Ha* made 1.852 (a), 4.379 (b), 8.00 (c), 12.254 (d), 16.524 (e), and 20.906 s (f) after impact and the image at *t*=-0.629 s. The largest cross shows the position of the brightest pixel at *t*=0.215 s, and a smaller cross, at current time.

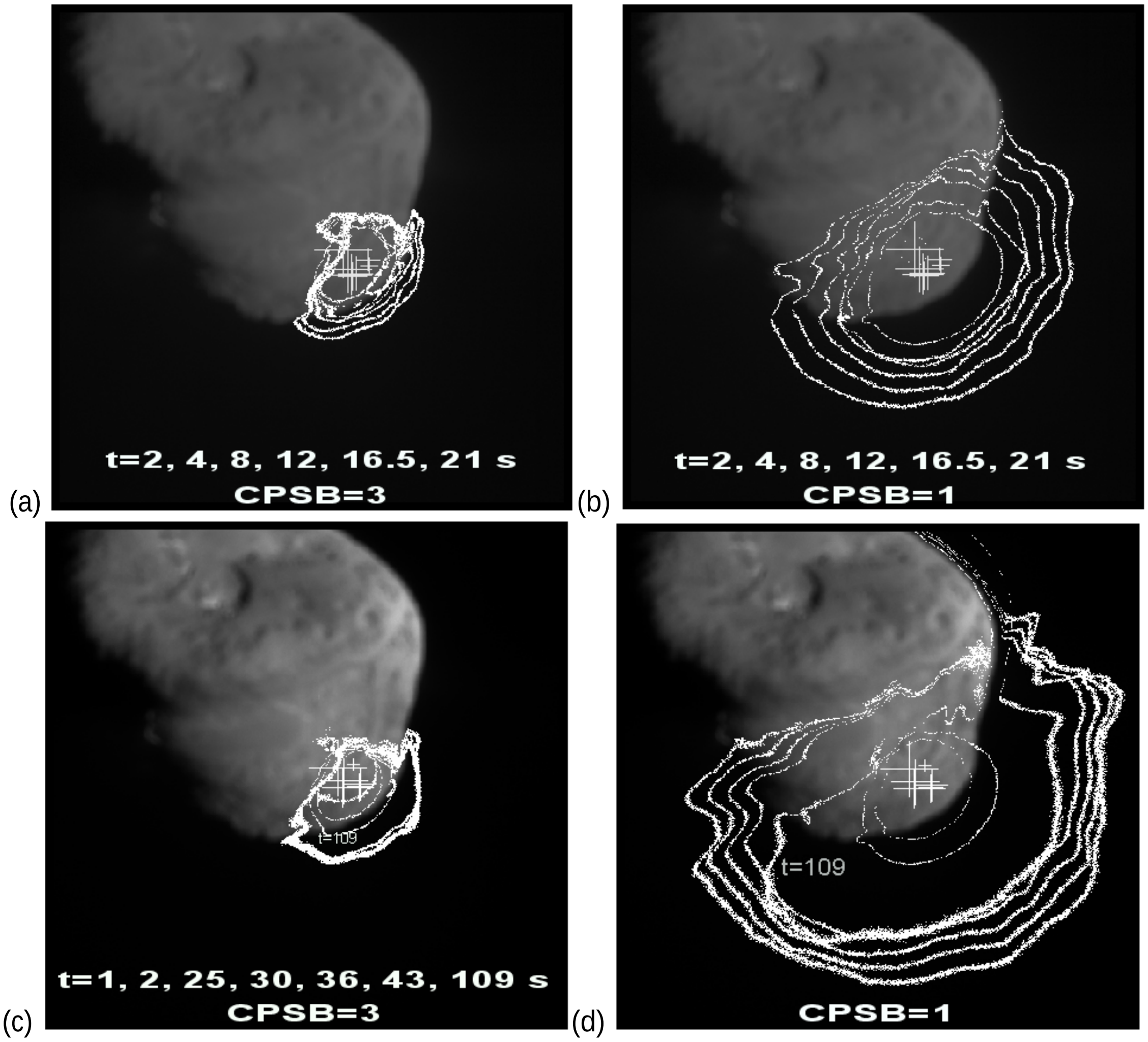

**Figure 6.** Contours corresponding to CPSB equal to 3 (a) and 1 (b), for the difference in brightness between HRI images from the series *Ha* made 1.852, 4.379, 8.00, 12.254, 16.524, and 20.906 s after impact and the image at *t*=-0.629 s. Contours corresponding to CPSB equal to 3 (c) and 1 (d), for the difference in brightness between *Ha* images made 1.008, 1.852, 25.332, 30.00, 35.715, 42.618, and 109.141 s after impact and the image at *t*=-0.629 s. The contour at *t*=109 s is the third from the place of impact, and other contours are larger for larger times. The largest cross shows the position of the brightest pixel at *t*=0.215 s. The size of a cross indicating the position of the brightest pixel at current time is smaller for a greater value of time.

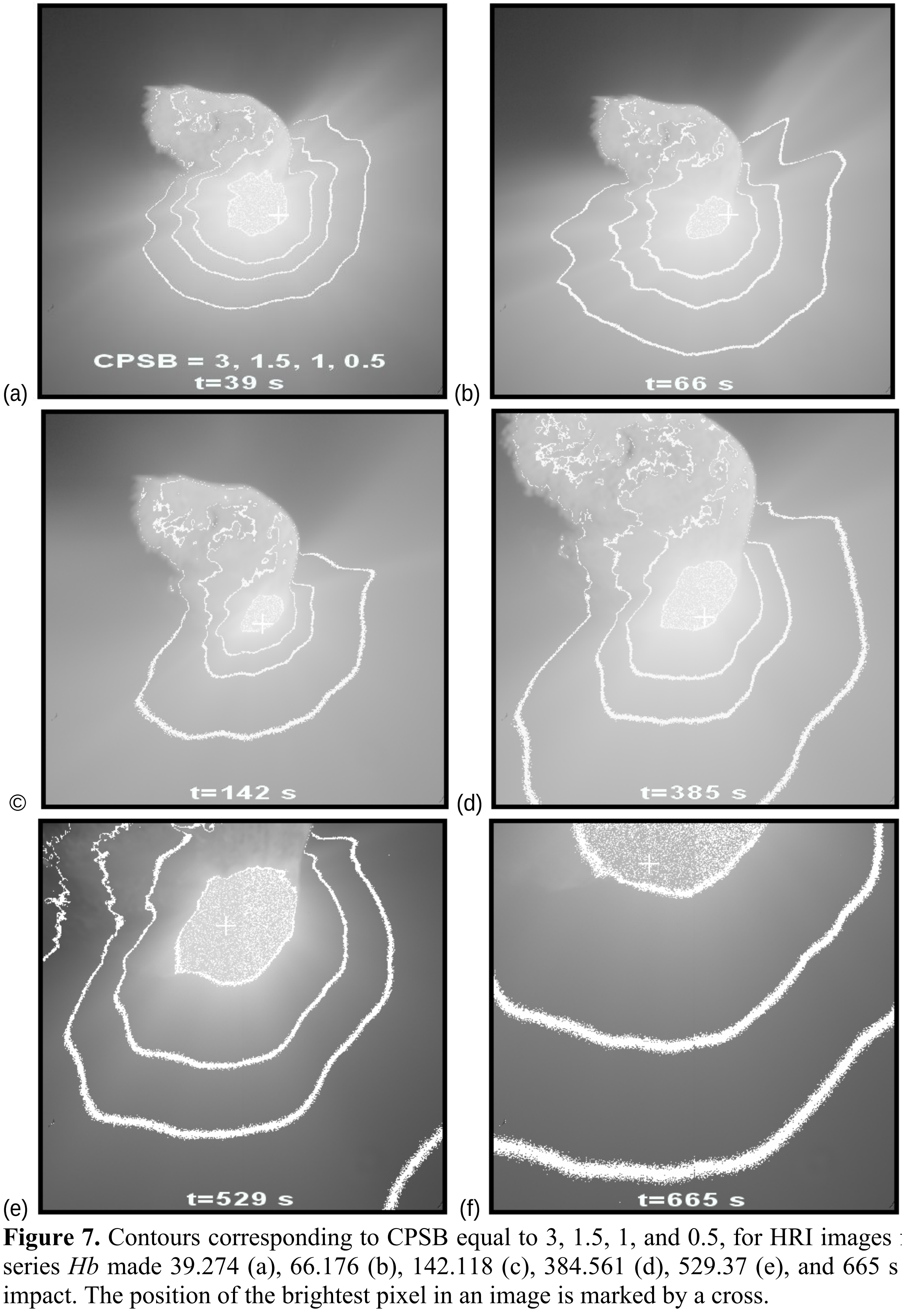

**Figure 7.** Contours corresponding to CPSB equal to 3, 1.5, 1, and 0.5, for HRI images from the series *Hb* made 39.274 (a), 66.176 (b), 142.118 (c), 384.561 (d), 529.37 (e), and 665 s (f) after impact. The position of the brightest pixel in an image is marked by a cross.

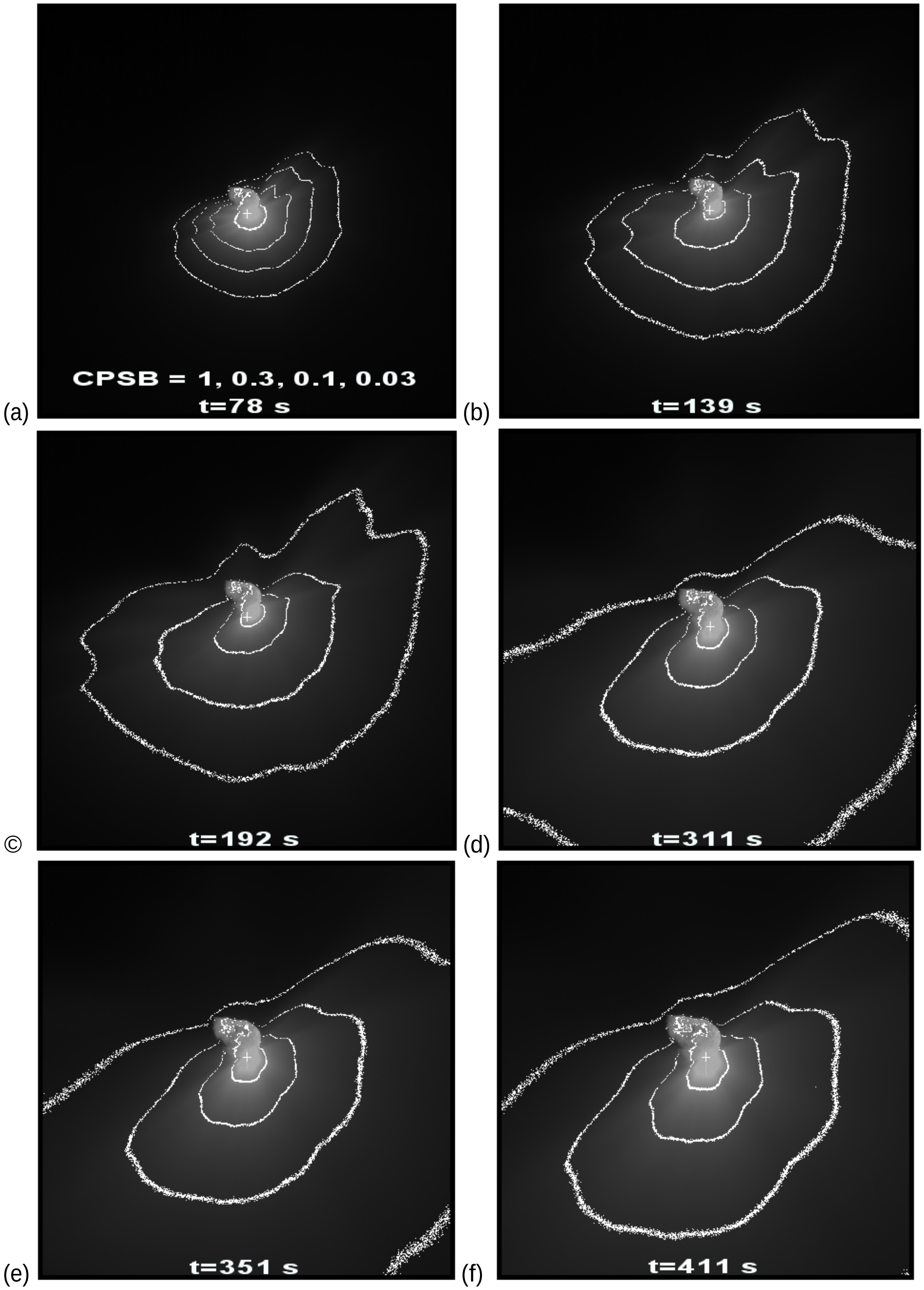

**Figure 8.** Contours corresponding to CPSB equal to 1, 0.3, 0.1, and 0.03, for MRI images from the series *Mb* made 77.651 (a), 138.901 (b), 191.53 (c), 311.055 (d), 351.043 (e), and 410.618 s (f) after impact. The position of the brightest pixel in an image is marked by a cross.

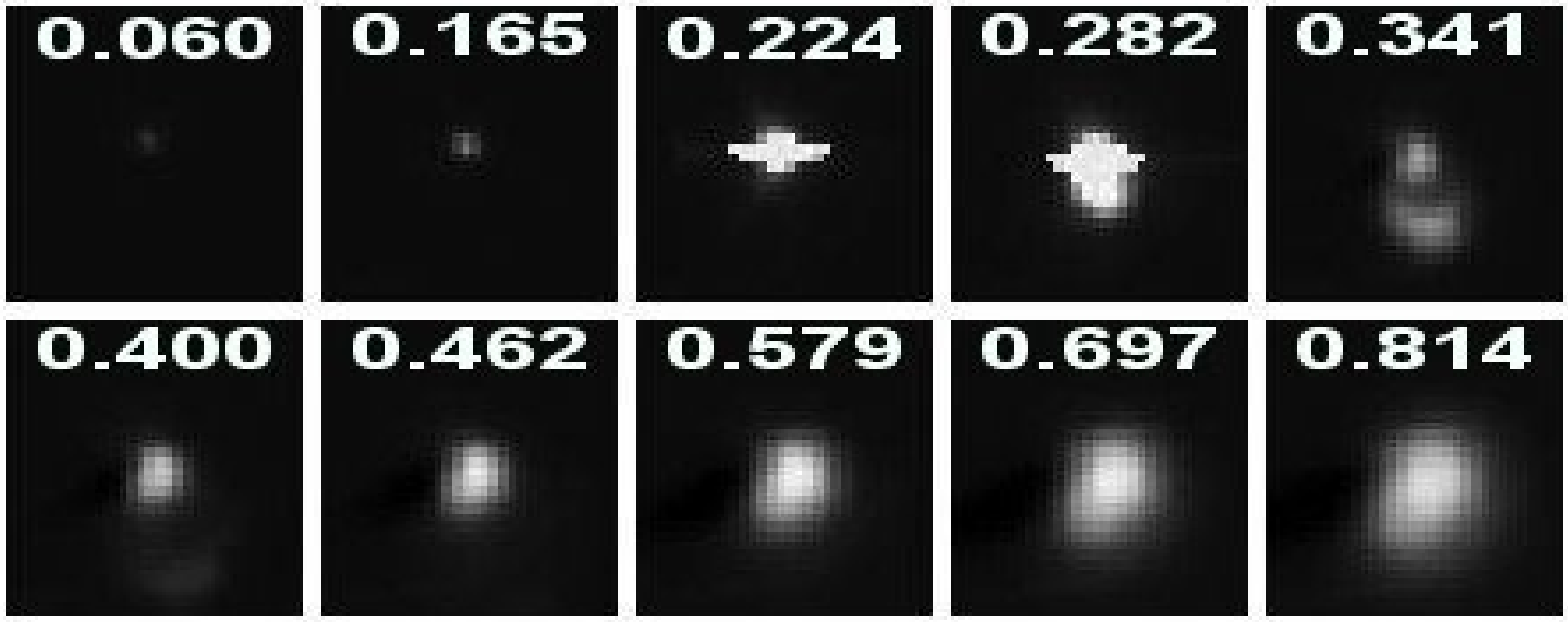

(a)

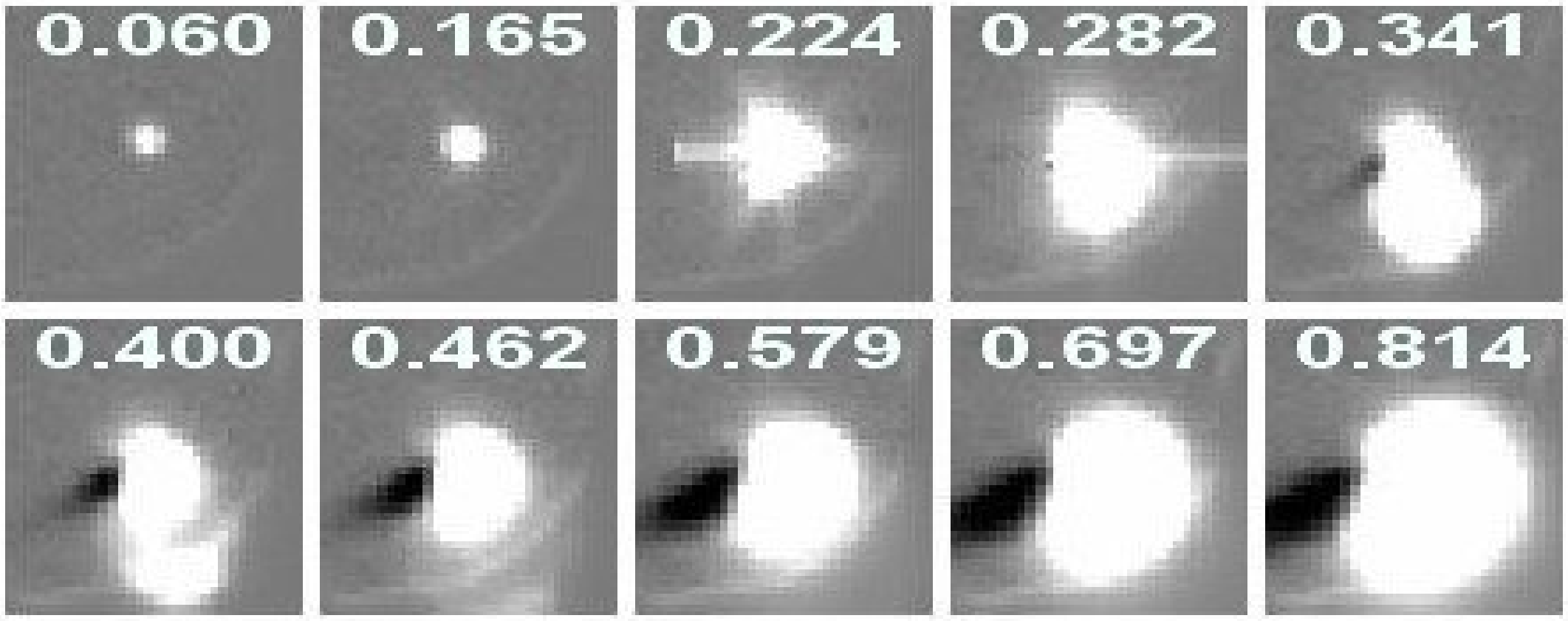

(b)

**Figure 9.** The difference in brightness between MRI (Medium Resolution Instrument) images (the series *Ma*) made 0.060, 0.165, 0.224, 0.282, 0.341 (upper row), 0.400, 0.462, 0.579, 0.697, and 0.814 s (lower row) after impact and the image at $t$=-0.057 s. In figure (a) a white region corresponds approximately to constant calibrated physical surface brightness CPSB≥3 (in W $m^{-2}$ $sterad^{-1}$ $micron^{-1}$), and in figure (b) it corresponds to CPSB≥0.5, but both figures present the same images.

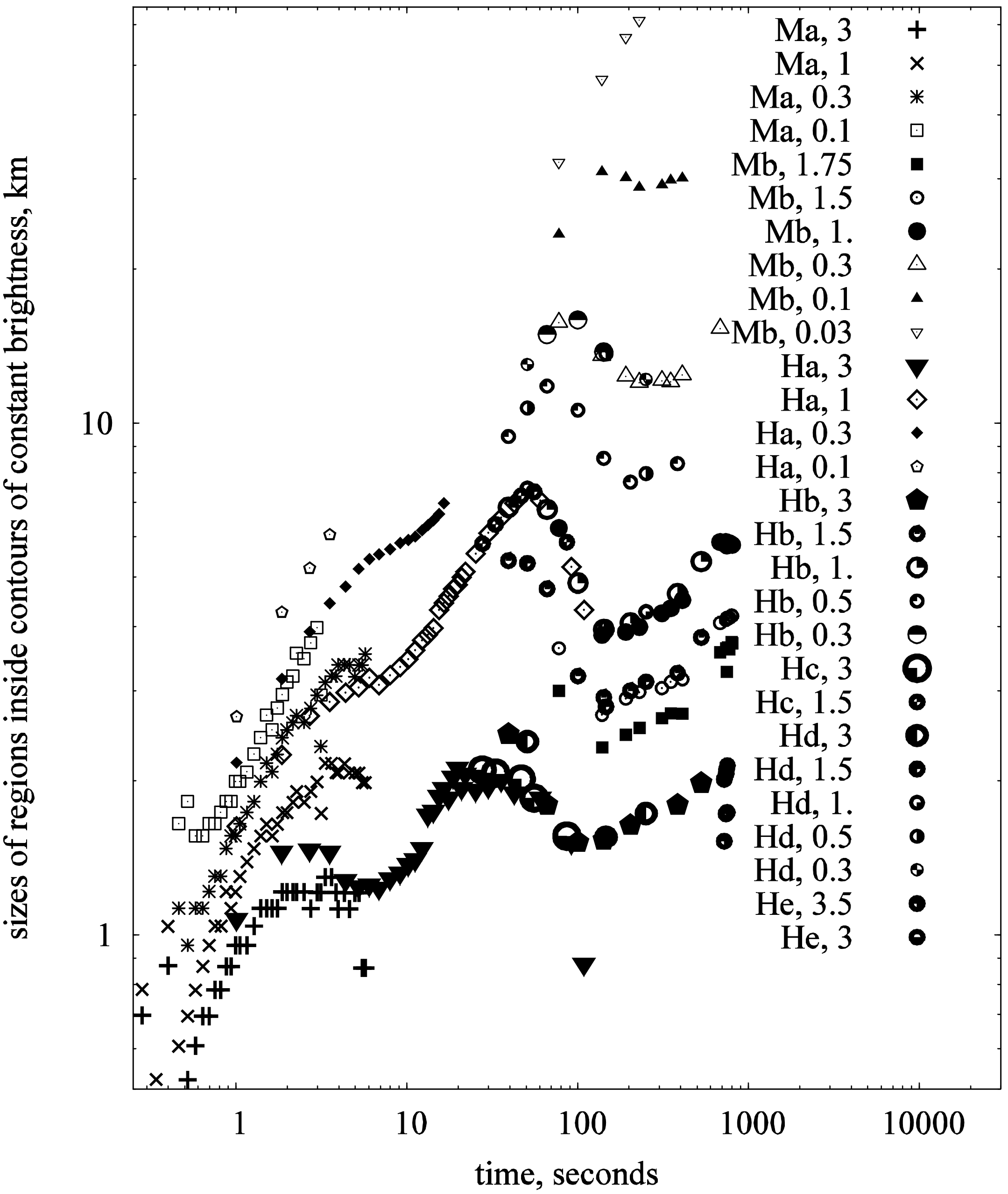

**Figure 10.** The variations in sizes $L$ (in km) of regions inside contours CPSB=const with time $t$ (in seconds). The series of images *Ma, Mb, Ha, Hb, Hc, Hd*, and *He* are described in Table 2. The number after a designation of the series in the figure legend shows the value of CPSB for the considered contour. In the series *Ma*, we considered $L$ as the distance from the place of impact to the contour down in $y$-direction. In other series, we analysed the difference between maximum and minimum values of $x$ for the contour.

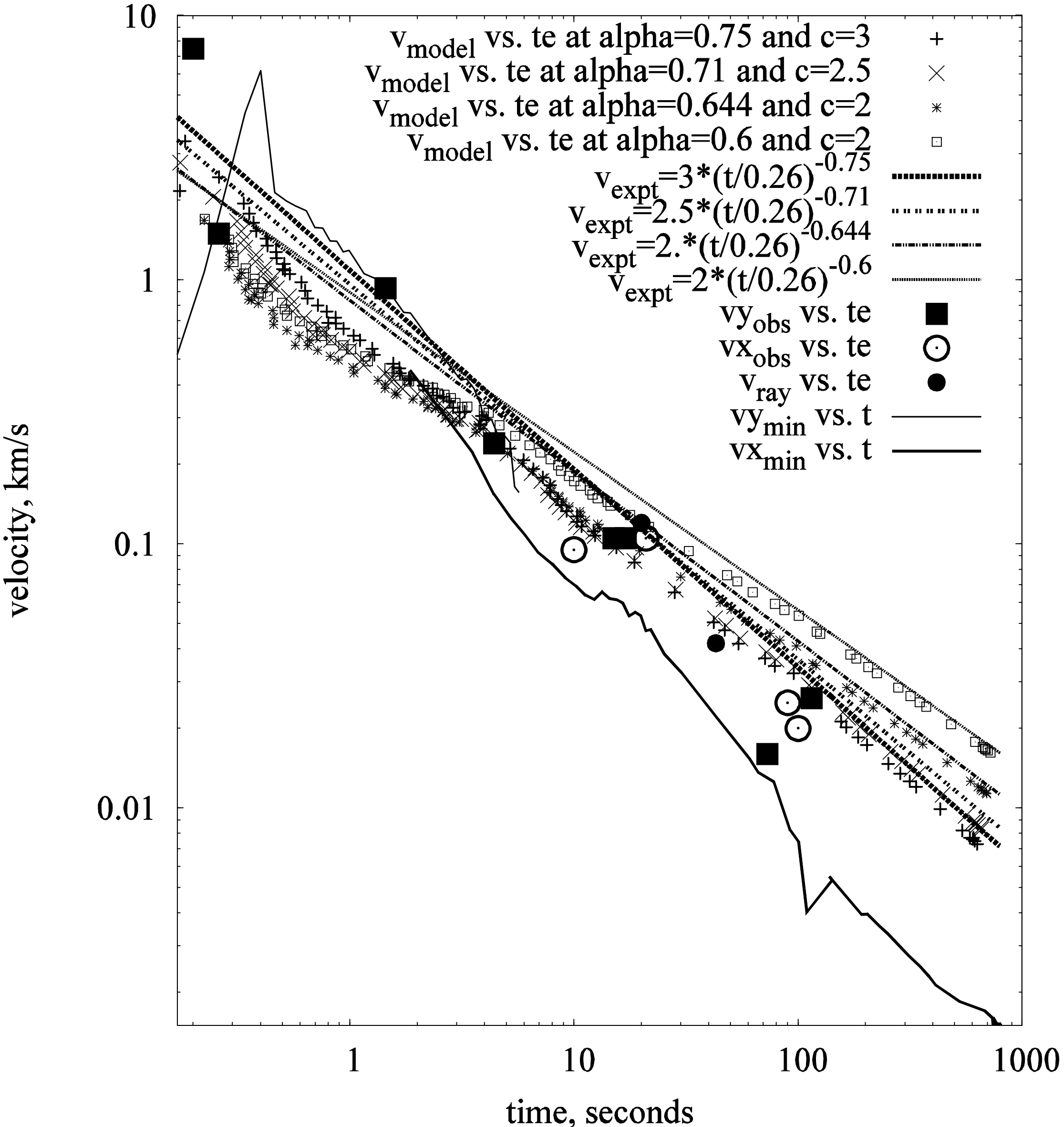

**Figure 11.** The typical projections $v_{\text{model}}$ of velocities (in km s$^{-1}$) onto the plane perpendicular to a line of sight at time $t_e$ of ejection for the model *VExp* for which velocities $v_{\text{model}}$ at $t_e$ are the same as velocities $v_{\text{expt}}=c\times(t/0.26)^{-\alpha}$ of particles at the edge of the bright region in an image made at time *t,* for four pairs of α and *c*. The distance from the place of ejection to the edge was used to find the dependence of *t* on $t_e$. The calculation of the size of the bright region is discussed in Sections 2.6 and 5.1. $vy_{\text{obs}}$ and $vx_{\text{obs}}$ are our estimates of the projection $v_p(t_e)$ of velocity that were based mainly on the analysis of minima and maxima of the plots presented in Fig. 10 for *y*-direction and *x*-direction, respectively. $v_{\text{ray}}$ shows two estimates of $v_p(t_e)$ obtained at analysis of rays of ejected material in Section 7. The values of $vy_{\text{min}}$ and $vx_{\text{min}}$ show the minimum velocities of particles needed to reach the edge of the bright region (in an image considered at time *t*) from the place of ejection moving in *y* or *x*-direction, respectively. Four curves $v_{\text{expt}}=c\times(t/0.26)^{-\alpha}$ with different values of α and *c* are also presented. Note that $v_{\text{expt}}$ and $v_{\text{model}}$ depend on *t* and $t_e$, respectively.

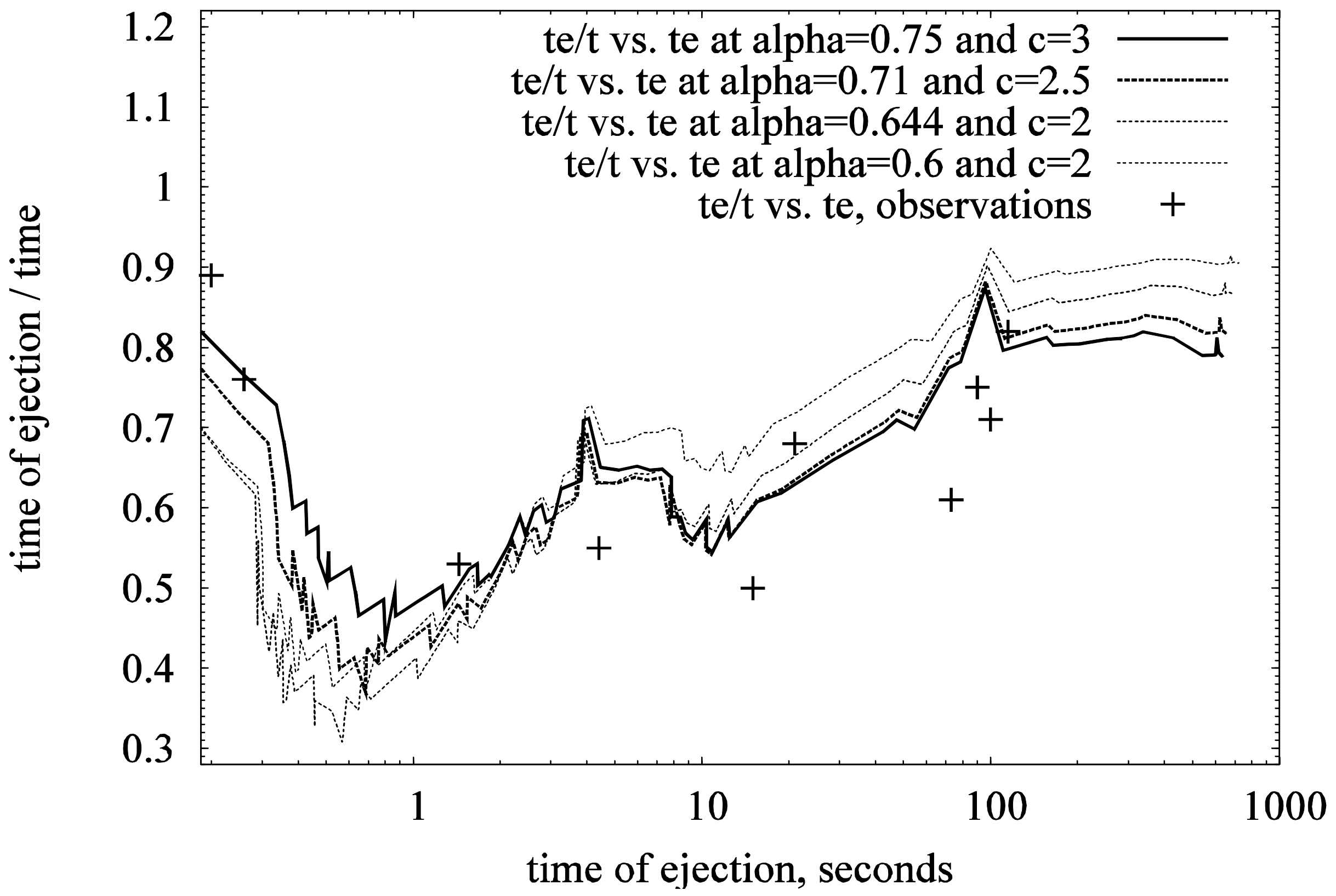

**Figure 12.** The ratio of time $t_e$ of ejection to time $t$ for four pairs of α and $c$. For the considered model *VExp*, velocities $v_{\text{model}}$ at $t_e$ are the same as velocities $v_{\text{expt}}=c\times(t/0.26)^{-\alpha}$ of particles located at the edge of the bright region in an image made at time $t$. Plusses "+" show the values of the ratio based on analysis of minima and maxima of the plots presented in Fig. 10.

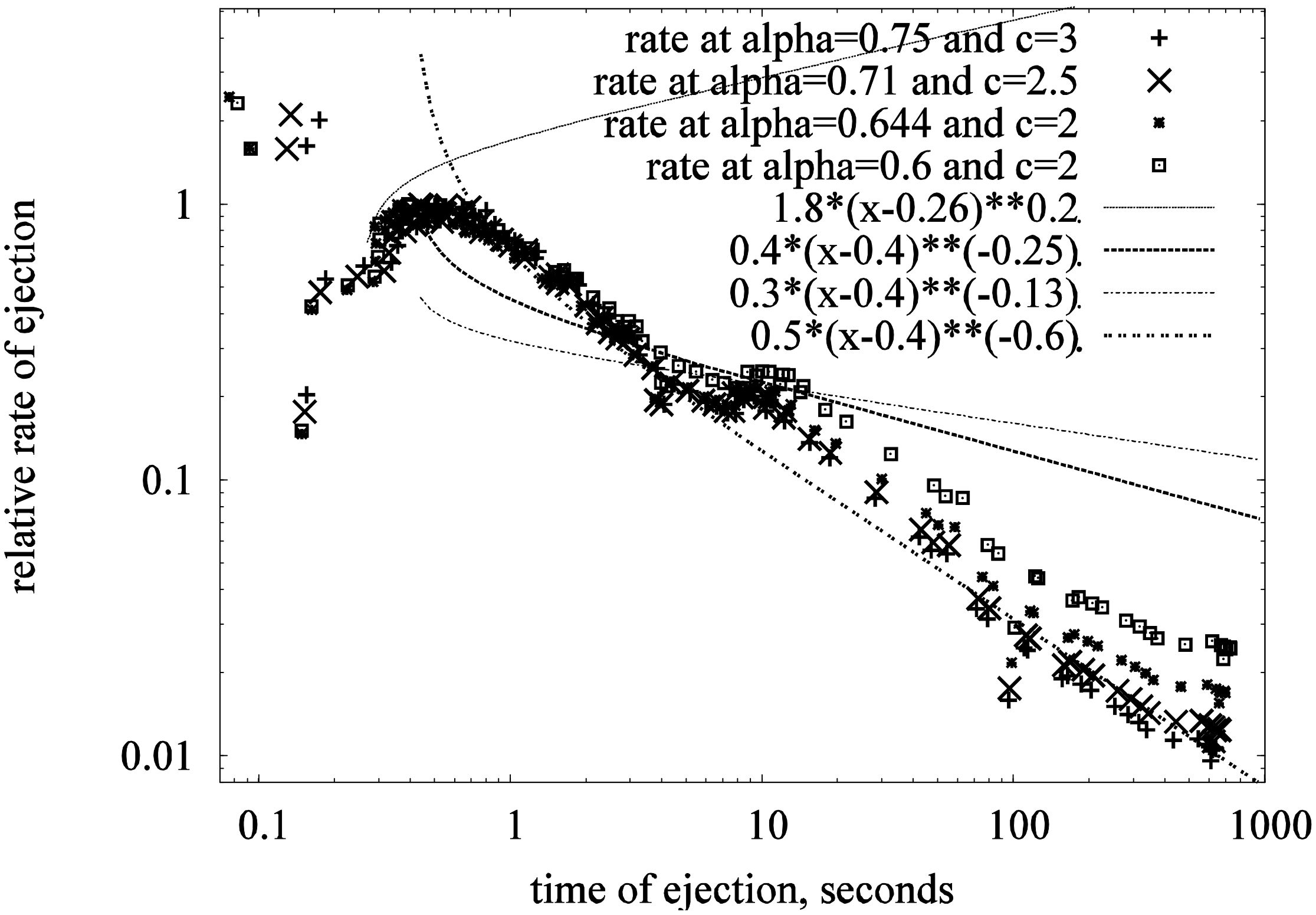

**Figure 13.** The relative rate $r_{te}$ of ejection of observed particles at different times $t_e$ of ejection for the model *VExp* for four pairs of α and *c*. The maximum rate at $t_e$>0.3 s is considered to be equal to 1. Four curves of the type $y=c_r\times(x-c_t)^\beta$ are also presented for comparison.

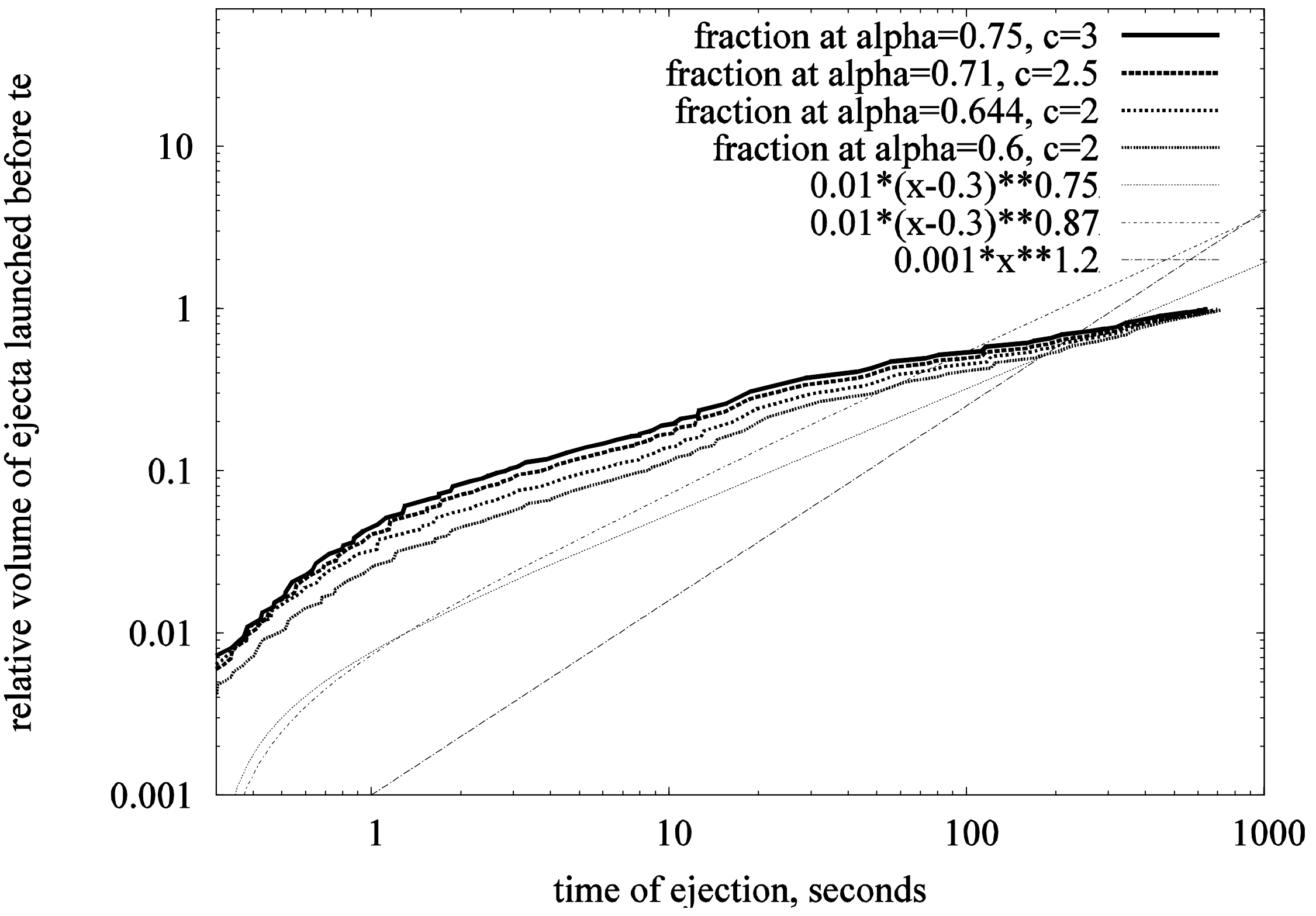

**Figure 14.** The relative volume $f_{et}$ of observed material ejected before time $t_e$ vs. $t_e$ for the model *VExp* for four pairs of α and *c*. $f_{et}$=1 for material ejected before the time $t_{e803}$ of ejection of particles located at the edge of the bright region in an image made at $t$=803 s. Three curves of the type $f_{et}=c_r\times(x\text{-}c_t)^{\beta}$ are also presented for comparison.

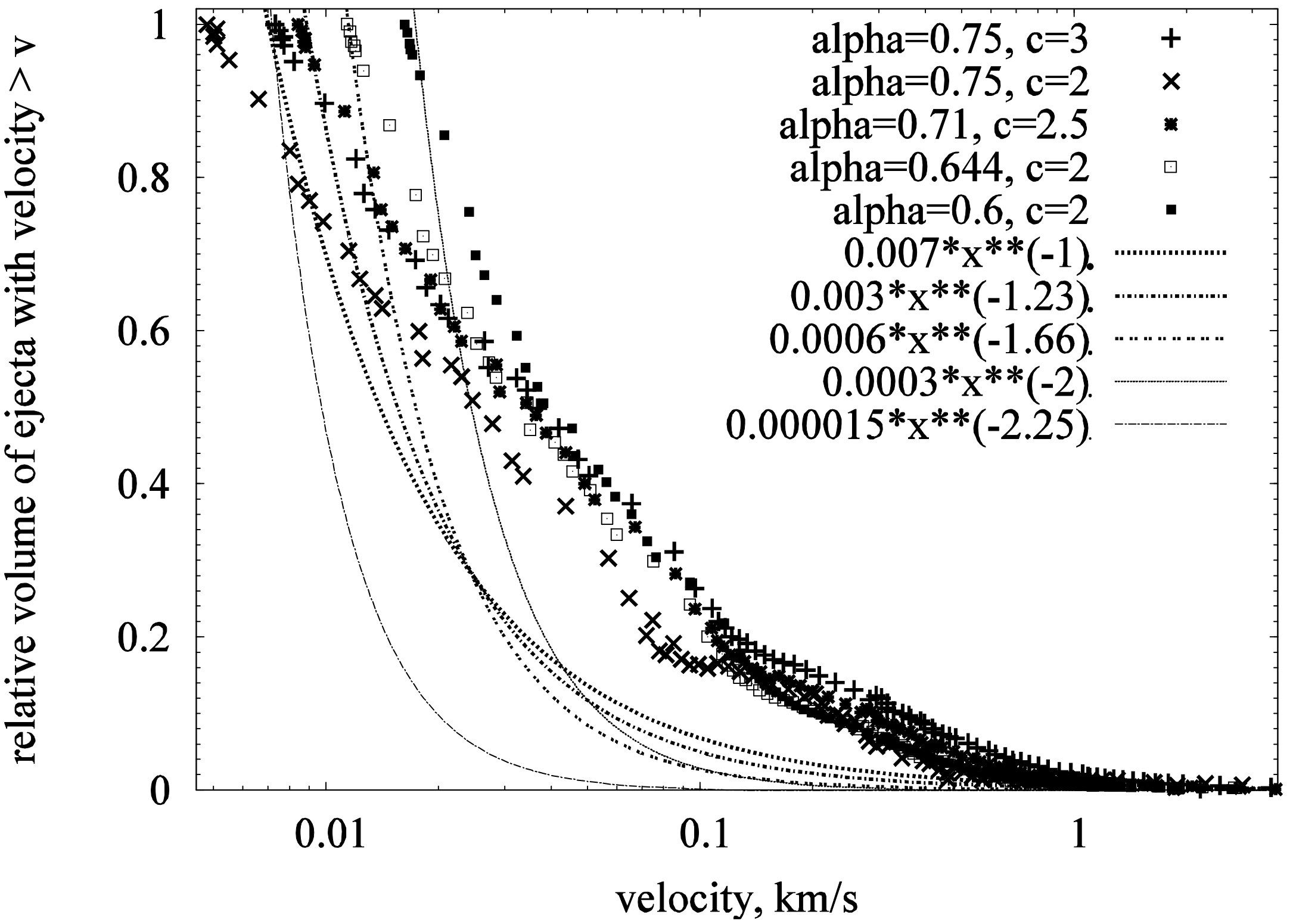

**Figure 15.** The relative volume $f_{ev}$ of observed material ejected with velocities greater than $v$ vs. $v$ for the model *VExp* for five pairs of α and $c$ (the values of $f_{ev}$ are presented by marks). $f_{ev}$=1 for material ejected before the time $t_{e803}$ of ejection of the particles located at the edge of the bright region in an image made at $t$=803 s. Five curves of the type $f_{ev}=c_r \times x^{\beta}$ are also presented for comparison.